\documentclass[a4paper,12pt]{article}
\usepackage{authblk,siunitx,hyperref,xcolor,ulem,graphicx,amsmath}
\usepackage[utf8]{inputenc}
\usepackage{csquotes}
\newcommand{\Herbi}{\textbf{Herbi}}
\newcommand{\Luna}{{\color{violet}{{Luna}}}}
\newcommand{\Jupi}{{\color{blue}{{{Jupi}}}}}
\newcommand{\Vita}{{\color{brown}{{{Vita}}}}}
\newcommand{\Mortis}{{\color{red}{{{Mortis}}}}}
\renewcommand{\H}[1]{\item[\textbf{Herbi}]#1}
\newcommand{\Lu}[1]{\item[\color{violet}{\textbf{Luna}}]{#1}}
\newcommand{\J}[1]{\item[\textcolor{blue}{\textbf{Jupi}}]{#1}}
\newcommand{\V}[1]{\item[\color{brown}{\textbf{Vita}}]{#1}}
\newcommand{\M}[1]{\item[\textcolor{red}{\textbf{Mortis}}]#1}
\newcommand{\LJ}[1]{\item[\textbf{\color{violet}{L}}\&\textbf{\color{blue}{J}}]{#1}}
\newcommand{\VM}[1]{\item[\textbf{\color{brown}{V}}\&\textbf{\color{red}{M}}]{#1}}
\newcommand{\R}[1]{\item[]\textit{(#1)}} 
\newcommand{\Regie}[1]{(\textit{#1})} 
\newcommand{\E}[1]{\textsc{#1}} 
\newcommand{\Musik}[1]{\textsf{#1 }} 

\newcommand{\Lyrics}[1]{{\color{teal}{#1}}}

\begin{document}
\title{\bf \texttt{\bf PLANETAMOS} \\
{\large A Physics Show Musical (Phyusical)}}
\author[1]{Lara Becker} 
\author[2]{Erik Busley} 
\author[3]{Jakob Dietl} 
\author[1,4]{Herbi K. Dreiner} 
\author[5]{Till Fohrmann} 
\author[3]{Kathrin Grunthal} 
\author[6]{Jana Heysel (n\'ee B\"urgers)}  
\author[1]{Finn Jaekel} 
\author[1]{Kristoffer Kerkhof}
\author[1]{Michael Kortmann}
\author[1]{Barbara Leibrock} 
\author[6]{Viola Middelhauve}  
\author[2]{Steffi Moll} 
\author[3]{David Ohse}
\author[7]{Johann Ostmeyer} 
\author[1]{Laura Rodr\'iguez G\'omez} 
\author[3,8]{Christoph Sch\"urmann} 
\author[6]{Anne Stockhausen}
\author[1]{Joshua Streichhahn}
\author[7]{Carsten Urbach} 
\author[7]{Heinrich von Campe} 
\author[1]{Alexandra Wald} 
\author[6]{Laura Weber} 
\author[1]{Inga Woeste} 
\affil[1]{\small Physikalisches Institut, Universit\"at Bonn, Nu{\ss}allee 12, 53115 Bonn, Germany}
\affil[2]{\small Institut f\"ur Angewandte Physik, Universit\"at Bonn, Wegelerstra{\ss}e 8, 53115 Bonn, Germany}
\affil[3]{\small Argelander-Institut f\"ur Astronomie, Universit\"at Bonn, Auf dem H\"ugel 71, 53121 Bonn,  Germany}
\affil[4]{\small Bethe Center for Theoretical Physics, Universit\"at Bonn, Nu{\ss}allee 12, 53115 Bonn, Germany}
\affil[5]{\small Institut f\"ur Geowissenschaften und Meteorologie, Universit\"at Bonn, Auf dem H\"ugel 20, 53121 Bonn,  Germany}
\affil[6]{\small Institut f\"ur Physikalische und Theoretische Chemie, Universit\"at Bonn, Wegelerstr. 12, 53115 Bonn, Germany}
\affil[7]{\small HISKP, Universit\"at Bonn, Nu{\ss}alle 14-16, 53115 Bonn, Germany}
\affil[8]{\small Max-Planck-Institut f\"ur Radioastronomie, Auf dem H\"ugel 69, 53121 Bonn,  Germany}
\maketitle
\begin{abstract}
We present a physics show musical with live physics experiments and live performed songs with live orchestral accompaniment. 
The musical was first put on stage in German at the Physikalische Institut, University of Bonn, on the 24th of March,  2019. Here 
we present the original German script as well as an English translation, including a translation of the songs. We also give brief 
descriptions of all the experiments employed, with photos as well as how we present the experiments.

The musical tells the story of a couple, Life and Death, nicknamed Vita and Mortis, who want to buy a planet where they can settle 
down and raise a family. They arrive at the store Planetamos, run by the manager Luna Callisto and the scientist Jupi Mercury. The 
latter present to Vita and Mortis what is required for life on a planet as well as all of the extra features for sale, to make it that very 
special place for their family. Through Vita and Mortis the audience learns about the physical requirements for life on a planet orbiting
a star.

The show involves 14 live physics experiments. The four actors sing 7 songs, which are to well-known tunes with new 
lyrics, in German. During the show the singers are accompanied by a small live orchestra, consisting of a 
flute, an oboe, a piano, a violin, a cello, a harp and percussion. The show was developed by the University of Bonn physics students, who 
have all been very active in the University of Bonn Physics Show (Physikshow der Universit\"at Bonn) the past few years.

To the best of our knowledge this is the first physics show musical with live experiments and live song and orchestra.
\end{abstract}

\tableofcontents

\section{Introduction}

\subsection{Physics Outreach}

There is a long tradition of presenting physics to a broader public, often now called outreach. For example in 1586, Simon 
Stevin (Stevinus) and Jan Cornets de Groot dropped various masses from the church tower in Delft, Netherlands,
\cite{Asimov}, an experiment often attributed to and possibly also performed by Galileo Galilei at the leaning tower of Pisa 
at around the same time \cite{Drake}. In 1656 and the following years the mayor of Magdeburg, Germany, Otto von Guericke, 
the inventor of the piston vacuum pump, publicly performed his famous experiment with several horses trying to pull apart 
two copper hemispheres where the air between them had been pumped out \cite{Guericke,guericke-vacuum-pump}. He also 
invented the now very popular public demonstration experiment, the vacuum canon, see for example chapter 29 of the third 
book of Ref.~\cite{Guericke}. Since 1851 there has been a Foucault pendulum in the Panth\'eon in Paris. In Britain the Royal 
Institution was founded in 1799 with the specific purpose of furthering scientific education and research. In 1825 at the the 
Royal Institution, the Christmas lectures series was founded by Michael 
Faraday \cite{Wiki-RI}, where he himself gave many famous lectures. The British Science Association (BSA) was founded in 
1831 with the goal of promoting and developing science. The South Kensington Science Museum in London was founded in 
1857. The Deutsches Museum in M\"unchen was founded in 1903, with the goal of communicating the science as directly as 
possible involving original objects. The Palais de la D\'ecouverte in  Paris was founded in 1937. Inspired by these, the 
Exploratorium in San Francisco was founded in 1969 by Frank Oppenheimer, with a strong emphasis on hands-on experiments 
and a direct link between visitors and the workshop developing new exhibit pieces. Here in Bonn a branch of the Deutsche Museum
was founded in 1995.

\medskip

Today outreach has become a feature of many science activities. There are now many more science museums. Large research 
laboratories such as DESY, CERN, and Fermilab have their own outreach divisions. In 2000, CERN, ESA, and ESO together
launched the activity ``Physics on Stage" with live physics performances  \cite{physics-on-stage}. The European Space Agency 
(ESA) has helped launch several national education and outreach centers (ESERO), see for example Ref.~\cite{Esero-Germany}. 
We are co-founders of ESERO Germany. There is also a European organization for science shows, Euro Science Fun \cite{esf}, 
of which we are a member and a co-founder (2005). Outreach has also become a central part of research funding, see for example 
\cite{DFG-Outreach}. Thus today there are very many aspects of science communication and outreach, including of course science 
entertainment. A well-known example in Germany are the Physikanten from Dortmund, founded in 2000 \cite{Physikanten}. The 
Bonn Physics Show was founded in Dec. 2001, see Section~\ref{sec:Bonn-Physics-Show}.

\subsection{Music and Physics Presentations}

As soon as science borders on entertainment, music can not be far. We can not even attempt to give a full history of science
entertainment and music, that would be a separate project. All the same we want to give a set of examples, of what we have found, 
to contrast it also with our own work. We believe we have developed the first physics musical with live music {\it and} with live 
experiments. We would however love to hear of any related activity and are happy to include it here in an update of this document.

\paragraph{Recent Science Songs:} These days there are numerous videos for example on YouTube with science songs, \textit{i.e.} a song is 
sung with science based lyrics. Well-known is for example the channel {\it A Capella Science} \cite{ACapellaScience} by Tim Blais, 
whose version of Queen's \textit{Bohemian Rhapsody} on string theory has received much deserved attention \cite{Bohemian-rhapsody}, 
as has his video about atomic and molecular physics based on the Ed Sheeran song \textit{Shape of You} \cite{Shape-of-You}. Another 
example is the group AsapSCIENCE \cite{AsapSCIENCE}, who sometimes also put out science songs such as this one on Red Dwarf 
Stars based on the song \textit{A Star is Born} \cite{RedDwarfStar}.  Lynda Williams, as {\it The Physics Chanteuse} is a ``multi-media 
musical solo-show on the nuclear state of the world". It calls itself a ``Cosmic Cabaret" and has been written up in the NY Times 
\cite{PhysicsChanteuse-NYT}. She even has a song about supersymmetry, see Ref.~\cite{PhysicsChanteuse-susy}

\paragraph{Earlier Science Songs:} 
Treating physics topics in songs goes much further back. A non-exhaustive historical collection of physics song sources can be found 
in Ref.~\cite{History-Source}. This includes songs sung at the Cavendish Laboratory, as early as the late 19th century, but also from
1920 the song entitled $h\nu$ by Gilbert Stead \cite{h-nu}, on early quantum mechanics, with very clever and rhyming lyrics.

In April 1932 the physicists at the Niels Bohr Institute, including Heisenberg on the piano, put on a musical version of Faust \cite{Farmelo-Dirac},
written mainly by Max Delbr\"uck. The main theme of the performance, as recounted by Farmelo in his book, was the story of the neutrino and how 
Pauli tried to persuade Ehrenfest of its existence. In the story God was represented (not played) by Bohr, Pauli by Mephistopheles, Ehrenfest by 
Faust and the neutrino by Gretchen. The three archangels of the prologue are represented by the astrophysicists Eddington, Jeans, and Milne.
This seems to have been a full fledged musical production with a story-line, just, we assume, no live experiments.

In 2018, the journal \textit{Physics Today} has published an article: ``Physics: The Musical?" \cite{PT-musical}. It is about Arthur Roberts, 
a physics professor at the State University of Iowa (now the University of Iowa). In the 1940s, while at MIT he composed several songs,
including lyrics. In 1947 when he moved to Iowa he and his new colleagues recorded several songs about physics, and remarkably about
funding for physics!

Tom Lehrer is well-known for many songs on maths and science related subjects. He is also the author of {\it The Physical Revue}, a 
musical revue performed at the Harvard Physics Department in 1951 and 1952  \cite{PhysicalRevue}. It was recorded by later Nobel 
laureate Norman Ramsay.

\paragraph{Summary:} This was just a sketch of physics related song activities. The point of this article is to present the Bonn Physics 
Show Musical: Planetamos. This combines a story line, songs mostly on physics, as well as live demonstration experiments, as are 
common in physics shows. We are not aware of any other group having done or doing this. 

\subsection{Outline}
This paper is organized as follows. In Section~\ref{sec:Bonn-Physics-Show} we present the Bonn Physics Show, which has founded
in 2001. In Section~\ref{sec:origins-planetamos}, we  discuss how we developed the Physics Show Musical: Planetamos. In 
Section~\ref{sec:script} we present an English translation of the full original script of the show, as performed in Bonn in the Wolfgang-Paul
lecture hall in July 2019. This
includes all the lyrics, as well as a list of the music in each case. In Appendix ~\ref{app:skript-German}, we present the full original
script in German. In Appendix~\ref{app:experimente} we  give details of the 14 demonstration experiments  used in the show,
as well as two related experiments, which did not make it into the show.

\section{The Bonn Physics Show}
\label{sec:Bonn-Physics-Show}
\subsection{The Beginning}

The Bonn Physics Show \cite{BonnPhysikshow} was founded  20 years ago, in  Dec.~2001 by Herbi Dreiner and Michael 
Kortmann at the Physikalische Institut, University of Bonn. From the start it has been first and foremost an education project 
directed at the University of Bonn physics students.  The goal is to improve the outreach capabilities of the students, while 
producing excellent outreach. The university physics students are given the opportunity to develop and present an on-stage 
live 2hr physics show. They can thus apply and communicate their new physics knowledge at a very early stage of their studies, 
all the while developing new outreach skills. 

In Dec.~2001, we recruited the first group of 25 students, who were at the beginning of their second year of physics studies 
in the 5-year Diplom Physik program.\footnote{Since 2006 in Bonn, we have phased in the 3-year Bachelor and 2-year Masters 
programs according to the Bologna process. This has significantly increased the course work of the students off-term
with laboratory courses. Therefore we have had to shift the first show to end of the first year of the Bachelor program when the 
students are unfortunately less experienced in physics.} 
After a preparation of over 11 months, in Nov. 2002 the first two performances of a 2 hour show were given in the Wolfgang-Paul 
Lecture Hall. It seats 550 people, but over 700 showed up! We introduced a booking system subsequently. The show was repeated 
with 4 performances in March 2003. In the following years a new show was developed, each year by a new class of about 20 second 
year students, with 3 performances in September and March, respectively. These shows mainly consist of classical physics, which 
are embedded in a storyline, basically in a play. Recent examples have been stories inspired by ``Sherlock Holmes", ``The Wizard of 
Oz", ``A Night in the Museum", etc.  The early years of the Bonn Physics Show are described in Ref.~\cite{Dreiner:2007cn}. 

\subsection{Shows on Modern Physics and other Show Activities}

We have been training about 20-25 new students in outreach every year, for 20 years now. In Germany physics students often stay 
at the same university for most of their education. Of the Physics Show students many have stayed in Bonn for several years and 
have been interested in continuing with show activity. Thus we continue to have a large pool of eager and experienced 
students to recruit from for scientifically and logistically more advanced Physics Show activities. 

\subsubsection{First Particle Physics Show 2004: 50 years of CERN}

In 2004, to mark the 50th anniversary of CERN, Herbi Dreiner, Michael Kortmann, Michael Kobel and Ewald Paul developed and 
performed for the first time a 2~hour show on elementary particle physics in Bonn. This was our first show on modern physics, and 
several new demonstration experiments were developed and built. We believe this was a first of its kind.\footnote{Please let us 
know if you are aware of any other related activity, earlier or later.} In 2008 a group in the physics department at Oxford 
University around Suzanne Sheehy developed a show called ``Accelerate", on particle physics, focussing mainly on how accelerators 
work \cite{bib:accelerate}.  In 2006 the group Euro Physics Fun (now Euro Science Fun) was founded, with activities in many different 
countries and  covering a wide range of topics across all fields of science \cite{ESF:2006abc}. 

In 2006 we started touring outside of Bonn with our shows. The first trip was to the Deutsche Museum, M\"unchen. See 
Ref.~\cite{Dreiner:2007xx} for a description of an early trip to the Deutsche Museum, M\"unchen and on the difficulties in touring.

\subsubsection{Annual Christmas Show}
Since 2007 we have been putting on an annual Christmas Show: The Physical Advent Calendar. The story line involves a physics
student, who through all the course work, has forgotten to open the doors on her Advent Calendar. Now shortly before Christmas and
with an exam coming up, she opens the first door. Miraculously Christmas elves come to life on-stage and for each door they bring a 
new experiment, helping her to understand her course work. This show has become a huge hit on campus, as well as on the road. We 
have performed it at  DESY Hamburg in December 2017 (and got stuck in quite a snow storm on the way back). We were scheduled to
travel to Durham, UK, to perform the show in English, in Dec.~2020. However the pandemic intervened. We still hope to take up this
invitation as soon as possible. Instead we made 25 YouTube videos, an intro and then one for each 
door~\cite{Bonn-YouTube}. The videos are in German, with English and Spanish subtitles.

\subsubsection{Second Show on Particle Physics}

In 2008 the core group of the Bonn Physics Show revitalized the Particle Physics Show, adding new demonstration experiments and 
making it more a show instead of a lecture with live experiments. With this show we first travelled to Berlin where we performed at the 
national exhibit, {\it Weltmaschine}, to mark the inauguration of the LHC accelerator. The exhibit and  our trip were funded by the 
BMBF (the German ministry for science and education) and was held in a brand new subway (metro) station in downtown Berlin, right 
in front of the Bundeskanzleramt, the office of the German Bundeskanzler, at the time Angela Merkel. We  travelled with this show
to the Deutsches Museum M{\"u}nchen (3 perf.; March 2009), DESY Hamburg (3 performances; Sept. 2009), and to Heidelberg 
University (3 perf.; Dec. 2009). In Sept. 2010, we took the show to CERN and performed it 3x in the Globe in French(!).\footnote{Here 
it was essential that two of the physics students spoke fluent French, after extended stays in France on the Erasmus program.} Finally
in Nov. 2010, we  performed the show in Bonn at the physics department.

\subsubsection{Third, Extended Show on Particle Physics: ``What's (the) Matter?"}
In 2014, supported by funding under the DFG grant CRC SFB-TR-110, we developed our current full-fledged particle physics show, 
which is again a play. See Ref.~\cite{Dreiner:2016enl} for a detailed description of the show.  With this show we travelled to Oxford, 
London, UK (2014); Padua, Trieste, Italy (2015); Copenhagen, Odense, Denmark (2016); Valencia, Barcelona, Spain (2017); Lisbon, 
Portugal, Madrid, Spain (2018); Mainz, Germany (2018); Amsterdam, Netherlands;  and Aachen, and Freiburg, Germany (2019).  See 
Ref.~\cite{Guardian-UK} for a description of the on-tour adventure.

\subsubsection{Other Short Shows}
We have developed various shorter formats on a variety of topics, with which we have travelled mainly throughout Germany. 
All this is to say that we have extensive experience in developing and performing physics shows at various levels. In particular, 
we have (had) students, who worked with the physics show for many years, and have a broad range of expertise. Furthermore
since Dec. 2018 we have received funding through ESERO. All this has led to the following Physics Show Musical: Planetamos,
which we describe in detail below.

\subsection{Awards for the Bonn Physics Show}
In 2006 the Physics Show Students were awarded the Alumni Prize of Bonn University.

In 2009 Michael Kortmann and Herbi Dreiner were awarded the European Physical Society High Energy Physics Outreach Prize
for the first particle physics show.

In 2020 the Physics Show Students were awarded the prize of the Bonn University Society for the ``best student initiative".
This was for the musical we present here as well as the set of 25 YouTube videos on the Physics Advent Calendar.

At the \textit{Science me} event in Geneva, Switzerland the Bonn physics show placed 3rd in 2016 and won in 2018.
At the \textit{Science me} event in Hannover, Germany, the Bonn Physics Show won the Special Prize in 2019.

\section{The Origins of the Bonn Physics Show Musical}
\label{sec:origins-planetamos}

As mentioned above, the idea of the Bonn Physics Show is to give the students a framework in which they themselves 
can develop and present physics shows on-stage in front of a large audience. The staff members Herbi Dreiner and 
Michael Kortmann encourage the students to develop new ideas and projects themselves. Over the years, on our physics 
show trips the students often got together for a sing along in the evening. Several have experience in choirs and can sing 
quite well. Furthermore this was a fun and cheap form of entertainment in the hotels, or after the bars had closed. Slowly 
the idea began to form to develop a Physics Show Musical. In 2018 they formed a core group to push the idea forward. 

A trademark of the Bonn physics shows is to have an overarching storyline for the entire show. Relatively early, one of the 
students specializing in astrophysics suggested a planetary store with four protagonists: two sales people and a couple as
customers. The science theme would be: ``what is needed for life on a planet?". The next step was to develop and build
demonstration experiments and bring them in a proper order coordinated with the logic of the storyline. 

After considering singing the songs to play-back accompaniment, the students were able to organize fellow physics students 
(not necessarily Physics \textit{Show} students), with enough instrumental experience to comprise a small orchestra. In order 
to discipline themselves, a deadline was set for a performance of a ``beta-version": March~2019.\footnote{Most of February 
and all of March are off-term in Germany.} This first version of the musical had 3 songs: the duet by Jupi \& Luna, the solo by 
Vita and the solo by Mortis.\footnote{See Section~\ref{sec:script} for the full lyrics of the songs translated into English and 
Appendix~\ref{app:skript-German} for the original lyrics in German.} The song ``Circle of Life" was played in the original 
version from a CD. The performance was for family and friends only, and was a success in the sense that the songs went well, 
as did the combination with experiments. 

A public performance date was set for July 2019. In the four months the entire production was further developed. The text was 
edited, and some experiments were  improved or swapped for new ones. Most importantly 4 new songs were added: the solo 
by Jupi, the duet by Vita \& Mortis, the solo by  Luna, and as the grand finale: the ``Circle of Life", now performed live, but a 
cappella. This was a huge success! 

Three further performances were planned for February 29th and March 1st, 2020, and were completely booked (550 people per 
show), but were cancelled at the last moment, as on the morning of February 29th, the first Covid-19 case in Bonn was reported. 
All the same an instrumental accompaniment had been written for ``Circle of Life", and an oboe as well as a harp were added to 
the orchestra and the arrangements were accordingly modified.

A few words on the participants. All on-stage performers are physics students, most of the orchestra was as well. Of the singers 
Jana Heysel (Luna Callisto), David Ohse (Jupi Mercury) and Johann Ostmeyer (Mortis) have sung in choirs. Johann has had 
extensive vocal training and has performed as a soloist. Laura Rodriguez Gomez has always enjoyed singing but was never in a 
choir. Jana, David and Laura all play an instrument. Jana Heysel, besides recently getting her Ph.D. in physics, also has a Bachelor's 
degree in musicology and music education. Over the years she arranged music in various smaller settings.

During the rehearsals the students received coaching and directorial support from two professionals: Karolin Biewald and Babette 
D\"ormer. The musical arrangements are discussed in Section~\ref{sec:musical-arrangements}.  A detailed list of credits is given in 
Section~\ref{sec:credits}.


\section{The Script (English)}
\label{sec:script}

The original script was written in German and is given in Appendix~\ref{app:skript-German}. Here we present our own translation
in English. As with all Bonn physics shows, the script evolves with each performance. This is the status of
Feb. 29th, 2020.

\subsection{Prolog}
\label{sec:prolog}
\textit{Describe stage, with objects.}
\begin{itemize}
	\R{Enter \Herbi\ front stage.}
	\H{\Regie{Using a hand microphone, directly addressing the audience.} Welcome, welcome! My name is Herbi Dreiner
	and I am a professor for \ldots theoretical elementary particle physics. [...] Let's go shopping for our own planet ...
			}
	\R{Enter Herbi stage left.}
	\R{Enter \Luna\ and \Jupi\ stage right.}
	\begin{figure}[t]
	\centering
	\includegraphics[width=0.75\textwidth]{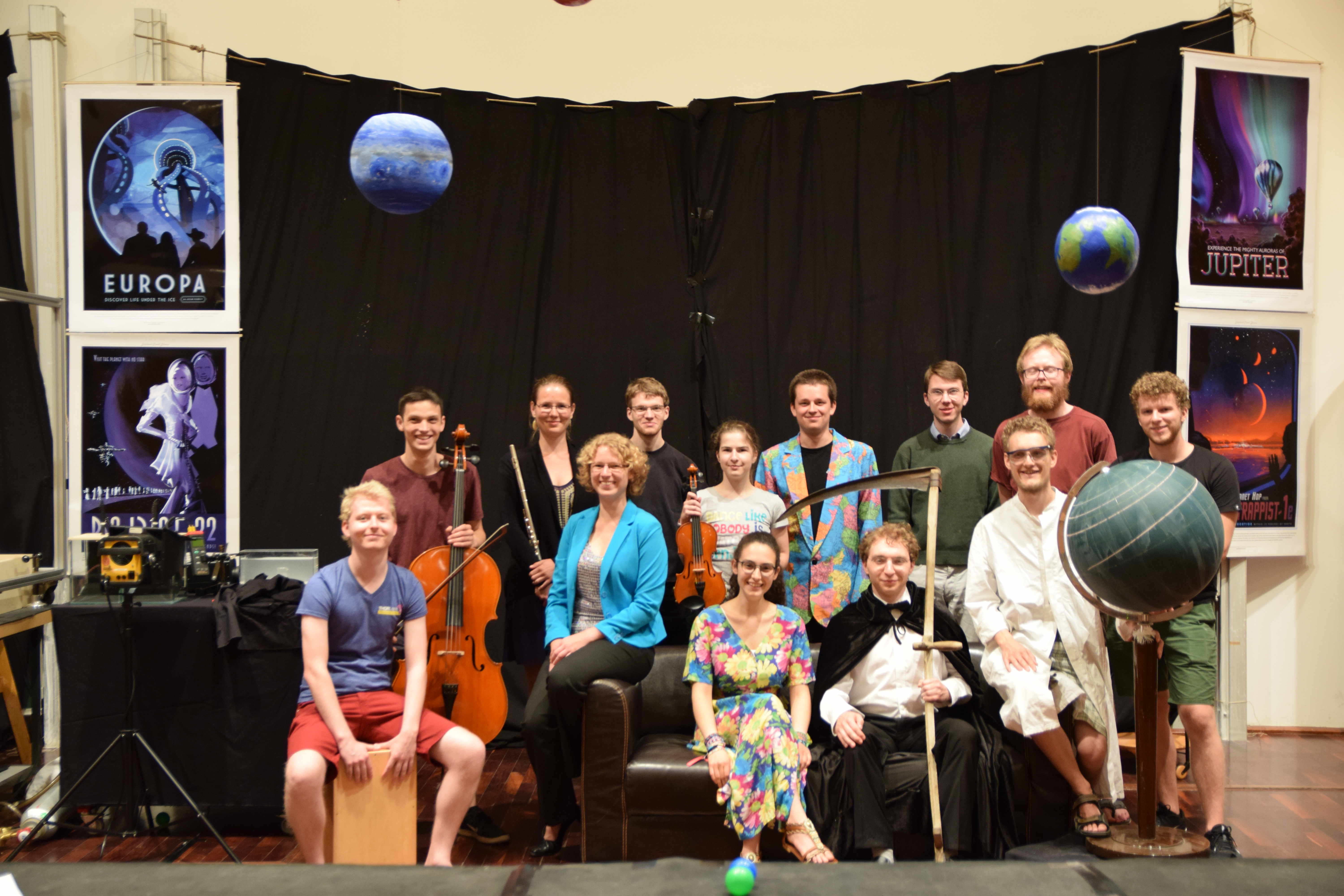}
	\caption{\small Actors and orchestra for the show in July 2019. On the sofa left to right the actors: Jana Heysel 	(n\'ee B\"urgers) 
	as Luna Callisto (wearing a blue jacket), Laura Rodr\'iguez G\'omez as Vita (wearing a colorful dress), Johann Ostmeyer as 
	Mortis (holding a sythe), and David Ohse as Jupi Mercury (wearing a lab coat). The others, the orchestra as well as stage technicians (camera,
	lighting, sound, projected presentations) from left to right: Erik Busley, Jakob Dietl, Viola Middelhauve, Peter Rosinsky, Lara Becker, 
	Joshua Streichhahn, Heinrich von Campe, Christoph Sch\"urmann and Till Fohrmann.}
	\end{figure}
	\Lu{\Regie{Does not notice \Herbi, addressing \Jupi.} \dots How could you forget where you put the comets?!?!
	}
	\J{\Regie{Follows her.} Hmm, yesterday I still knew where they were.
	}
	\Lu{\Regie{Searching the stage.} Did you already look by the asteroids?
	}
	\J{They weren't there.
	}
	\begin{figure}[t]
	\centering
	\includegraphics[width=0.75\textwidth]{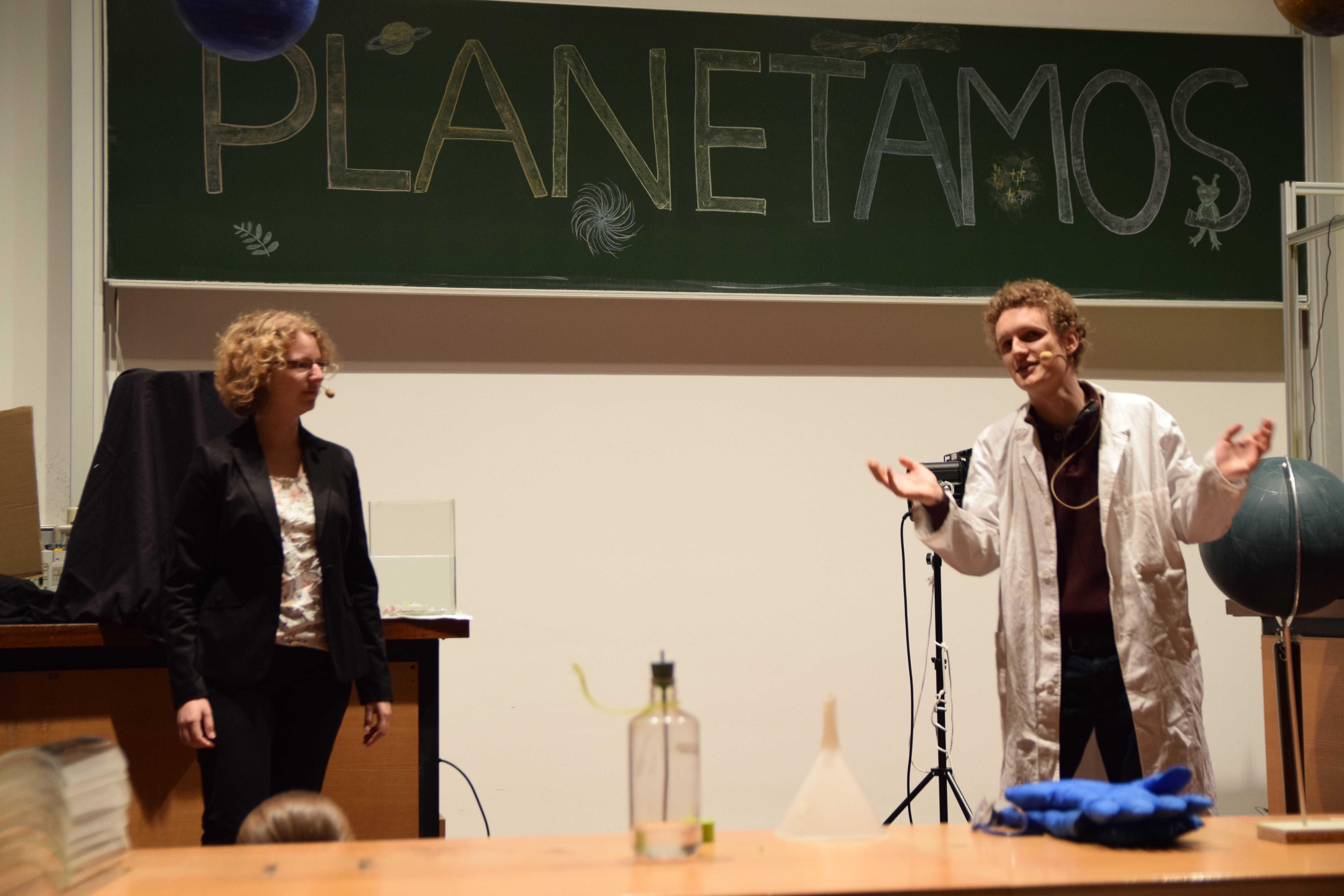}
	\caption{\small Luna  Callisto (left, Jana Heysel) and Jupi Mercury (right, David Ohse).}
	\end{figure}
	\Lu{And with the meteorites?
	}
	\J{No, not there either.
	}
	\Lu{With the shooting stars?
	}
	\J{\Regie{Slaps his forehead.}  Oh. Yes. Now I remember, the fog is lifting from my mind. I filed
	the comets with the nebulae.  \Regie{Remains stage front right by the catalog.}
	}
	\Lu{\Regie{To-once-work-with-professionals:}  Come on, Jupi, you have to enter it in the catalog, otherwise they'll be lost!
	\Regie{Sees \Herbi, walks over to him and welcomes him mid-stage.} Oh, hello. Can I help you?
	}
	\J{\Regie{Retrieves the Messier-catalog and enters the comes.}
	}
	\H{My name is Herbi. I would like to shop here.
	}
	\Lu{My name is Luna Callisto. \Regie{Shakes his hand.} Are you looking for something specific?
	}
	\H{I would like to buy a planet. My home planet is unfortunately a bit on the wrong track 
		\dots
	}
	\Lu{What were you thinking of, ``Herbi"?
	}
	\H{Well. You know, I  \dots
	}
	\Lu{A gas giant, or an ice giant, or maybe even a rocky planet?
	 And {\it (ambiguously)} do you prefer it \textit{hot} \ldots or \textit{cold}?
	}
	\H{\Regie{Embarrassed.} Uh. Well it is a bit hot back home. Hmm, but that's actually not why I am here.  
	So, it's a bit embarrassing, but you're an expert, no? So, \ldots do you also sell disc worlds? Or hollow planets?
		}
	\Lu{\Regie{Laughs out loud, then turns serious.} No! Do you even have enough change?
	}
	\H{Well, I'm only a professor \dots
	}
	\Lu{\Regie{Places her arm on \textbf{Herbi}'s shoulder and gently escorts him to stage left, towards  
	his seat, but herself remains on-stage.} Have a seat, and relax. Here at \texttt{PLANETAMOS} you are in good hands.
	}
	\Lu{It's maybe best if we first show you our catalogues, maybe you will find something appropriate, ``Herbi".  My colleague
	Jupi Mercury will be delighted to show you our portfolio.
	}
	\J{\Regie{Excited.} Oh yes, our product variety.  I can sing you a thing or two about it!  
	\Regie{Moves center stage.}
	}
	\R{\Herbi\ exit.}
	
	\vspace{0.7cm}
	
	\Musik{Orchestra music: \parbox[t]{9.2cm}{"A Whole New World" from the movie "Aladdin" (Music by A. Menken)}}
	
	\medskip

	\J{{\color{teal}  Tell me, what's the world worth?\\
	We create what you long for:\\
	Mountains, valleys and oceans\\
	in a fancy way combined.\\\\
	We can show you the world.\\
	Take a look, get excited:\\
	We provide you with planets,\\
	in a modern way designed.\\\\
	A whole new world,\\
	a precious \texttt{PLANETAMOS} brand,\\
	will be the perfect place,\\
	the living space\\
	for all your future beings. }
	}
	\Lu{{\color{teal} A whole new world\\
	A dazzling place you never knew,\\
	can be your living space,
	a wondrous place,
	just consider what gives rise to live.}
	}
	\J{{\color{teal} There are just a few necessities...}
	}
	\Lu{{\color{teal} First you orbit a star.\\
	For a pleasant existence,\\
	you should keep the right distance,\\
	not too close and not too far.\\
	An iron core...}
	}
	\J{{\color{teal} ...provides a magnetic field.}
	}
	\Lu{{\color{teal} An atmosphere around your world.}
	}
	\J{{\color{teal} ...creates a nice feel-good climate.}
	}
	\Lu{{\color{teal} And for the genesis\\
	of life, it is\\
	essential to have water in the world.}
	}
	\J{{\color{teal} Some fireworks...}
	}
	\Lu{{\color{teal} ... from volcanos for free.}
	}
	\J{{\color{teal} ... with ashes, lava, sulfur, smoke.}
	}
	\Lu{{\color{teal} ... they are spewing around you.}
	}
	\LJ{{\color{teal} If you're in buying spree,\\
	you'll get for free\\
	a full-moon that will orbit 'round your world.}
	}
	\J{{\color{teal} We build a world...}
	}
	\Lu{{\color{teal} We build a world...}
	}
	\J{{\color{teal} ...so brave and new...}
	}
	\Lu{{\color{teal} ...so brave and new...}
	}
	\J{{\color{teal} ...so full of grace...}
	}
	\Lu{{\color{teal} ...a wondrous place...}
	}
	\LJ{{\color{teal} ...just made for you.}
	}
	\J{\Regie{Throws confetti and stretches his arms upwards.}
	}
	\Lu{\Regie{crosses her arms.}
	}
	\R{Freeze.}
\end{itemize}

\subsection{Celestial Mechanics}
\label{sec:himmelsmechanik-e}
\begin{itemize}
	\R{Enter \Vita\ stage left. \Vita\ is wearing  a hat, sun glasses and a hand bag.  Door bell sound from the piano as she enters
	store. \Luna\ and \Jupi\ reawaken after freeze.}
	\V{\Regie{Lively} Hello? Good day to you! Is this \texttt{PLANETAMOS}, yes? \Regie{Hangs her hat on the clothes rack 
	at the entrance.}}
	\J{\Regie{In thought, looking towards the audience.} Oh. Ah. Yes! \Regie{Turns to his left to the globe, tinkers with it, 
	washes the surface, etc. ... \footnote{The surface of the globe is like a blackboard, easy to draw on with chalk.}}}
	\Lu{Welcome to \texttt{PLANETAMOS}, THE address for all your planetary needs. \Regie{Walks over to \Vita.} 
	My name is Luna Callisto. \Regie{Offers her hand.}}
	\V{\Regie{Struts to \Luna\ and enthusiastically shakes her hand.} Glad to meet you!  I am Life. But you can call
	me Vita. \Regie{Places her handbag on a table. Removes her shades and searches for the glasses case in her bag.}
	}\Lu{Wow, that I could live to see this. Wonderful to have so much \textit{life} in our little planet store. \Regie{Short laugh.}
	My colleague Jupi Mercury and I are happy to help you.}
	\J{\Regie{Raises his head, short wave and nod, with a screw driver between his teeth.}	}
	\V{\Regie{Finds her glasses case. And swaps her glasses.} Wonderful!}
	\Lu{What can we get  you, Vita?}
	\V{\Regie{Cleans her glasses with a cloth.} My husband, Mortis, and I would like to purchase a planet, but we have not 
	yet agreed on what kind. It's probably best, if we wait for him. He still has to park our space ship. But maybe you can 
	already show me how planets are formed.}
	\Lu{I'd be happy to.
	}\V{\Regie{Grabs \Luna's arm.} Do you also have \ldots baby-planets?
	}\Lu{We are very proud of our in-house planet production. We give our planets extra time to develop, and let them
	ripen several million years in starlight. This ensures absolute top quality. 
	}\V{How exciting! What  are the ingredients for planet production? I must admit, I have no idea.}
	
	\begin{figure}[t]
	 \centering
	 \includegraphics[width=0.75\textwidth]{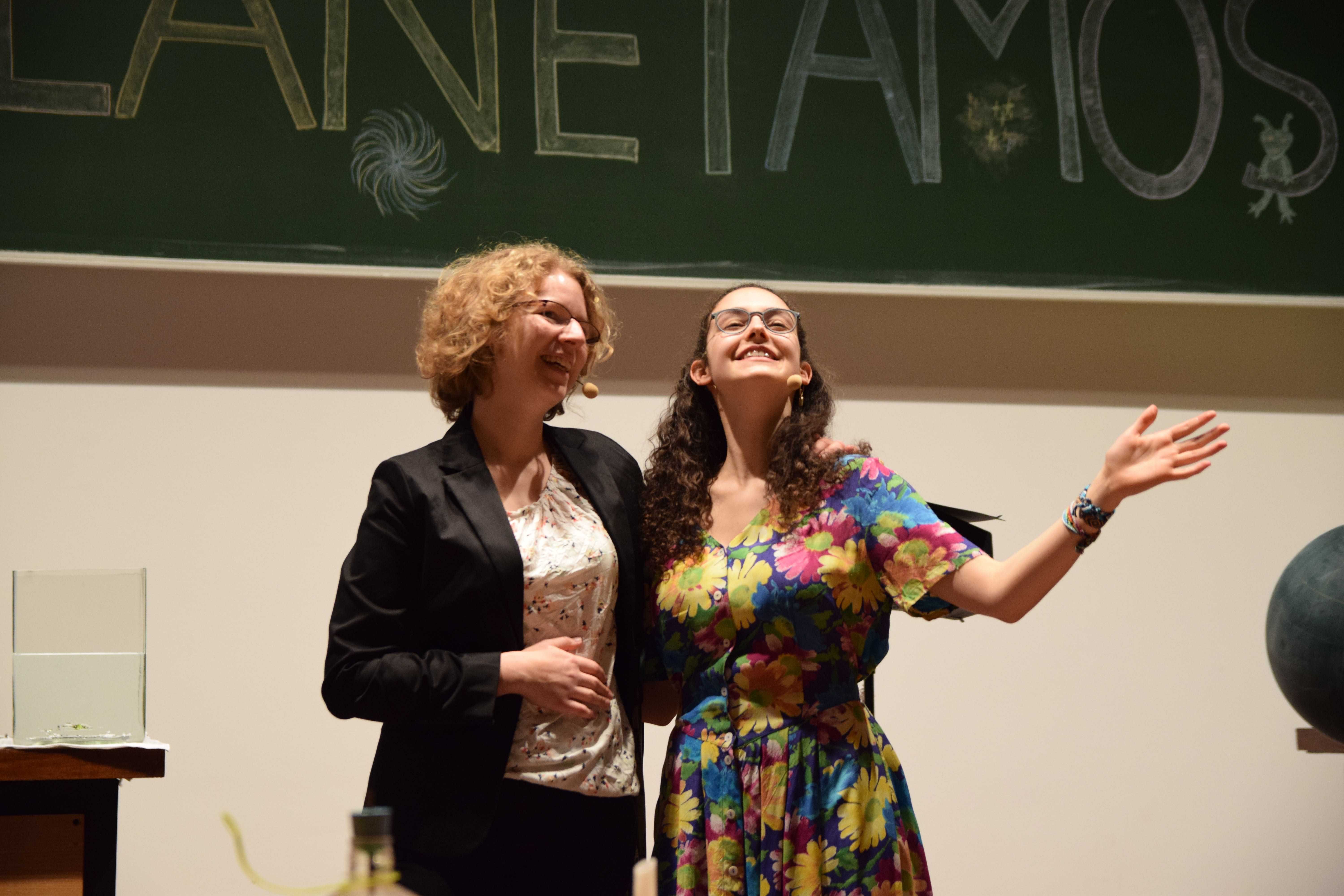}
	 \caption{Luna (left, Jana Heysel) and Vita (right, Laura Rodr\'iguez G\'omez).}
	 \label{fig:play-acting}
	 \end{figure}
	
	\Lu{We only use the best ingredients from near young stars in the protoplanetary disk. We have outsourced the star 
	production to star-t-ups, they have the expertise. We focus on our strengths and buy the young stars whole sale.
	}\V{So your work starts in the early hours of a star's life?}
	\Lu{Exactly. I am not an expert on planet formation, for this we better ask my colleague, Jupi Mercury. He is
	a planetologist. }
	\J{\Regie{Startled to hear his name.} Oh. Yes? \Regie{Cleans the chalk off of his hands.}
	}\V{Oh, that's exciting.  \Regie{Runs to Jupi.} I've always wanted to meet a planetary scientist!
	\Regie{Grabs \Jupi's right hand and shakes it vigorously.} My name is Vita, or Life. I am so happy to meet you,
	Dr. Mercury \dots
	}\J{Jupi, please. I only have a degree as Master of the Universe. It's in the stars, when I will finish my Ph.D. thesis.
	Experiments in planetology take very long, you know.}
	\Lu{\Regie{Goes upstage, right, pages through a catalogue.}}
	%
	\V{Please tell me, how do baby-planets develop?}
	\J{Well, to produce planets, we begin with a protoplanetary disk, around a young star. Here you can see a picture of
	such a disc, full of dust.  \Regie{Figure shown by projector.} As you can see, near the star at the center, the disk is very
	bright, and gets darker and darker towards the edge.}
	\V{Very interesting. And how do the planets emerge from the disk?}
	\J{First of all, the protoplanetary disk, is not the same everywhere: close to the central star it is hot, further away it gets
	cooler. Due to the heat, the matter close to the star is a gas, further away it consists of larger solid pieces. Consider this 
	balloon, which is filled with carbon dioxide, a gas that also occurs in the disk.}
	
	\medskip
	
	{\bf Experiment: Balloon in liquid nitrogen, condensation}, \textit{cf.} App.~\ref{app:condensation}.
	
	\medskip
	
	 \J{At the relatively warm temperature here in our planet store, carbon dioxide is a gas, just like the bubbles in our 
	 drinks. In this container here, I have liquid nitrogen, which is very cold, namely $-200\,^\circ\mathrm{C}$. 	If I 
	 submerge the balloon in the liquid nitrogen, it cools the carbon dioxide. \Regie{Jupi submerges the balloon in the 
	 liquid nitrogen. The balloon shrinks. Jupi then cuts open the balloon.} The carbon dioxide solidifies and becomes
	 what we know as dry ice. You can see what looks like a fine snow powder.}
	\V{So on the outskirts of he disk the carbon dioxide is solid, and towards the center it is a gas?}
	\J{Yes, exactly. The same is true for water. Rocks and metals are however in solid form throughout the disk, in little
	pieces, which form a dust cloud, like this here. \Regie{The sugar for the cotton candy.}}
	
	\medskip
	
	{\bf Experiment: Cotton Candy}, \textit{cf.} App.~\ref{app:cotton-candy}.
	
	\medskip

	 \J{Together the grains of dust in the disk circle around a young star. Wenn two grains hit each other...they stick together.}
	 
	 \begin{figure}[t]
	 \centering
	 \includegraphics[width=0.75\textwidth]{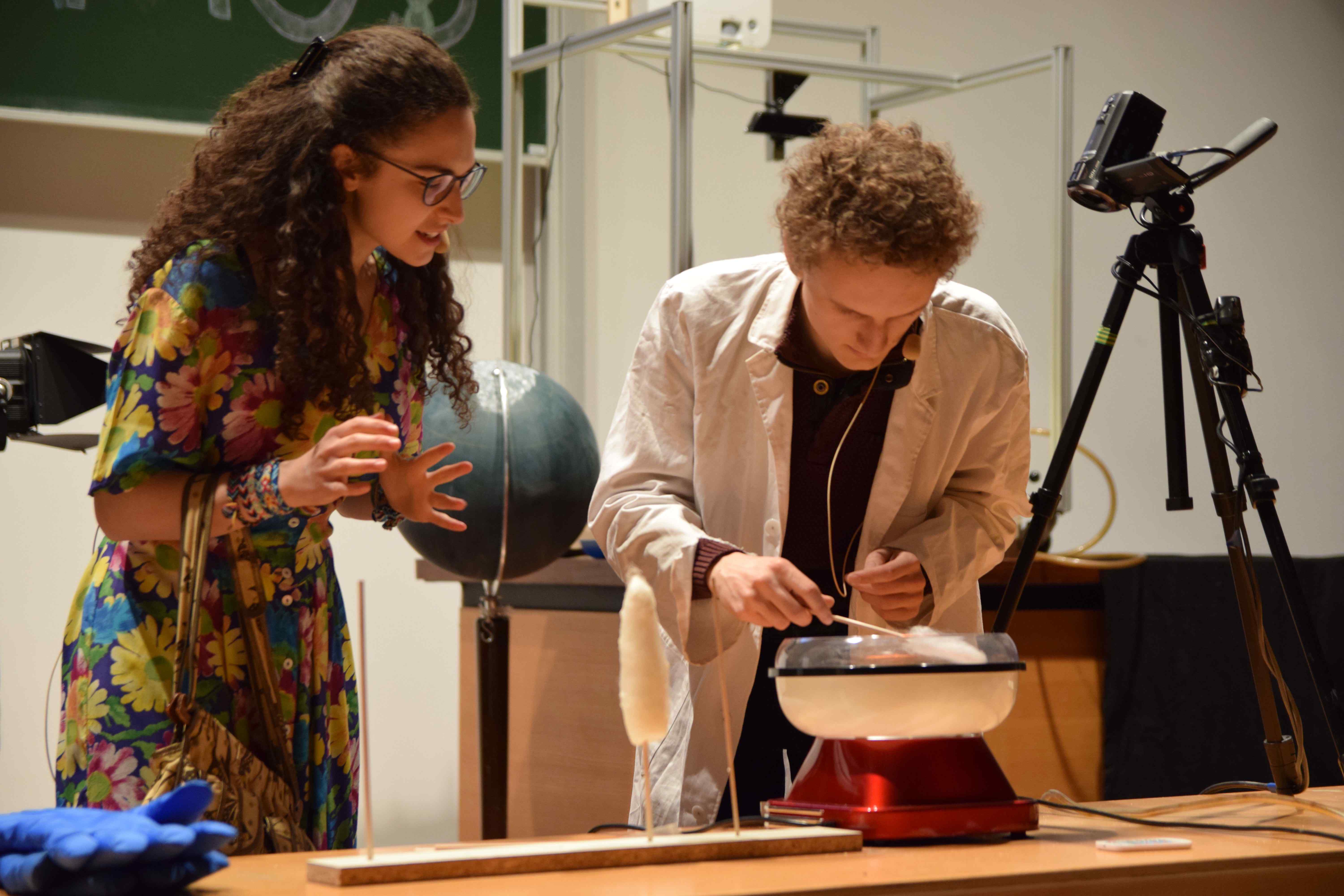}
	 \caption{Vita (left) (Laura Rodr\'iguez G\'omez) and Jupi (David Ohse) and the cotton candy experiment.}
	 \label{fig:cotton-candy}
	 \end{figure}

	\V{Like the dust under my bed!}
	\J{Oh, yes. Thus close to the star larger metal and rock clumps form. Further out, where the carbon dioxide and water are
	frozen, ice is also added. You see?}
	\V{\Regie{Hands to her cheeks.} Ooooh, how cute. Oooh, wie süß. And what is the crux of the matter?}
	\J{I do not understand this question. Matter has no crux.}
	\V{Well, what keeps a world, or the world, together in its heart, \Regie{Embraces herself.} eh.. its core.}
	\J{Ah yes. So the small clumps stick together after randomly bumping into each other. When the clumps get bigger
	gravity keeps them together and ensures that they keep growing.}
	\V{\Regie{Shocked.} You mean the big eat the small? Are planets cannibals?}
	\J{I wouldn't say it like that.... Anyway, when a planet exceeds a certain size it can even capture and retain 
	a lot of gas. It is then called a gas giant. \Regie{Hands \Vita\ the cotton candy.}}
	\V{\Regie{Tries the cotton candy.} Ooh, lovely! So, what planet sizes do you offer?}
	\Lu{\Regie{Looks up, stands to the right of \Vita\ and shows her the corresponding pages in the catalogue.}
	We have all sizes, starting from a a few thousand to even a hundred thousand kilometers diameter. Don't forget
	Vita, if you double the diameter you have four times the surface!
	\R{\Luna\ tries to convince \Vita\ to buy a larger, more expensive planet. \Jupi\ doesn't understand ...}}
	\J{But also gravity is twice as strong!}
	\Lu{But a larger planet can retain an atmosphere.}
	\J{But if it's too large, it will become a gas giant!}
	\V{Hmm. But a gas giant doesn't have a firm surface, you said?}
	\J{Then you can't keep your feet firmly on the ground. A solid planet crust gives you a lot more possibilities.}
	\Lu{So how about a large rocky planet?}
	\V{Yes, that sounds good. What diameter can you recommend? \Regie{Places the cotton candy with the other baby 
	planets which are already on the table.}}
	\Lu{The model Terra Firma with a diameter of 12,700 kilometers is very popular.}
	\V{Amazing, 12,700 is my favorite number!}
	\Lu{An excellent choice.}
	\V{\Regie{Lays her arm around \Luna\ and waves around her other arm.} You know, Luna, I just simply love sunbathing.
	I have always dreamt of lying in a flowering meadow, enjoying the warm radiation on my skin.}
	\Lu{Oh yes, we know this feeling from our quarterly reports. We like to bath in the glory of our success. \Regie{Giggles}}
	\J{Oh. Yes!}
	\V{What do recommend in order to have a flourishing planet?}
	\Lu{Jupi, how can we ensure that Life will feel good on a planet?}
	\J{\Regie{Stands next to \Luna.} We need a constant source of energy nearby, which provides light and warmth. To achieve this, we let the planet circle 
	around a star. \Regie{Adressing \Vita.} Life is quite pleasant on a planet orbiting orderly around a star.}
	\V{But can't we even take two stars? \Regie{Pose: The Thinker}}
	\Lu{Of course, Vita. For just a small extra fee, we are happy to place you in a binary star system.}
	\J{Yes. But ... in that case it would be difficult to place your planet on a stable orbit close to the stars.}
	\Lu{\Regie{Rebukes \Jupi\ with a stare.} During the day you would have two shadows!}
	\J{\Regie{Not noticing \Luna's stare.} Such a binary setup would lead to considerable fluctuations in the radiant flux on your planet.}
	\Lu{\Regie{Rams her elbow into \Jupi's ribs.} My colleague would just like to emphasize that life with two stars 
	is much more varied and diverse.}
	\V{\Regie{Looks from one to the other and considers her options.} Hmm. Maybe we will start with just one star. 
	We need to leave room for improvement.}
	\Lu{\Regie{Contrite.} As you wish.}
	\V{What kind of stars do you have available?}
	
	\begin{figure}[t]
	\centering
	\includegraphics[width=0.47\textwidth]{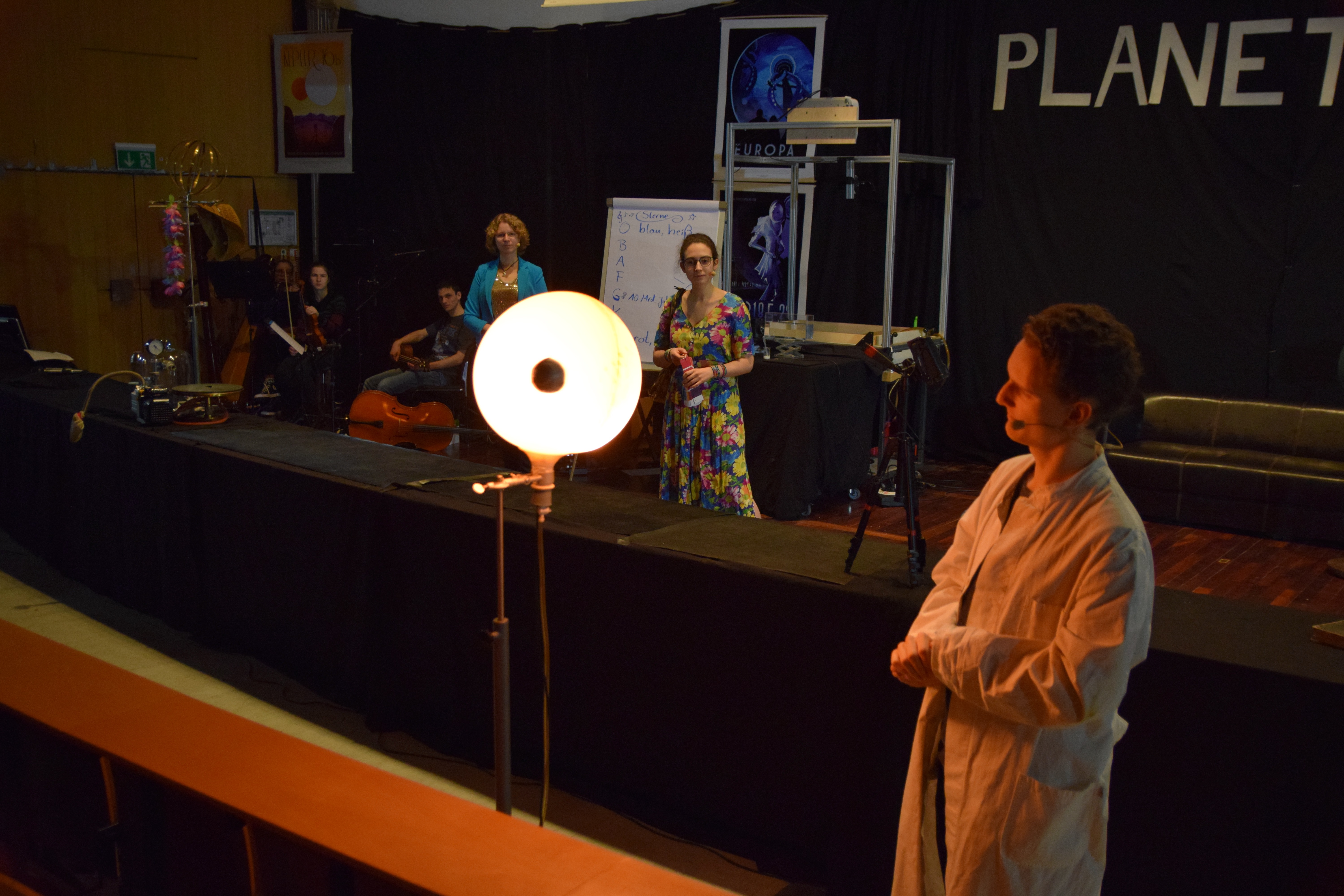}\,	\includegraphics[width=0.47\textwidth]{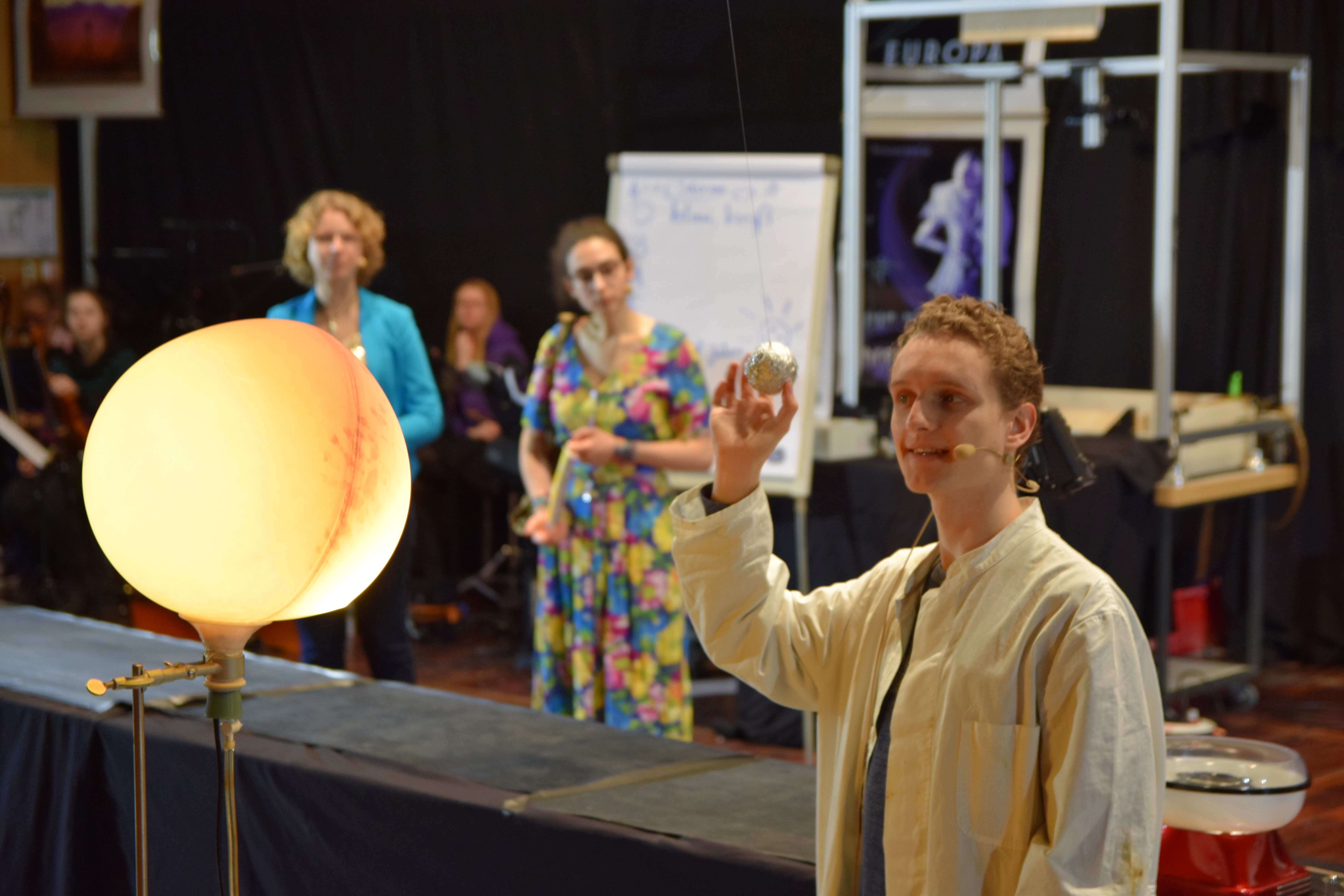}
	\caption{\small David Ohse as Jupi presenting the luminous star and the orbiting planet. In the background 
	Jana Heysel as Luna and Laura Rodr\'iguez G\'omez as Vita, as well as the members of the orchestra.}
	\label{fig:orbiting-planet}
	\end{figure}

	\medskip
	
	{\bf Experiment: shining balloon-star with orbiting planet}, \textit{cf.} App.~\ref{app:planet-pendulum}.
	
	\medskip

	\J{\Regie{Goes downstage to the experiment and adjusts the ``star" (balloon).}}
	\Lu{\Regie{Back on her game.} Oh, we have the full range of possibilities!   O, B, A, F, G and even K or M-Stars.}
	\V{Excuse me?}
	\J{With this sequence of letters we denote the various star models. {\bf O}bviously {\bf B}old {\bf A}stronomers 
	{\bf F}ind {\bf G}reat a{\bf K}ronyms,  {\bf M}ostly.}
	%
	\Lu{\Regie{Writes the sequence of letters  OBAFGKM from top to bottom on a flip chart.}}
	\V{What is the difference?}
	\Lu{Well, Vita, the star models' main difference is their mass, and thus their temperature and brightness. The lightest
	and coolest stars are called M-stars.}
	%
	\J{\Regie{Turns on illumination inside balloon to red.}  At these low temperatures of about 3000$^\circ$~Celsius the
	stars begin to glow red. We call them red dwarfs.   \Regie{Hangs the planet by a thread.}  For such cool and dim 
	planets we must put the planet on a small radius orbit, so that it is warm enough for life to develop.
	\Regie{Launches the planet on a small orbit.}}
	
	\medskip
	
	\Musik{Music: "Jupiter" from the orchestra suite "The Planets" (G. Holst)}
	
	\medskip

	\J{In the habitable zone around a star, the temperatures are just right for sunbathing.}
	\V{Oh yes, I can imagine.}
	\Lu{Due to its proximity, the star appears very large in the sky. Do you like the M-star, Vita?}
	\V{\Regie{Puts her hands on her hips.} It looks cool.}
	\J{Yes, but beware, M-stars tend to have irregular eruptions of plasma. Furthermore, the planet is so close to the star,
	that it is firmly in the grip of the latter's gravitational pull. The planet can no longer spin freely about its axis. It 
	always presents the same side to the star. This is called tidal locking. The side facing the star has permanent
	day time and it gets very hot. On the dark side it is very cold and always night time.}
	\V{How dreary. No! I need more, variety in my life. Without a day-night rhythm I would lose all sense of time.}
	\Lu{I understand. How about a hotter O-star?}
	\J{\Regie{Inflates the balloon with compressed air and switches from red to orange, then white and finally blue.}
	Oh yes! For that I must add a lot of mass. An O-star weighs about 1000 times more than an M-star. Correspondingly
	it is very bright, hot and shines blue.}
	\Regie{Extra spot on balloon.}
	\V{The habitable zone you talked about, is that now far away from this bright O-star?}
	\J{Exactly.  \Regie{Launches the planet on a large radius orbit.} By the way, the strong radiation from the star
	blasts away most of the remaining dust. Then your planet always has a free path ahead.}
	
	\medskip
	
	{\Musik{Music: "Jupiter" from the orchestra suite "The Planets" (G. Holst)}}
	
	\medskip

	\Lu{I can warmly recommend this O-star to you, Vita.}
	\V{I think I might warm to the idea. How long will it last?}
	\J{Well, these O-stars do not live very long. That's only a short pleasure. After a few million years they explode.}
	\V{Just imagine! That is horrible. The star expires just as life is reaching its full bloom. No way, I do not want an O-star.}
	\Lu{How about a middleweight G-star, Vita?}
	\J{Oh! Yes, that should work. Let me reduce the mass.  \Regie{Let's air out of the balloon and turns the color to blue and white.}}
	\V{And G-stars live  long and don't eject matter all the time, Jupi?}
	\J{Yes, exactly. You can relax for 10 billion years. That's how long the G-star stays stable. \Regie{Launches the planet on its 
	closed orbit.} If we place the planet at the habitable radius it takes exactly pi times ten to the seven seconds to circle the
	star once. We also call that one YEAR.}
	
	\medskip
	
	{\Musik{Music: "Jupiter" from the orchestra suite "The Planets" (G. Holst)}}
	
	\medskip
	
	\V{Could you please recapitulate that for me?}
	\J{Yes, I would love to.}
	
	\medskip
	
	{\Musik{Orchestra music: \parbox[t]{9.8cm}{"Bare Necessities" from the movie "The Junglebook" (Music by T. Gilkyson)}}}
	
	\medskip
	
	\J{\Lyrics{Look for a cold and dark red star,\\
	but choose an orbit not too far,\\
	the M-type star has tidal locking trait.\\
	That means your planet has one side\\
	in darkness, and the other bright.\\
	No chance for life to live. How sad a fate!\\

	You need more distance, you like it hot?\\
	A blueish O-type you'll like a lot!\\
	But you have only little time\\
	before your star will stop to shine:\\
	The star explodes, you might be irked,\\
	eliminates your whole new world.\\
	That's not what you long for, is it?\\
	No, no, an O-type star is not a proper choice,\\
	you won't rejoice.\\
	
	The best choice for you is by far\\
	a mid-weight yellow G-type star.\\
	A G-star offers everything you need!\\
	It lives for sev'ral billion years\\
	and all in all the star appears\\
	to be the perfect choice for life, indeed!\\
	
	You get some warmness, you get some light.\\
	you'll love to live there. You'll see I'm right!\\
	The bees are buzzing in the tree,\\
	the flowers blossom lovely.\\
	Well, evolution, sounds bizarre,\\
	depends completely on the star!\\
	So let me tell you:\\
	With such a G-star for your world you will rejoice.\\
	A perfect choice!}}
	\R{Freeze.}
	\V{Fantastic. The G-star it is. That's what I need.}
	\Lu{Our pleasure. We will get you a G-star, at a triple-star great price!}
	\V{Oh, I just realized, I do like some variety. Can you build in proper seasons?}

	\medskip
	
	{\bf Experiment: Blackboard-Globe and Spot -- Seasons}, \textit{cf.} App.~\ref{exp:Mondphasen}.
	
	\medskip
	
	\J{Seasons are no problem. We must simply tilt the rotational axis of your planet. Let me show you with this globe.
	\Regie{Jumps over to globe on a stand. The surface  of the globe is blackboard material.}}
	\V{\Regie{Watches from the right, somewhat impatiently.}}
	\Lu{\Regie{Moves the spot, representing the star, according to the seasons.}}
	\J{The rotation axis of the planet always points in the same direction. It is fixed in space even when the planet follows 
	its orbit. This half  \Regie{Points at the upper half.} of the planet we call the Northern hemisphere. The other half  
	\Regie{Points at lower half.} is called the Southern hemisphere. The equator \Regie{Spins the globe and draws a line
	between the two hemispheres.} separates the two halves. If the star is vertically above the equator  and the rays just glance
	over the two polar regions, both hemispheres get the same amount of heat and it is Spring in the North.}
	\R{All three freeze on-stage.}

	\medskip

	\Musik{Orchestra music: "Spring" from the violin concerts "The four seasons" (A. Vivaldi)}
	
	\medskip

        \R{After the music all three continue normally.}
	\J{A quarter of a year later the planet is here, but the axis still points that way. \Regie{Moves the globe to the middle upstage.} 
	Now the area  around the south pole has the dark polar night. At the north pole the sun doesn't set for many days. Overall the
	Northern hemisphere gets a lot more radiation than the Southern hemisphere. On the Northern hemisphere it is summer.}
	\Lu{\Regie{Moves to where the globe was in Spring.}}
        \R{All three freeze onstage.}
        
        \medskip
        
	\Musik{Orchestra music: "Summer" from the violin concerts "The four seasons" (A. Vivaldi)}
	
	\medskip
	
	\R{After the end of the music all three protagonists continue normally.}
	\J{As the planet continues on its orbit, the days on the northern hemisphere become shorter again, on the southern hemisphere shorter.
	\Regie{Moves the globe further to the right.} At some point both hemispheres again receive the same amount of light. Throughout days and nights
	have equal length. This point of day-night-equality marks the beginning of autumn.}
	\R{All three protagonists onstage freeze.}
	
	\medskip
	
	\Musik{Orchestra music: "Autumn" from the violin concerts "The four seasons" (A. Vivaldi)}
	
	\medskip
	
	\R{After the end of the music all three protagonists continue normally.}
	\Lu{Exactly, and afterwards we have winter on the northern hemisphere. \Regie{Goes and stands at the appropriate point.}}
	\R{All three freeze onstage.}
	
	\medskip
	
	\Musik{Orchestra music: "Winter" from the violin concerts "The four seasons" (A. Vivaldi)}
	
	\medskip
	
	\R{After the end of the music all three protagonists continue normally.}
	\V{And then we have Karneval!}
	
	\medskip
	
	\Musik{Music: "Viva Colonia" (H\"ohner)} 
	
	\medskip
	
	\R{In the Rheinland area around Bonn and Cologne Karneval is often called the 5th season.}
	
	\R{All three protagonists onstage dance to the music.}
	\V{Wow, wonderful! My own planet! And it will be full of life!!}

	\medskip
	
	\Musik{Orchestra music: \parbox[t]{11.7cm}{"For the first time" from the movie ``Frozen" \\(Music by K. Anderson-Lopez and R. Lopez)}}
	
	\medskip

	\V{\color{teal}Finally, a dream comes true!\\
	So many planets, fair and new.\\
	I haven't seen so many worlds before!\\
	I've dreamed of creation for so long,\\
	for sure I've sometimes done it wrong.\\
	But finally I bring it to my world!\\
	Magnetic field and orbit,\\
	water and atmosphere.\\
	But there's much more to do, that's crystal-clear:\\
	
	Come along and let me show you,\\
	see how gorgeous life can be!\\
	Think of trees and scented flowers,\\
	all these plants belong to me.\\
	Here the giant mountains, and there the endless sea...\\
	I will create the home for many,\\
	here in my new world.\\
	
	I can't wait to start! What shall I create first?\\
	
	Stars are shining bright at night,\\
	the moon pours forth its silver light,\\
	the owl is shouting far across the land.\\
	The sun gives rise to all this life,\\
	thanks to its power you survive!\\
	Just see the forest, meadows, hills and dales!\\
	Here the swarm of birds is rising,\\
	the bees are diligent.\\
	The butterfly will touch the firmament.\\
	Come along and let me show you,\\
	see how gorgeous life can be!\\
	Think of puppy-dogs and kitten,\\
	even they belong to me.\\
	All creatures of flora and fauna are beautiful and great!\\
	I will create the home for many,\\
	just the home for every creature\\
	here in my new world!}

	\R{All three protagonists freeze.}
\end{itemize}

\subsection{Moon and Oceans}
\label{sec:mond-engl}
\begin{itemize}
	\R{Enter \Mortis\ from the back of the auditorium. Dressed as Death, with a black cape and a scythe. Walks down the left aisle, along the front of the stage,
	enters stage right.}
	
	\medskip
	
	\Musik{Music: \parbox[t]{13.3cm}{"Harmonica" from the movie "Once Upon a Time in the West" (Music by E. Morricone)}}
	
	\medskip
		
	\M{\Regie{Hangs the scythe on the hatstand, and takes off his headphones. At that moment the music stops. \Mortis\ slowly walks to the piano
	and has the Schicksalsmotiv played, the motif consisting of the first four notes of the first movement of Beethoven's 5th symphony.}}
	\R{\Vita, \Luna\ and \Jupi\ revive from their rigor mortis.}
	\V{\Regie{Radiant with happiness.} Mortiii! At last, you are here! \Regie{Rushes to \Mortis, dances around and hugs him.} 
	Was it hard to find a parking spot?}
	\M{Argh!}
	\LJ{\Regie{Exchange fearful glances.}}
	\V{Are you sure you locked the spaceship?}
	\M{\Regie{Presses the spaceship key.} \Musik{Sound of a car locking}}
	\V{Luna, Jupi, may I introduce? My husband: Mortis. \Regie{She links arms with him.} Dear, these are Luna and Jupi.}
	\Lu{ \Regie{Fearful.} Welcome at \texttt{PLANETAMOS}. Dead certain, we have the right planet for you. May I offer you both something to drink?}
	\V{How attentive of you. For me a fruit tea please, with all possible fruits!
	\Regie{Addressing \Mortis.} You are probably dead tired, dear. What would you like?}
	\M{Coffee}
	\Lu{Certainly. Sugar? Maybe a shot of primeval soup?}
	\V{Yes, please!}
	\M{Black.}
	\Lu{Naturally. Jupi, could you please prepare the hot drinks for our highly esteemed customers?}
	\J{\Regie{Startled.} Oh! Yes. Fruit tea. Coffee ... Black. On its way! \Regie{Exits stage left.}}
	\V{Just imagine, dear. I have seen baby-planets! Fascinating. I have already chosen a planet model 
	and a suitable star \ldots for it to orbit. We will get a rocky planet accompanied by a wonderful G-star. 
	By tilting the planet's rotational axis we will get four seasons for free. Isn't that great?}

	\begin{figure}[t]
	\centering
	\includegraphics[width=0.75\textwidth]{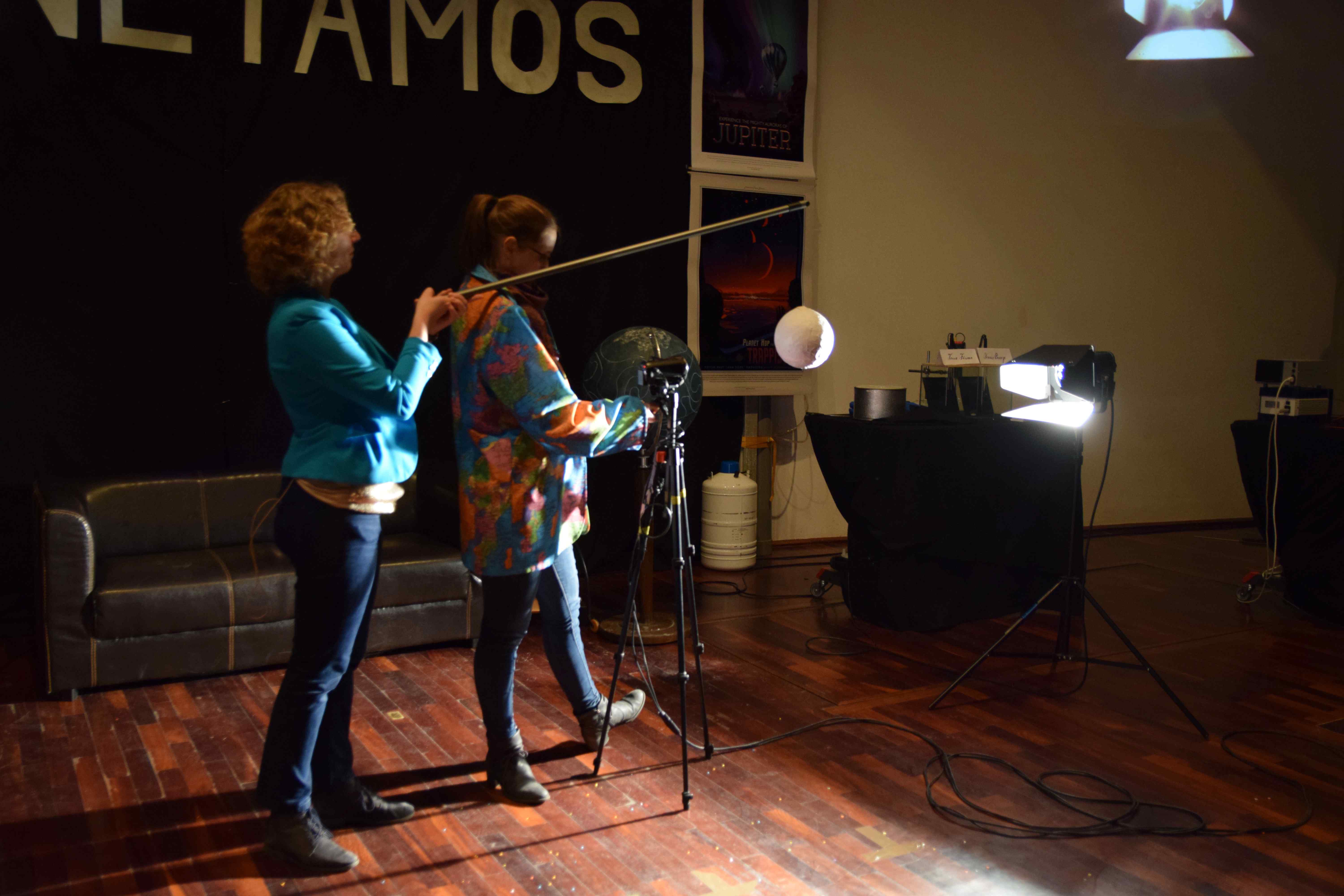}
	\caption{\small Experiment demonstrating the phases of the moon. On the left Jana Heysel (Luna Callisto) but here assisting
	with the experiment.  On the camera Anne Stockhausen also wearing a map of the world jacket.}
	\label{fig:moon-phases}
	\end{figure}

	\M{Ahgh.}
	\Lu{A propos rotational axis and seasons, may I make you a special offer? How about one or two moons for your planet?
	 \Regie{Brings forth a moon hanging by a string from a stick.} Our Super-Silver-Satellite-Special includes an extra large moon.}
	\M{Unnecessary.}
	\V{But dear, a moon is sooo romantic! We can go for long walks together by moon light.}
	\M{Argh.}
	\Lu{Moons are not only nice for a romantic honeymoon, but are also useful. They stabilize the rotational axis of your planet. That protects the seasons
	from 	disruptions due to fluctuations of the axis.}
	\V{That's interesting. Please tell me more about this moondane phenomenon.}
	\Lu{With the Super-Silver-Satellite-Special the distance between the moon and your planet is 60 times the planet radius. This leads to tidal locking. Vita, do 
	you remember what my colleague Jupi said about tidal locking with M-stars?}
	\V{Oh, yes, I do! Here I think tidal locking means that the moon always shows the planet the same side.}
	\M{Deadly boring.}
	\Lu{I would love to take you for a spin on this moon.  \Regie{Spins the moon expectantly.} You could even build your own house there, Mortis. \Regie{No reaction.}
	Anyway, independently of the tidal locking, in all editions the moon regularly changes its shape.}
	\R{The camera films the crescent of the moon, which is projected onto the screen.}
	\Lu{ \Regie{Holds the moon at the corresponding spot relative to the light and the camera and points to the screen.}  There you can follow how the moon 
	as seen from the planet changes its shape. If the moon is between the star and the planet, the moon is dark, we call this a new moon. As it travels along 
	the crescent waxes slowly to a full moon. Afterwards it wanes back to a new moon.}
	\V{I'm sure that looks unearthly!  \Regie{Addressing \Mortis.} We can set up a lunar calendar in our world, Mortis!}
	\M{Argh.}
	\V{Do you deliver the moon separately?}
	\Lu{No, we make it easy for ourselves. When the planet is still young we shoot a somewhat smaller planet at it. In the huge impact the two planets melt 
	together and hurl rock rock fragments into the orbit. Over time these automatically clump together to a moon which orbits the planet.}
	\R{Animation of the formation of the moon.\footnote{We used the film: \texttt{motion2\_25.avi} from the website: 
	\url{https://www.boulder.swri.edu/~robin/moonimpact/}.}}
	\V{Clever!}
	\Lu{Our innovative Super-Silver-Satellite-Special has a further advantage: the tidal forces the moon exerts on the planet. One the one side, the 
	moon's gravity  is pulling at the planet. However, they circle a joint center of mass, thus the resultant centrifugal force also pulls at the planet's 
	other side. This leads to wonderful, spectacular effects. We thus recommend a liquid for your planet surface.}
	\V{A liquid on the solid planet crust? How about  \dots}
	\M{\dots Sulphuric acid?}
	\V{No, dear! No corrosive acid. That would destroy our wonderful planet. I would prefer \ldots}
	\R{Enter Jupi with a tray.}
	\M{\dots mercury.}
	\J{Mercury? Did I just hear my name?}
	\V{No, I would prefer something non-toxic. But  \dots}
	\M{\dots hot red molten lava?}
	\V{No, no. Water, \ldots yes, water, that would be nice.}
	\M{Argh.}
	\J{Oh yes! Water offers many possibilities. \Regie{Fearful glance towards \Mortis.}}
	\Lu{\Regie{Offers the drinks.} Please help yourself.}
	\M{\Regie{Walks to \Jupi, takes the cup of black coffee and inhales a deep breath.} Coffee!  \Regie{Enjoys 
	his coffee and warms a bit towards the others.}}
	\Lu{The tea must steep a bit to reach its full vitalizing effect. Please have a seat. \Regie{Offers the sofa.}}
	\R{\Luna, \Vita\ and \Mortis\ sit.}
	\J{\Regie{Places the tray in front of the others.} Unfortunately the water on a planet's surface is typically not liquid.
	In the habitable zone it is usually a gas. In the outer reaches of a solar system it immediately turns to ice. You might
	remember that out there ice planets can develop. due to the very low pressure in outer space you can not have liquid 
	water.}
	\Lu{However we can offer you the ultimate solution!}
	\Regie{Conspiratorial glance between \Luna\ and \Jupi.}
	\LJ{An atmosphere!}
	\M{\Regie{Looking at his coffee.} Argh. Black. Hot. Delicious.}
	\Lu{A gaseous layer surrounding the planet produces the necessary pressure to condense the water. This produces 
	a nice  atmosphere.}
	\J{May I demonstrate this with your tea?}
	\V{Please, use your creativi-tea, Jupi.}
	
	\medskip
	
	{\bf Experiment: full mug of tea in the glas vacuum chamber},  \textit{cf.}  App.~\ref{exp:tea-vacuum}.
	
	\medskip

	\J{\Regie{Places the tea mug in the vacuum chamber and pumps out the air.} Please don’t chai this at home. 
	Here in \texttt{Planetamos} we have normal atmospheric pressure and a pleasant room temperature. Obviously
	water - and also tea - is liquid. Now I place the tea mug under this glas dome and pump out some of the air.
	The temperature stays the same, but the pressure is gradually decreased. Not as many air molecules push
	against the tea, and we get an airless space, a vacuum. This is more like in outer space}
	
	\medskip
	
	\Musik{Music: \parbox[t]{13.3cm}{"Aquarium" from the Suite "The Carnival of the Animals" (Music by C. Saint-Saëns)}}
	
	\medskip
	
	\J{Look, the water is boiling off and turning into a gas. This is steam, but at room temperature!
	\Regie{Opens valve and lets air enter vacuum chamber again.  Lifts the glass dome and offers \Vita\ her tea
	again. }}
	\Lu{Before your tea has boiled off, you may of course drink it, Vita.}
	\V{Hones-tea, a tea-rriffic experiment. So without an atmosphere all the water would  evaporate?}
	\J{Cer-tea-nly. Or it would freeze, depending on the temperature. By the way, you need a very dense atmosphere.
	For example, our model Mars-Krasni  has a too low atmospheric pressure to have liquid
	water on its surface.}

	\Regie{Foto of Mars-Krasni on screen.}

	\V{I was wondering, from where do you get all the water for our planet?}
	\M{Good question. No water, no coffee.}
	\Lu{That is correct. To create large oceans on  your planet we gather lumps of ice from the outer reaches of
	your solar system and let them ``hail" down on your planet. Logistically that is the easiest way of doing this.}
	\VM{Argh.}
	\Lu{Please forgive me if I am prying, but may I ask how you two met?}
	\V{\Regie{Lovingly looks at \Mortis.} Do you remember, Mortiii?  Back then at the restaurant 
	at the end of the universe, our first date at the Big Bang Burger Bar. It was there that we \ldots}
	\M{\Regie{Smiles.} \ldots fell madly in love. }
	\VM{And I am over the moon for you!}
	
	\medskip
	
	\Musik{Orchestra music: \parbox[t]{9.cm}{"If I never knew you" from the movie "Pocahontas" (Music by J. Secada und Shanice)}}
	
	\medskip
	
	\M{\Lyrics{Since the day I met you,\\
	since you came into my life,\\
	I have seen how beautiful the universe can be.}
	}\V{\Lyrics{And I can rest and calm down,\\
	when you hold me in your arms.\\
	At the end I find in you the shelter that I need.}
	}\VM{\Lyrics{What I now can clearly see,\\
	what I sensed for long,\\
	that the missing part of me\\
	I can now find in you.\\
	'Cause you are my companion,\\
	I have endless faith in you.\\
	And I promise I'll be always true.}}
	\M{\Lyrics{We will create a world that's liveable,\\
	homelike and cozy place to be.}
	}\V{\Lyrics{Can you imagine that all beings in the world\\
	will experience our love with ev'ry breath.\\
	Each existence is controlled by life and death.}
	}\VM{\Lyrics{And we create together our world with woe and joy.}
	}\M{\Lyrics{Without form, and void.}
	}\V{\Lyrics{Soon full of delight.}
	}\VM{\Lyrics{We will tell the tale of death and life.}
	}\R{Freeze.}

	\Lu{\Regie{Spoken.} A wonderful plutonic relationship! 
	}\M{None of your business.
	}\R{Black. 15min intermission.}
\end{itemize}

\subsection{Atmosphere}
\label{sec:atmosphaere-engl}
\begin{itemize}
	\item[] \Musik{Orchestra music: \parbox[t]{11.7cm}{Overture: Instrumental medley of the songs ``A whole new world", 
	``If I never knew you" and ``For the first time in forever".}}
	
	\medskip

	\Regie{All four sitting together.}
	\M{\Regie{Addressing \Luna.} Small piece of advice from me.  \Regie{Addressing \Jupi.} For a longer life:
	\Regie{Adressing audience} Eat more vegetables. \Regie{Bites into a radish and throws more into the
	audience.}}
	\J{Oh. I see.}
	\V{Mortiii \Regie{smiles}, when did you start freely distributing such wisdom? \Regie{Giggles, addresses \Jupi\ 
	and \Luna.} The coffee agreed with him. But you should only follow his advice, if you want to die healthy.}
	\Lu{Sometimes all you need is the air that you breathe (and to love her) \Regie{Lyrics: The Hollies}, sometimes \ldots not.
	Which brings us back to the atmosphere: What gases would you like in your atmosphere? We can offer \ldots }
	\M{Nitrogen?
	}\V{Oxygen?
	}\M{Argon?
	}\V{Carbon dioxide?
	}\M{Neon?
	}\V{Helium?
	}\M{Methane?}
	\Lu{We can also create a gas mixture for you, if you wish.}
	\V{Jupi, what do we have to take into account with our atmosphere? \Regie{Puts down her tea mug.}}
	\J{Naturally oxygen, but it is also essential to include the right amount of greenhouse gases. Especially $\mathrm{CO_2}$, carbon 
	dioxide is important. Given the fixed radiation from the star,  an atmosphere with $\mathrm{CO_2}$ heats up quicker, than one without.}
	\Lu{The amount of $\mathrm{CO_2}$ can lead to heated discussions.}
	
	\medskip
	
	{\bf Experiment: Green house effect in glas containers}, \textit{cf.} App.~\ref{app:treibhaus}.
	
	\medskip

	\J{If you like, I can explain the green house effect using these two glas containers. The two equal stone slabs represent two rock planets.
	Both planets have their own atmosphere, contained in the two aquaria. The right atmosphere is the same as the air in our lovely store, here.
	So mainly a mixture of nitrogen and oxygen.}
	\Lu{That is our standard atmosphere for the model terra-Firma.}
	\J{The left atmosphere is pure CO$_2$  \Regie{Pipes in the CO$_2$.}}
	\Lu{We also call the $\mathrm{CO_2}$-rich atmosphere: Venus-Breeze.}
	\J{As you can see, a match burns normally on Terra-Firma, while ... in Venus-Breeze it immediately extinguishes. \Regie{In turn places a match in each 
	container.} Now I shall irradiate each container with starlight and measure the air temperature with these two thermometers. Both atmospheres are 
	transparent for the visible starlight, which thus reaches the rock slabs, and they warm up. The rocks then themselves emit thermal radiation, which
	is invisible to our eyes and which is in the infrared.  This thermal radiation passes through the air in Terra-Firma, largely unhindered, whereas with 
	Venus-Breeze it is absorbed and partially re-emitted towards the rock slab.  Thus Venus-Breeze heats up more. This is the green house effect. The
	resulting average global temperature is very sensitive to the amount of CO$_2$.}
	\VM{\Regie{Both lean forward to observe the thermometers better.}}
	\J{\Regie{A bit embarrassed} Oh, sorry, this will take a while, before we see a difference.}
	\Lu{How would you like a romantic walk on the beach, between high and low tide?}
	\V{\Regie{Takes \Mortis' hand and looks deeply into his eyes.} That would be wonderful, just the two of us, you and I ... on the beach.}
	\M{At sunset?}
	\V{Yes, with the sky as red as fire!}
	\Lu{Ah, how deep is the ocean -- Love! We are happy to arrange a flaming sky, a wildly romantic sunset .}
	\J{Absolutely. Since the sky does not automatically turn red and orange. We must add a small ingredient.}
	\Lu{For just an Ursa-minor surcharge you can have a spectacular evening afterglow, and you get a wonderful sunrise as well, for free.
	And during the day -- a fabulously bright blue sky! What do you think?}
	\J{All we have to do, is bombard your atmosphere with numerous small boulders.}
	\V{\Regie{Horrified?} What?!?! Are you serious, Luna?!?!}
	\Lu{\Regie{Looking at \Mortis.} Deadly serious.}
	\J{Not to worry, we mainly use small chinks of rock, about 100 meters in diameter. They only cause local damage. That's the best and easiest
	way to get dust into the atmosphere and it's the small dust particles which give the beautiful red sunsets and the deep blue sky during the day.}
	\V{Huh? What's the connection to the sunsets and the blue sky?}
	
	\begin{figure}[t]
	\centering
	\includegraphics[width=0.75\textwidth]{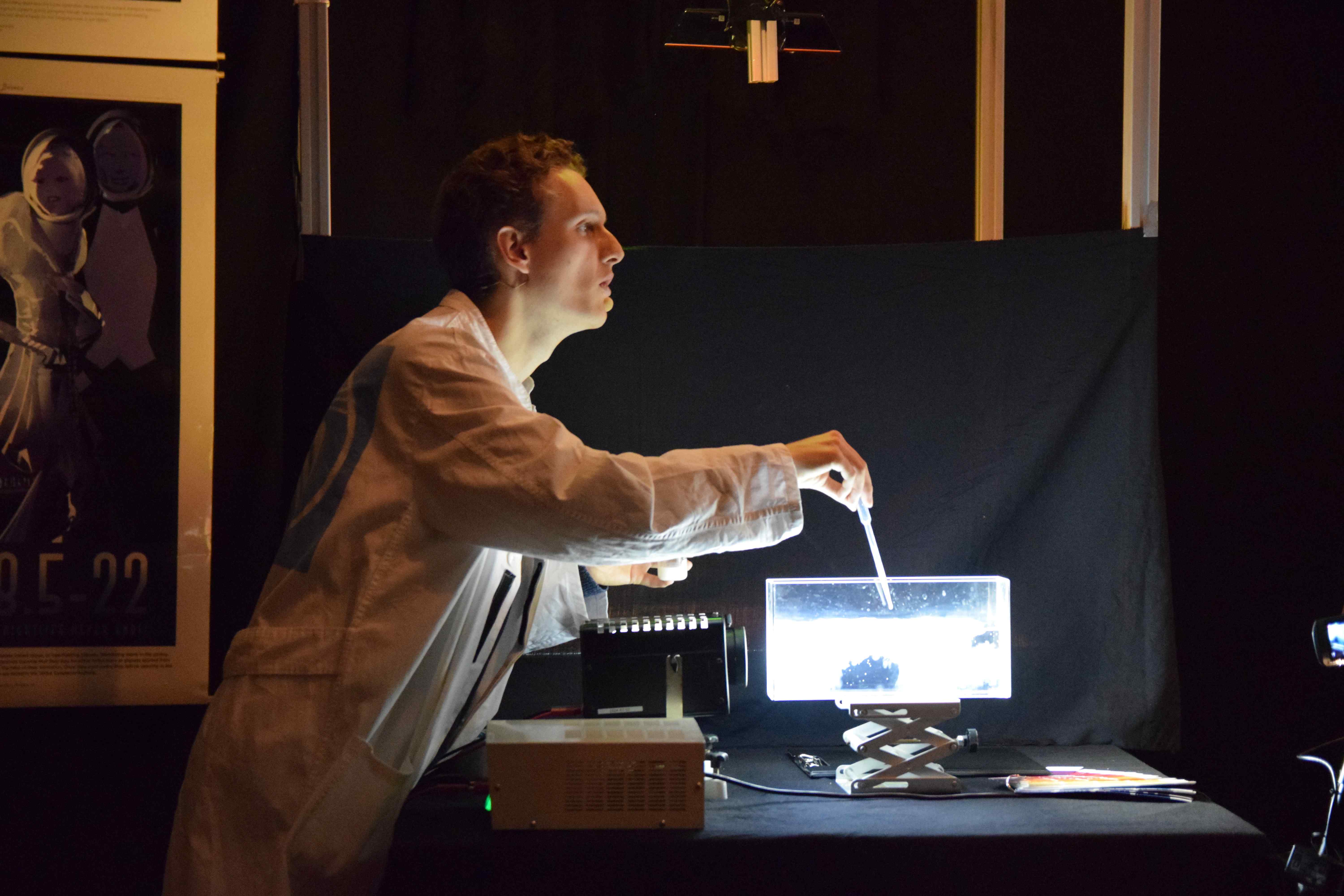}
	\caption{\small David Ohse as Jupi Mercury putting some droplets of milk into the aquarium. There is a hint
	of blue light coming out the side.}
	\label{fig:milk-sunset}
	\end{figure}
	
	\medskip
	
	{\bf Experiment: Rayleigh Scattering with an Aquarium and a bright light}, \textit{cf.} App.~\ref{app:rayleigh}.
		
	\medskip

	\J{Oh, I can demonstrate that with this setup.  Let the aquarium filled with water represent the atmosphere. 
	And this bright lamp shall be our sun.  \Regie{Presents the \E{aquarium} and points the lamp at one end.} To 
	start out with the atmosphere contains no dust, and the light shines through uninhibited. The is now evening 
	red or bright daytime blue. Now let me add some ``dust", for which I will use a small amount of milk. \Regie{Pours 
	some milk into the aquarium and stirs it.}}
	\VM{\Regie{Turn their backs to the audience and put their arms around each  other.}}
	
	\medskip
	
	\Musik{Music: "Force Theme" from the movie "Star Wars" (Music by J. Williams)}
	
	\medskip
	
	\J{As you can hopefully see, the dust, that is the milk, scatters blue light out the sides and straight ahead, at the end of the aquarium, we have red light.}
	\V{But where do the colors come from?}
	\J{They are all contained in the white light from our lamp. White light is simply a mix of all the colors in the rainbow! We just need something
	to separate the colors. The little dust particles scatter the blue light in all directions, unlike the red light.}
	\V{I see. But why is it only red in the evening and the morning? The atmosphere is always the same, no?}
	\J{That is due to the different paths the sunlight takes. Close to the horizon the light must go much further through the atmosphere to reach us.
	Along the way the blue light is scattered aside and only the red light remains. }
	\VM{\Regie{Turn with their interlocked arms to \Jupi.}}
	\J{\Regie{Pours more milk into the aquarium.} Oh, now I have gone to far. The size of the bombarding chunks is very important. At 400 meter radius 
	or more they can cause tsunamis.}
	\V{Oh No!}
	\M{Argh. I like it.}
	\J{At a few kilometer diameter the impact throws up so much dust the planet actually markedly cools.}
	\VM{\Regie{Release their arms.}}
	\V{How unpleasant.}
	\M{Totally cool.}
	\J{For a large rock of 20 kilometer diameter the impact ignites a huge fire, which darkens the entire sky.}
	\V{Terrible!}
	\M{Perfect!}
	\J{With a 400 kilometer meteorite all the oceans evaporate. then it is best to stay underground.}
	\M{Argh. \Regie{Poses as The Thinker.}}
	\V{Aiiiih! \Regie{Hand gesture: Monkey does not see, covers her eyes.}}
	\Lu{Stop! \Regie{Reproachful to \Jupi;  Hand gesture: Monkey does not hear, covers her ears.}}
	\J{Oops. \Regie{Hand gesture: monkey does not speak, covers his mouth.}}
	\R{All lower their arms.}
	\Lu{\Regie{Stands between \Mortis\ and \Vita.} Let me make you a special deal. First we will shoot a large planet 
	like object at your world and from the debris we will make your moon. Then we will drop some medium sized lumps 
	of ice from the edge of your solar system on  it, creating the oceans. Afterwards we will only throw in some much 
	smaller chunks to create the gorgeous sunsets. Oh, and especially for you, these small objects will enter your 
	atmosphere as shooting stars.}
	\V{\Regie{Longingly.} We'll have shooting stars?}
	\VM{Agreed.}
	\Lu{Excellent, we shall prepare our special all inclusive great bombardment for you. \Regie{Suddenly recalls.} 
	How is our greenhouse effect coming along, Jupi?}
	\J{Oh! \Regie{returns to experiment.} Yeeees. have a look. The facts speak for themselves.}
	\VM{\Regie{Bend over to better read the thermometers.}}
	\V{\Regie{Reads off the Thermometers.} The atmosphere with $\mathrm{CO_2}$ is now \_ degree Celsius warmer 
	than the store air we used in the other aquarium.}
	\J{You've said it. By the way, also methane and normal water vapor also contribute to the greenhouse effect.}
	\V{What does it mean for our world? \Regie{Pulls out her notebook.}}
	\J{If your inhabitants burn too much oil and coal, then the  $\mathrm{CO_2}$-levels in your atmosphere will rise, 
	and thus also the average temperature. if it goes too high, this can lead to mass extinctions and other effects.}
	\M{Wonderfully cruel this greenhouse effect. A gift of the heavens!}
	\V{\Regie{Hits \Mortis\ with her notebook.} No, Mortis, stop! I must take notes: NOT TOO MUCH $\mathrm{CO_2}$ ...}
	\J{Yes, it is a very delicate balance. If you instead cover the entire planet with plants, which extract $\mathrm{CO_2}$ 
	from the atmosphere, the temperature will go down again. If you however go to far you will an ice age, covering 
	everything in a thick layer of ice. The ice reflects the sunlight more effectively, making it even colder \ldots}
	\M{\dots deathly cold! Winter is coming! \Regie{Grins diabolically.}}
	\V{\Regie{Hysterical.}  My wonderful daisies! How can we save them? How do we get just the right amount of 
	$\mathrm{CO_2}$ for a natural greenhouse effect?}
	\Lu{No need to panic, Vita. Of course we have the perfect solution for you.}
	\R{Conspiratorial glance between \Luna\ and \Jupi.}
	\LJ{Volcanoes!}
	
	\Regie{A la Luis de Funes:}
	
	\V{No!}
	\M{Yes!}
	\V{Oooh!}
	\Lu{Volcanoes are an efficient way to inject enough $\mathrm{CO_2}$ into the atmosphere.  They can even lead 
	to the melting of a global layer of ice.}
	\M{I like volcanoes.}
	\V{Is that vital? Volcano eruptions cause so much dirt and also trouble.}
	\J{Yes, true. In the short term, volcanoes lead to some mortal danger. However, in the long term volcanic activity 
	is essential for life. Volcanoes are an important ingredient in developing an atmosphere. They accelerate the 
	outgassing of the planet. They also liberate the CO$_2$ trapped in  weathered rock. Please, let me demonstrate 
	this for you.  \Regie{Goes over to the \E{Volcano}-experiment and puts on gloves.}}
	
	\begin{figure}[t]
	\centering
	\includegraphics[width=0.5\textwidth]{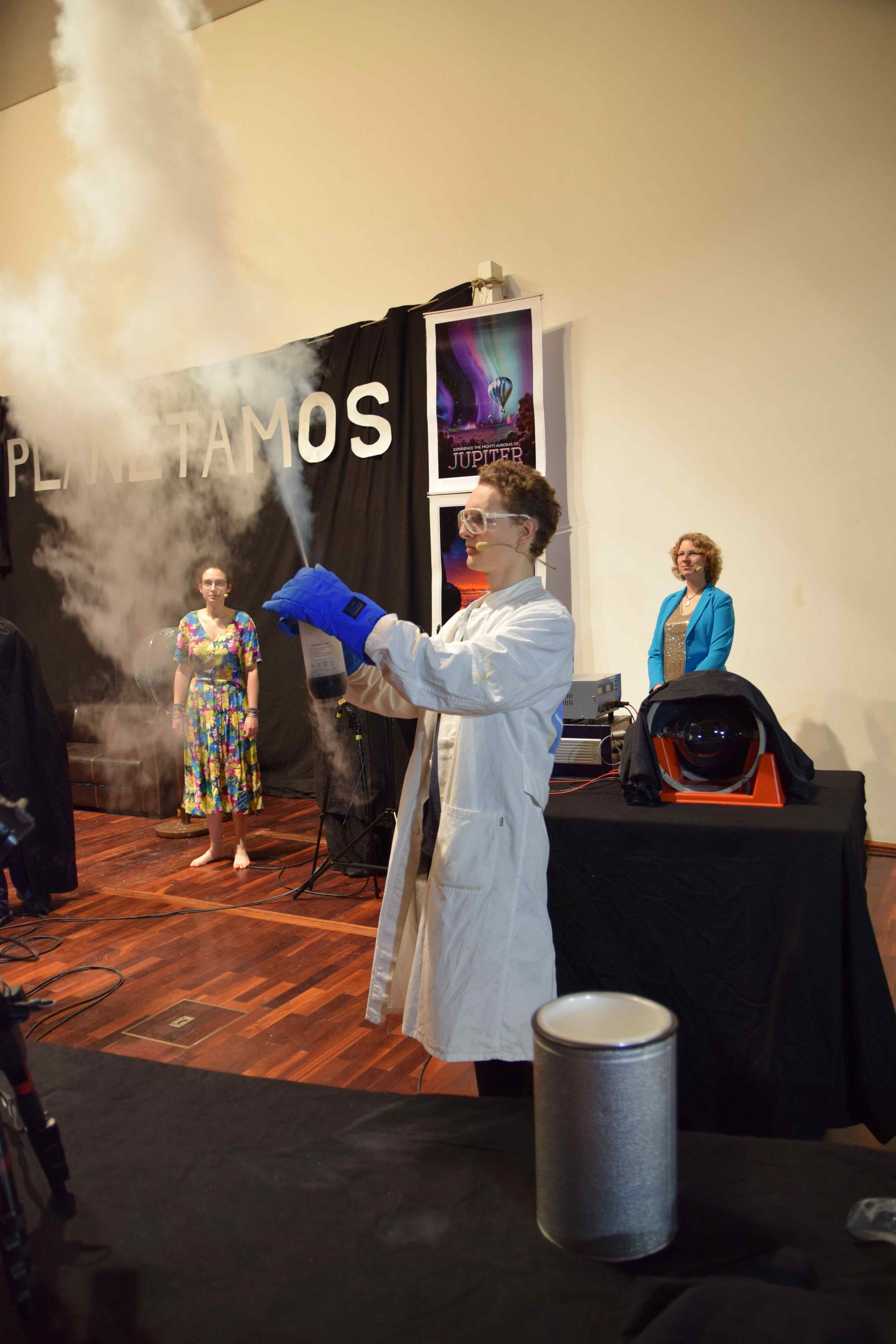}
	\caption{\small David Ohse as Jupi Mercury with a capped bottle of liquid nitrogen. The cap has a small tube allowing
	the gas to escape. On the left Laura Rodr\'iguez G\'omez as Vita and on the right Jana Heysel as Luna Callisto.}
	\label{fig:nitrogen_volcano}
	\end{figure}
	
	\medskip
	
	{\bf Experiment:} \parbox[t]{12cm}{{\bf Volcano-experiment, liquid N$_2$ in a bottle with a small opening}, \textit{cf.} App.~\ref{app:vulkan}.}

	\medskip
	
	\Lu{Deep inside a rocky planet many chemical substances are bound, for example carbon. You could make good
	use of that, Vita, no?}
	\V{\Regie{Nods, gets a nail file out of her bag and focuses on her nails.}}
	\J{In order to demonstrate the effect, I can inject the gas into the rock layers. \Regie{Pours liquid nitrogen into a
	marked glass beaker. On the back side rock layers are depicted.}}
	\Lu{As you can see the gas remains in the rocks. However through cracks and crevices, liquid rock flows can 
	transport the gas closer to the surface, where volcanoes can eject it into the atmosphere.}
	\J{\Regie{Places a small tube with a cork into the glass beaker and awaits a liquid nitrogen fountain.}}
	
	\medskip
	
	\Musik{Music: Symphony No. 9 "From the New World", 4th movement (A. Dvorak)}

	\medskip
	
	\M{Volcanoes are wonderful. I want volcanoes.}
	\Lu{Of  course, as many as you like, Mortis}
	\M{Many.}
	\V{\Regie{Briefly chews her nails.} But not too many, Darling.}
	\M{\Regie{Laughs.}}
	\Lu{\Regie{Addressing \Jupi\ conspiratorially.} I believe we have hit the right note with him.}
	\M{\Regie{Points at the pianist with outstretched arm.}  Hey, you! Play the song of death for me.}
	
	\medskip
	
	\R{The pianist plays the  \Musik{"Imperial March" from the movie "Star Wars" (Music by J. Williams)}}
	
	\medskip
	
	\M{No, no. The other one.}
	
	\medskip
	
	\Musik{Orchestra music: \parbox[t]{11.7cm}{"What a Wonderful World" \\(Music and lyrics by B.~Thiele as ``G.~Douglas" and G.~D.~Weiss)}}
	
	\medskip
	
	\M{\Lyrics{I see deserts of sand, hot and sere,\\
	white frosted land, cold and severe,\\
	and I think to myself:\\
	What a wonderful world.\\\\
	I see storm and flood, a hurricane,\\
	a volcano brings death, fire and pain.\\
	And I think to myself:\\
	What a wonderful world.\\\\
	In darkness of the ocean, in brightness of the sky,\\
	in each spot of the planet there's grief, you can't deny.\\
	A virus, that small, quite hard to see,\\
	still you imagine how fell it can be.\\\\
	I see drought in the fields, smoke and blaze,\\
	fear, oppression, dark lonely days.\\
	And I think to myself:\\
	What a wonderful world.\\
	Yes, I think to myself:\\
	What a wonderful world.}}
	\end{itemize}

\subsection{Planetary Surface}
\label{sec:planetenoberflaeche-engl}

	\begin{figure}[t]
	\centering
	\includegraphics[width=0.75\textwidth]{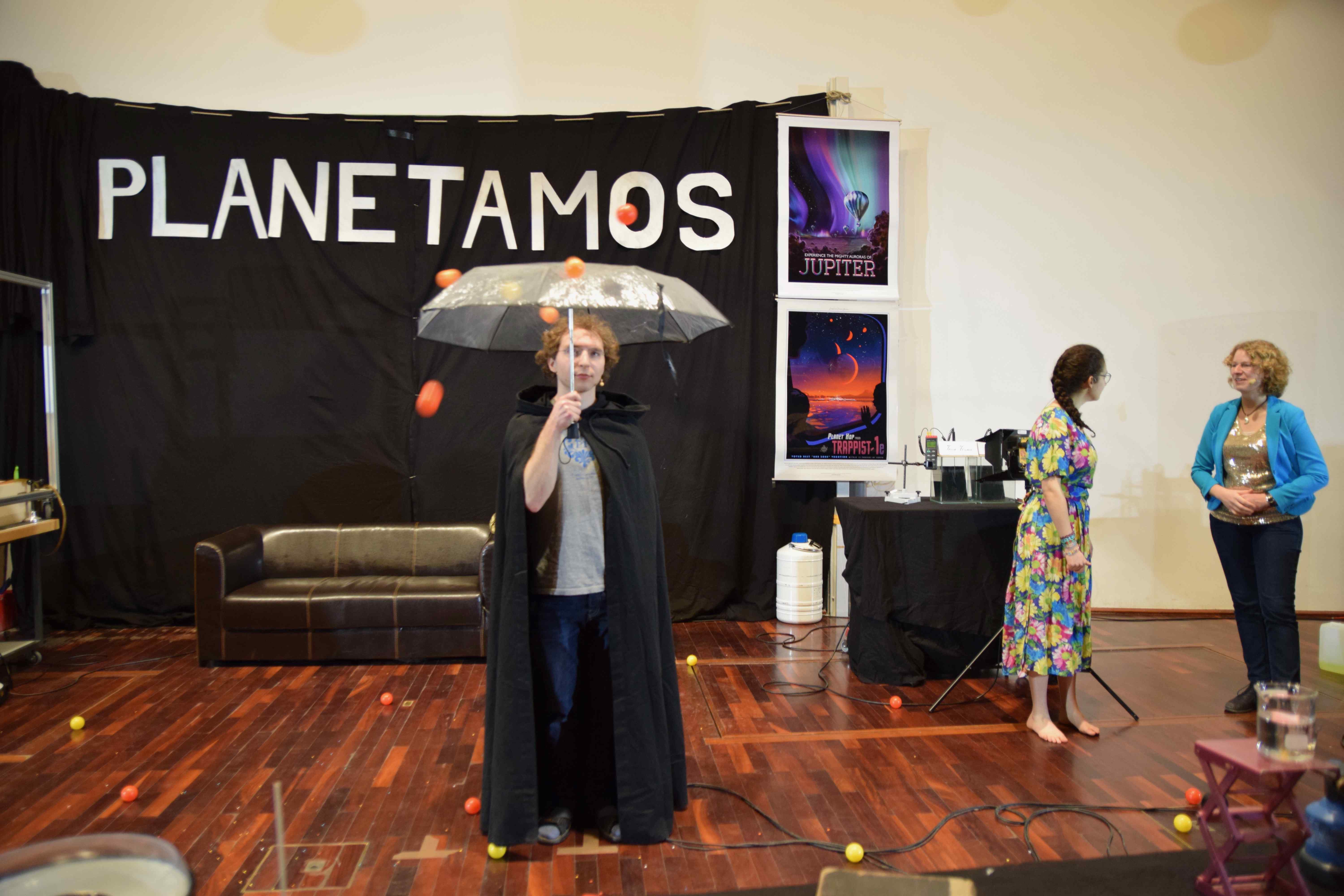}
	\caption{\small Johann Ostmeyer as Mortis, bombarded by ``cosmic rays''.  On the right Laura Rodr\'iguez G\'omez
	as Vita and Jana Heysel as Luna.}
	\end{figure}

\begin{itemize}
    \R{\Mortis\ finishes his song. The audience is excited. As the thunderous applause subsides, ping pong balls begin
    falling on him from the ceiling.}
    \M{Ouch. \Regie{Pulls on his hood.}}
    
    \R{\Mortis\ takes a side step. The balls follow him.}

    \J{Oh. Those are cosmic rays. I guess the roof is leaking ...}
    \V{Comic rays? Not very funny.}
    
    \J{{\it Cosmic} Rays. These are energetic charged particles from outer space, which hit the planet. The strongest
    source is the solar wind from the star which the planet orbits. The {\it cosmic} rays are damaging for life, too much
    can be fatal.}
    \M{Deathly rays. I like it. I'll have two!}
    \V{But Mortis, you must protect yourself.}
    \Lu{No problem. We have the perfect product for you: a global magnetic field, represented here by this umbrella.
    \Regie{\Luna\ hands \Mortis\ an umbrella.}}
    \V{Excellent! We need one.}
         \Lu{I'm glad we could convince you. Let us fix the details in the contract. I will start preparing it. \Regie{Exits.}}
	\V{\Regie{Addressing \Jupi} How does such a magnetic field work?}
         \J{It deflects the radiation.}
   
	\medskip       
    
       {\bf Experiment: Helmholtz Coils}, \textit{cf.} App.~\ref{exp:Helmholtz}.
    
	\medskip       
    
         \J{I can demonstrate that nicely here. This apparatus creates a beam of charged particles. That is the blue line you
         can see. And now watch out what will happen, when I turn on the magnetic field.}
         \V{Ah, so we simply steer the cosmic rays away from our planet with the magnetic field?}
          \J{Yes,exactly!}
          \M{\Regie{Disappointed} A magnetic field protects the planet from deathly rays? Very sad.}
	\J{Well, even with a magnetic field you don't get all around protection. In the polar regions  the energetic particles
	can often penetrate deeply into the atmosphere. That leads to beautiful Northern Lights. In-between you can also
	turn off the magnetic field for a few years, or reverse the polarity. The cosmic rays can then enter for a limited time
	and kill some life forms.}
	\M{\Regie{In thought.}  Cosmic rays. We could let \textit{them} do it.}
	\V{So you will install a gigantic bar magnet into our planet?}
	\J{No, not quite, hihi. How would we reverse the polarity? And furthermore, the magnet would not stay magnetic.
	The core of the planet is so hot, that it is not possible to permanently magnetize it.}

	\begin{figure}[t]
	\centering
	\includegraphics[width=0.45\textwidth]{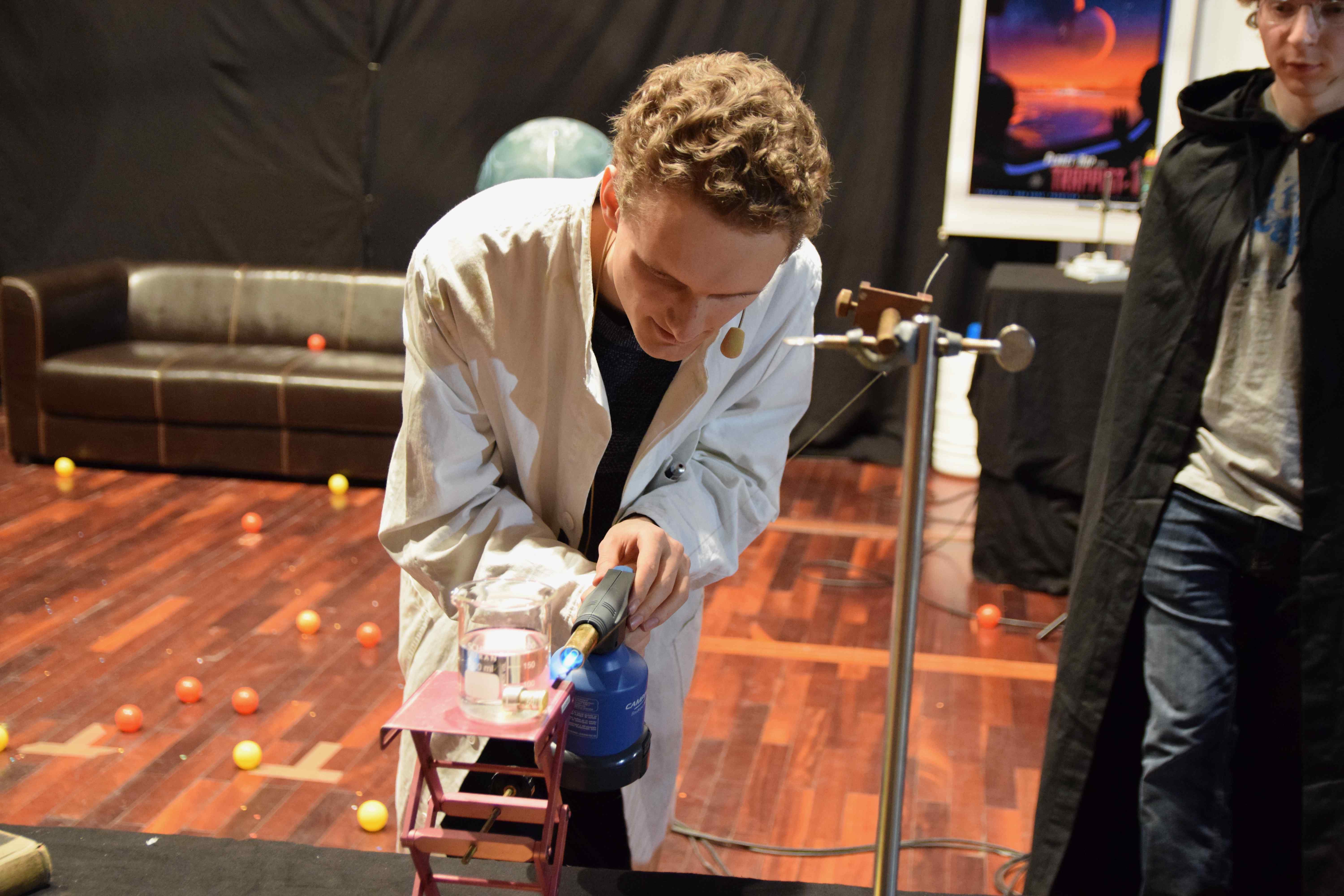} \includegraphics[width=0.45\textwidth]{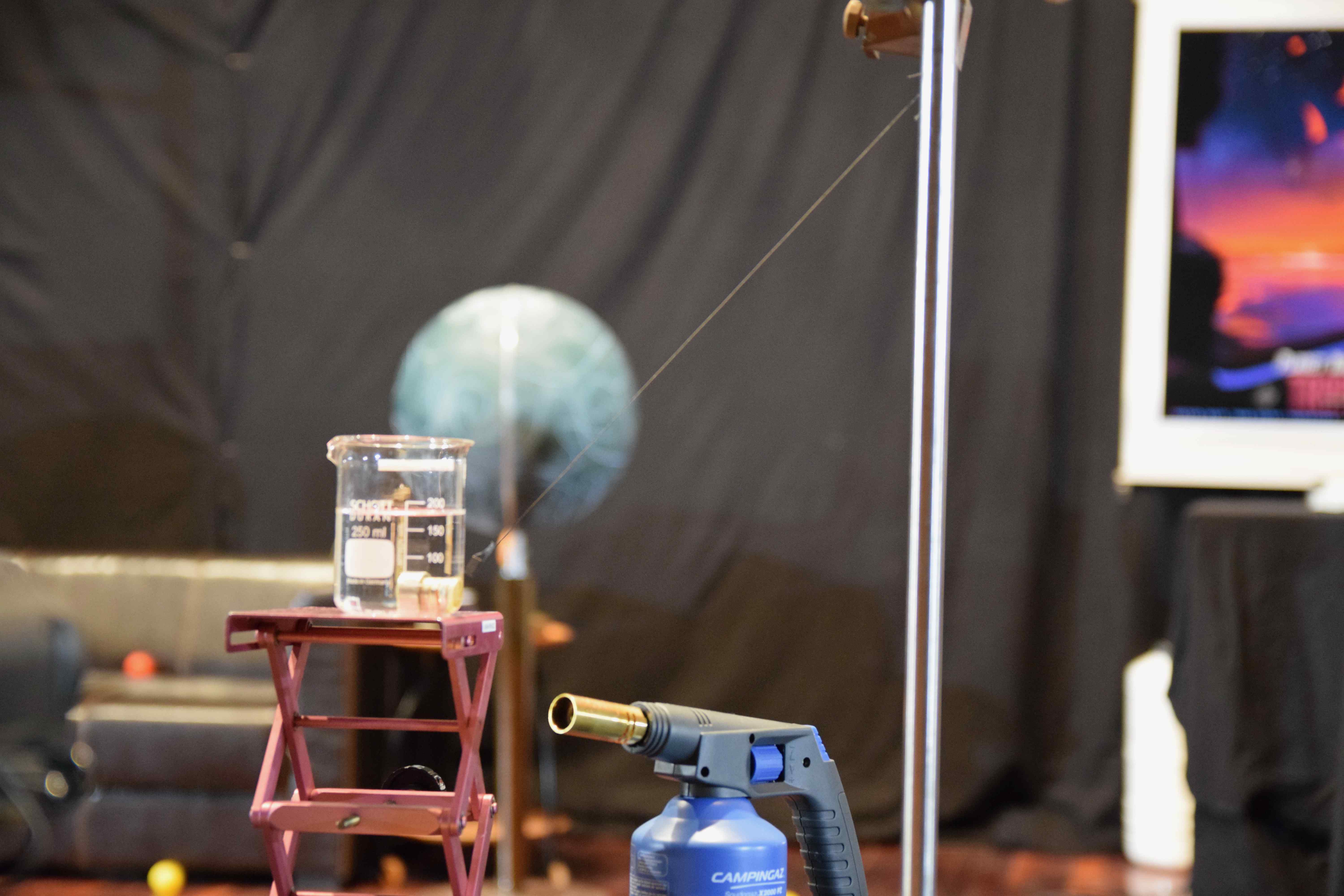}
	\caption{\small On the left, David Ohse as Jupi performing the Curie-Temperature experiment. Standing on the right Johann Ostmeyer
	as Mortis. On the right a close-up  of the experiment.}\label{fig:curie_temperature}
	\end{figure}
	
	\medskip
	
	{\bf Experiment: Curie-Temperature}, \textit{cf.} App.~\ref{exp:Curie}.
	
	\medskip

	\J{Consider this piece of nickel, it is magnetic and is attracted to the permanent magnet you see in the beaker filled 
	with water. Now we heat the nickel, it suddenly loses its magnetism and falls away. This is true for all magnets, if
	they become too hot, they are no longer magnetic.}
	\V{And when the piece of nickel has swung back and forth, has it cooled enough to be magnetic again?}
	\J{Yes, precisely.}
        \V{So how do you create a global magnetic field, if a simple permanent bar magnet doesn't work? And how do you
        turn it on?}
        \M{And off again?}
	\J{Actually, I am not supposed to tell you. Luna is very strict with company secrets. Please promise me you won't tell
	anybody what I'm about to explain to you. Don't breathe a word!}
	\V{Promised!}
	\M{I'm as silent as a grave.}
	\J{Okay. Deep inside the planet the temperature is very high and we obtain a plasma. The electrons are separated 
	from the atoms by energetic collisions and we have freely moving electrically charged particles.}
	
	\medskip
	
	{\bf Experiment: Jacob's Ladder}, \textit{cf.} App.~\ref{exp:Hoerner}.
	
	\medskip

	\J{We can create a small plasma here in our laboratory. The lightning bolts you see, that is a hot plasma. As you
	can see, because it is so hot it rises. The same happens inside the planet's core. Thus the plasma is always moving.
	The moving plasma leads to large electrical currents. These currents create the magnetic field.}
	\Lu{\Regie{Enters the stage with the contract. Hears \Jupi's last words; angry} Jupi did you reveal how we create the
	magnetic field?}
	\J{No!}
	\Lu{Yes!}
	\J{Ooh.}
	\Lu{We have discussed this many times. The details of the magnetic field are not discussed with customers for 
	operational reasons.  \Regie{Gives \Jupi\ a reprimanding look.}}
	\J{But why?}
	\Lu{\Regie{Addressing \Vita\ and \Mortis.} Only \texttt{PLANETAMOS} can offer you a global magnetic field. We are the
	sole planet store that can offer full global magnetic protection. \Regie{turning to \Jupi.} And we would like to keep 
	it that way.}
	\J{So strange that the competition doesn't understand global magnetic fields. Maybe I should give them some pointers
	\ldots }
	\Lu{\Regie{Addressing \Jupi\ authoritatively.} Don't you dare!}
	\J{Oh, well. Okay.}
	\Lu{\Regie{Addressing \Vita\ and \Mortis.} My apologies. Please, let us go through your contract together. 
	\Regie{Opens the folder.}}
	
	\vspace{3cm}
	
	\Musik{Orchestra music: \parbox[t]{9cm}{"Dream a little dream of me" \\(Music by F. Andre \& W. Schwandt)}}
	
	\medskip

	\Lu{\Lyrics{You want a world to live in,\\
	remember that not each is life-giving.\\
	Your world could be unique in the space,\\
	a life-affirming, joyful place.\\\\
	A G-star is required.\\
	Your planet orbits it as desired.\\
	A slightly tilted axis is set\\
	for the seasons that you get.\\\\
	Take care that your axis is stable,\\
	to swaying immune.\\
	For this and for tides in the oceans,\\
	you need a moon.\\\\
	The atmosphere can build up\\
	air pressure for the liquid in your cup.\\
	For climate reasons we have for you\\
	a close look at the CO2.\\}}
	\Lu{\Regie{Spoken.} We take care of purple romantic sunsets by bombing your planet with some little rocks. 
	By doing this, the atmosphere gets dusted. Additionally, you get innumerable shooting stars for free!}
	\M{How about my volcanos?}
	\Lu{\Lyrics{For temperature control you should have\\
	volcanoes in store.\\
	We established them on the planet.\\
	That's what you asked for.\\\\
	Magnetic field protection\\
	is crucial for the cosmic rays deflection.\\
	That's all you need, your dream can come true.\\
	Now this world belongs to you!}}

	\Lu{What kind of weather would you like on your joint planet?}
	\V{Rainbows!}
	\M{Tornadoes!}
	\J{Given the atmosphere you have chosen, with all the water and the solar radiation from your G-star, you will
	have those automatically.}
	\Lu{Well, for a small star light special supplement.}
	\J{To get a rainbow, you just need a few water droplets in the atmosphere.}
	\V{Argh. \Regie{Gets a hand fan out of her purse, fans herself, follows the sales talk getting angrier.}}
	\Lu{With respect to cyclones we have some extra special offers. You will like this, Mortis.}
	\J{You probably mean the \dots}
	\LJ{Fire tornadoes.}
	\Lu{These are not your run of the mill whirl  winds.}
	\J{No, not your every day dust devils.}
	\Lu{We are talking about flaming cyclones.}
	\M{Oh yeah, light my fire! Please show me your fire tornadoes.}
		
	\medskip
	
	{\bf Experiment: Fire tornado}, \textit{cf.} App.~\ref{exp:firetornado}.
	
	\medskip

	\J{Of course, over here, please.  \Regie{Walks over to the \E{Fire tornado}.}}

	\medskip

        \J{\Regie{Ignites the  \E{Fire tornado}.}}
	
	\medskip
	
	\Musik{Music: "Sauron's Theme" from the movie "Lord of the rings" (Music by H. Shore)}
	
	\medskip
         
         \J{To maintain a cyclone we need 
	warm air, which rises. With a fire tornado the air is heated by a ... fire. The hot air rises, sucking in new colder 
	air from the sides at the bottom. The rotating cage rotates the in-flowing air, which rotates faster,  as it reaches 
	the middle  axis, life a figure skater pulling in her arms.}
	\Lu{How about that, Mortis? Does it ``turn you on''? }
	\M{The explanation makes sense, but the tornado is very small.}
	\Lu{This is just a demonstration model. We can offer all sizes, from miniature to gigantic.}
	\V{\Regie{Totally upset and pissed off, freaking out, closes her hand fan and throws it against \Mortis.} 
	Absolutely not! No Way! Auf keinen Fall! Are you trying to destroy our lovely planet?! There will be no 
	fire tornado in this world. You always only think of yourself, Mortis \ldots}
	\R{Everybody is talking and screaming:}
	\M{Just a moment. It is only a small whirl wind. \ldots}
	\Lu{Please Luna, we were just thinking out loud \ldots }
	\J{We also have harmless small fire tornadoes. We'll figure something out.}
	\V{You're just trying to off load your junk on us. I will not sign your contract. Mortis. We are leaving.}
	\M{Please calm down, darling. Think of your  puppies.}
	\Lu{Outrageous. This is the best planet store in the galaxy. I should ban you from our house.}
	\J{Good idea. Then I can focus on my collection of comets, again.}
	\M{\Regie{Very loud:} {\bf SILENCE!} No fire tornadoes.}
	\R{All fall silent, and lower their arms. Now they talk individually again.}
	\V{Oh Darling. \Regie{Embraces \Mortis\ sobbing.}}
	\Lu{\Regie{Crosses her arms.}}
	\M{I will sign. The price is irrelevant.}
	\Lu{\Regie{Counting money in her mind's eye, a wide grin on her face.} Always a pleasure! \Regie{Hands \Mortis\ 
	the folder with the contract.}}
	\V{Thank you. \Regie{Stands on her own again.}}
	\M{\Regie{Sighs, mumbling.} Mors, Mortis f.}
	\J{You know what? I will make a glorious fjord landscape on the Norther Hemisphere for you. You planet will be
	the most beautiful of all!}
	\Lu{Exactly. \Regie{Hands \Vita\ a tissue.}}
	\V{\Regie{Blows her nose thoroughly.}}
	\J{Would you like to watch the landscape planning}
	\VM{We'd love to.}
	\J{Let me show you over here.}

	\begin{figure}[t]
	\centering
	\includegraphics[width=0.5\textwidth]{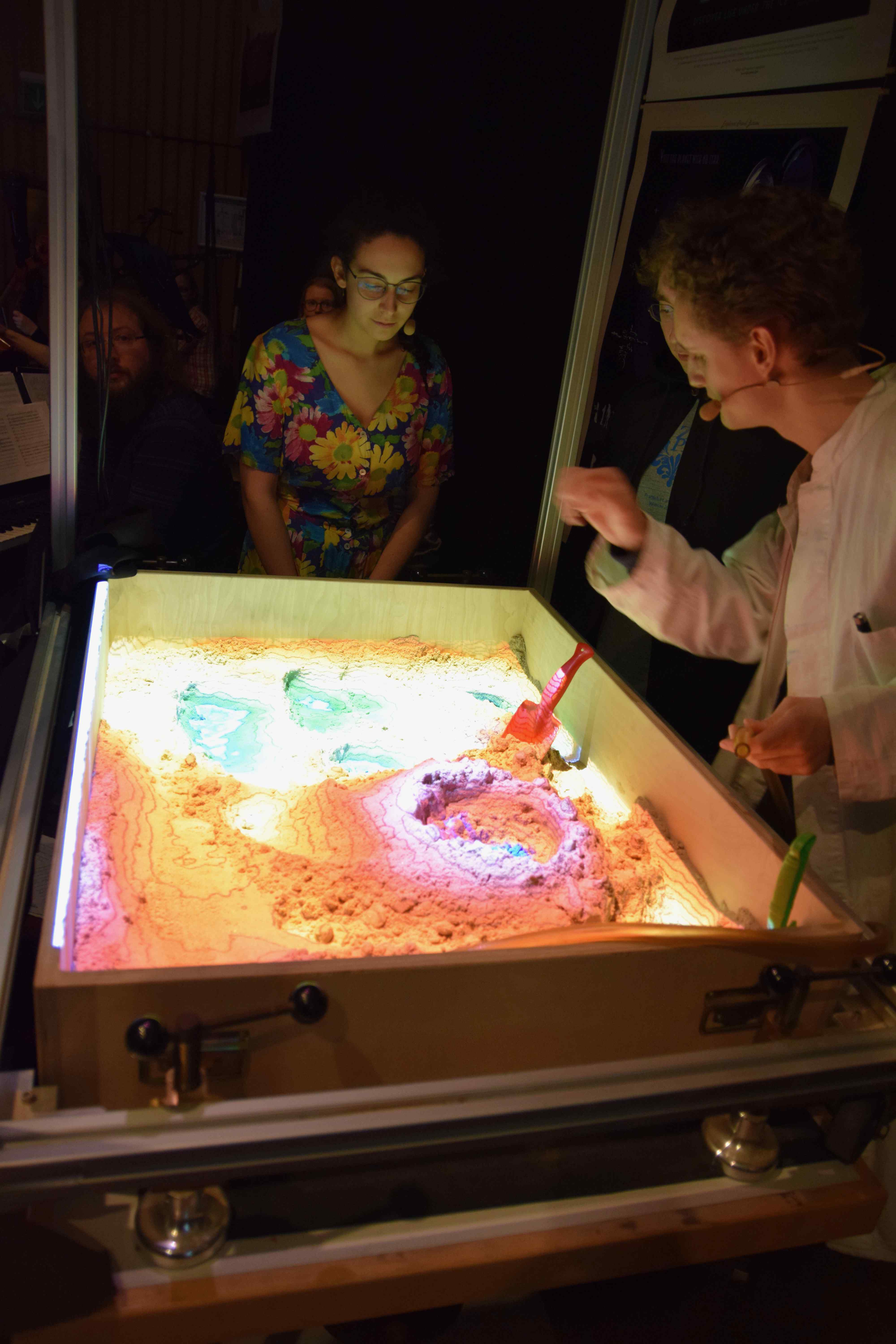}
	\caption{\small The sandbox experiment. On the left  Laura Rodr\'iguez G\'omez
	as Vita and on the right David Ohse as Jupi Mercury.}
	\end{figure}
	
	\medskip
	
	{\bf Experiment: Sandbox}, \textit{cf.} App.~\ref{exp:Sandbox}.
	
	\medskip
	
	\R{They all gather around the \E{Sandbox}.}
	\Lu{In the beginning the planet is desolate and empty. With various techniques we can combine mountains, 
	valleys, and oceans to a magical combination.  On the one side, the wind and weather influence the landscape 
	development. On the other side, also inner processes of the planet help to fix the surface structure.}
	\J{For example: under the planetary surface large bubbles can fill with molten rock. If such a bubble becomes too
	large you get a volcano cone at the surface. \Regie{Inflates a balloon.} When the volcano erupts, the bubble empties,
	and a crater remains. That can then fill with water, leaving a lake, a maar.}

	\M{How about a Great Deluge?}
	\J{Oh, just a second. We can arrange that. \Regie{Floods the \E{Sandbox}.}}
	\V{Oh, that is inspiring. An underwater volcano seems to be the ideal socio-cultural environment for my first living
	creatures.}
	\Lu{One more thing. \Regie{Lays her arms over \Vita's and \Mortis'   shoulders.} Vita, Mortis, we need a name 
	for your New World.}
	\V{Oh, yes! Of course. We need a name for our baby-planet, darling!}
	\M{Argh.}
	\Lu{How about a meaningful acronym? An abbreviation?}
	\V{Hmm. \Regie{Ponders.} Something with  \Regie{Ponders} {\bf E}volutionary!  \Regie{Ponders}  \dots}
	\M{\Regie{Ponders} {\bf A}dventure \dots {\bf R}esort \Regie{Ponders.} and \dots}
	\Lu{{\bf T}errestrial  {\bf H}abitat. {\bf E}.{\bf A}.{\bf R}.{\bf T}.{\bf H}. Earth!}
	\V{Eaaarth. Yes, that sounds good, lovely.}
	\J{Oh, yes. The name \enquote{{\bf E}volutionary  
	{\bf A}dventure {\bf R}esort and {\bf T}errestrial {\bf H}abitat} emphasizes the down-to-earth nature of
	your world as well as the sensitive balance of life.}
	\Lu{Let's be honest, life is always a death trap.}
	\VM{\Regie{Hold each other and laugh.}}
	\M{I'm laughing my head off!}
	\V{Long live Evolution. Viva la evolución!}
	
	\medskip
	
	\Musik{Orchestra music: \parbox[t]{10cm}{"Circle of Life" from the movie "The Lion king" \\(Music by E. John)}}
	
	\medskip

	\V{\Lyrics{And today we receive our planet,\\
	it will orbit around the bright sun.\\
	Such a glorious work, that has never been seen!\\
	We esteem the effort you've done.}}
	\Lu{\Lyrics{All the life in this world just has started.\\
	There is more to be found, to prospect.\\
	Take a look at the sky, see the clouds passing by,\\
	think of life and behave with respect.}}
	\J{\Lyrics{In the circle of life,\\
	with the world proceeding,\\
	with the sun and the stars\\
	and a silver moon.}}
	\V{\Lyrics{Our world is a part\\
	in a complex cosmos\\
	and we circle\\
	in the circle of life.}}

	\R{Black. Spot on \Mortis.}
	\M{\Regie{Cut throat gesture.} THE END!}
	

	%

\end{itemize}

\centerline{\bf The End}

\section{Musical Arrangements}
\label{sec:music}

Music is an essential part of the physics show musical. While several experiments are 
accompanied by recorded music, and as indicated in the complete text above in Sec.~\ref{sec:script}, 
the main songs of the play are performed live on stage by the actors and a small live orchestra. This is also 
explicitly indicated in the script. Each of the four actors sings a solo, a duet and a quartet, which in total makes 
seven songs over the whole play. All songs have been adapted from well-known tunes, whereas the instrumentation 
has been implemented for the available instruments. The songs have been chosen to confirm the respective 
mood of the singing characters. To relate the music to the story, the lyrics have been rewritten completely such that 
they summarize the (astrophysics) plot and/or match the atmosphere. The orchestra consists of a flute, an 
oboe, a violin, a cello/bass ukulele, a harp, a piano and percussion (caj\'on and cymbal).
The instrumentation varies from minimal instrumentation (piano and percussion) to full instrumental accompaniment, 
depending on the song. For most of the songs the piano player is vamping, gearing to the given chords, the same 
holds for the percussionist. All other instruments have their own melody parts, which are partly arranged from the 
original orchestra melodies, partly new compositions. This leads to a polyphonic orchestral accompaniment.

The musical arrangements have been written specifically for this physics show musical. One reason was the consideration 
of the lyrics which partly deviate from the original verse structures. In some songs the stanzas have been shortened or the 
order of interludes have changed. The main aspect for the arrangements was the instrumentation. All instrumentalists in the 
play are students and musical laypersons, which makes it unsuitable to use original orchestra sheet music. Additionally, the 
unconventional instrumentation required a musical arrangement, which considered the typical timbre and volume of each 
instrument. This enables a well-balanced sound within the orchestra and in combination with the singers. During the 
performance all instruments and singers are recorded by individual microphones to guarantee a good sound-mixing.

\subsection{The Individual Songs}
The seven songs are equally distributed over the full play. The first song in the prologue is a duet with Luna Callisto 
and Jupi  Mercury. They present the portfolio of their galactic planet shop \texttt{PLANETAMOS} and describe the 
requirements for the evolution of life on a planet. The original tune of the song is ``A whole new world'' from the movie 
``Aladdin'', the orchestra arrangement includes all instruments. 

The next song is a summary of three different star classes where Jupi explains the respective characteristics and 
gives Vita advice, which star to choose for her planet. This song is adapted from ``Bare Necessities'' from the movie 
``The Jungle Book'' and instrumented with flute, bass ukulele, piano and percussion. 

At the end of the first act Vita sings a song about her plans of making the new planet worth living on, describing the 
beauty of the landscape and animals she aims to create. Here, the original was the song ``For the first time in 
forever'' from the movie ``Frozen'', again arranged for full instrumentation. 

Right before the break at the end of the second act, Vita and Mortis sing a duet about their love. Both realize that 
their relationship - the interplay between life and death - will rule every being on their planet. This song, again with 
full instrumentation, is based upon the song ``If I never knew you'' from the movie ``Pocahontas''. 

The fifth song at the end of act three is sung by Mortis and accompanied by piano and percussion. It is arranged 
from the song ``What a wonderful world'' and characterizes Mortis'  ideas of a ``wonderful'' world full of hazard, 
natural disasters and diseases. 

In the fourth and last act Luna sings a song about the features of the new planet Vita and Mortis are going to buy. It 
recaps all phenomenons which have been described in experiments before and is structured as the contract of sale. 
The melody is taken from the song ``Dream a little dream of me'' and the instrumentation includes violin, bass ukulele, 
piano and percussion. 

The last song of the play is a quartet sung by all four actors and accompanied by the whole orchestra, with the melody 
from ``Circle of life'' from the movie ``The Lion King''. The planet is addressed, which is now orbiting the sun and ready 
for life to evolve.

\subsection{Instrumental Pieces}
Besides these seven songs the physics show musical also contains several instrumental pieces played by the orchestra. 
When Jupi explains the occurrence of seasons on the planet, each season is accompanied by the respective theme 
from A. Vivaldi's ``Four seasons''. The accompaniment was arranged for violin and cello. Additionally the whole orchestra 
plays a medley which was arranged from melody fragments of the songs ``A whole new world'', ``If I never knew you'' and 
``For the first time in forever''. This medley is performed right after the break in the play and thus represents an overture 
to the second half.

\label{sec:musical-arrangements}

\section{Credits}
The Bonn  Physikshow operates largely as a collective.  The final on-stage product typically evolves
through lots of input during rehearsals. In the  case of this musical there was a clearer distribution
of some of the tasks, in particular involving musical skills. Due to the unusual nature of this production,
we want to point out these special contributions.

\begin{itemize}

\item The first draft of the text for the story was written by David Ohse, who strongly pushed for  this
project from the beginning. The German text was subsequently modified during the rehearsals and
even after the performances. The final text was translated into English by Herbi Dreiner.

\item The texts for the songs were written by Jana Heysel who was joined by Laura Rodriguez Gomez and Johann Ostmeyer
on two songs. The songs were translated into English, also by Jana Heysel.

\item The orchestral arrangements of the songs for the instrumentation, were written by Jana Heysel.

\item The following people played in the orchestra in various performances (alphabetically by instrument):

{\it Cello/Bass-Ukulele}: Jakob Dietl\\
{\it Flute}: Viola Middelhauve, Laura Weber\\
{\it Harp}: Inga Woeste\\
{\it Oboe}: Carsten Urbach, Laura Weber\\
{\it Percussion}: Erik Busley, Kathrin Grunthal\\
{\it Piano}: Heinrich Von Campe\\
{\it Violin}: Lara Becker

\item The sound engineering in particularly for the live songs and the orchestra, as well as the lighting was performed by Till Fohrmann, Christoph Sch\"urmann,
and Peter Rosinsky.

\item The technical support for all experiments was provided by Kristoffer Kerkhof and Michael Kortmann.

\item The scenic design using the jet propulsion laboratory ``Visions of the Future" posters \cite{jpl-Posters},
 was developed by Lara Becker. The planetary mobile hanging from the ceiling was developed by Till Fohrmann.

\item The plan containing the distribution of the experiments on-stage and detailed information for everyone where and when to enter and exit the stage was developed by Jana Sch\"uller-Ruhl.

\item The on-stage camera work to project the experiments was performed by Anne Stockhausen and Joshua Streichhahn.

\item As mentioned, the Bonn Physics Show operates as a collective, and thus many people give feedback during 
rehearsals and there is no true director. The professional artists Karolin Biewald and Babette D\"ormer  gave 
invaluable advice and feedback during several hours of rehearsals, and should be considered assistant directors.

\end{itemize}

\label{sec:credits}

\section*{Acknowledgments}
We would like to thank ESERO Germany for financial support. We thank Katharina Brand (n\'ee Hortmanns) for helping to develop the original idea of a physics show 
musical, for many inspiring sing-alongs and discussions on our numerous travels throughout Europe, which in the end led to this production. Katharina, too bad you 
had left Bonn before we got this far. We thank Peter Rosinsky for help on the sound engineering of the live performances. We thank Philipp Knaus for his consultations 
on paleontology and for providing the earth map suit. We thank Oliver Schl\"omer for providing us with the 
special sand for the sandbox and for geological consultation on the sandbox. We thank Daniel Papendorf for singing instructions, and more! During the rehearsals we 
received coaching and directorial support from two professionals: Babette D\"ormer and Karolin Biewald, for which we are very grateful! One of us, HKD, would like to 
thank Corinne Heath for encouragement to explore all aspects of show activity. We thank Patricia Z\"undorf and Petra Weiss for coordinating the booking process of 
the performances.

\begin{appendix}

\section{The Original Script in German}
\label{app:skript-German}

\subsection{Prolog}
\label{sec:prolog-d}
\begin{itemize}
	\R{Auftritt \Herbi.}
	\H{\Regie{Ins Handmikrophon, zum Publikum.} Herzlich Willkommen. Ich bin Professor Herbi Dreiner \dots theoretische Teilchenphysik \dots Physikshow der Uni Bonn \dots Amerika \dots Blablakeks \dots \Musik{Königin der Nacht} \dots
	}
	\R{Auftritt \Luna\ und \Jupi\ von links.}
	\Lu{\Regie{Bemerkt \Herbi\ nicht, zu \Jupi.} \dots Du kannst doch nicht einfach vergessen, wo du die Kometen hingestellt hast!
	}
	\J{\Regie{Folgt ihr.} Gestern wusste ich noch, wo sie sind.
	}
	\Lu{\Regie{Sucht die Bühne ab.} Hast du schon bei den Asteroiden nachgeguckt?
	}
	\J{Da waren sie nicht.
	}
	\Lu{Und bei den Meteoriten?
	}
	\J{Da auch nicht.
	}
	\Lu{Bei den Sternschnuppen?
	}
	\J{\Regie{Haut sich gegen die Stirn.} Oh. Ja. Ich hab's. Es fällt mir wie Schnuppen aus den Wolken. Ich habe die Kometen bei den ganzen Nebeln einsortiert. \Regie{Bleibt vorne links beim Katalog stehen.}
	}
	\Lu{\Regie{Einmal-mit-Profis-arbeiten:}  Mensch Jupi, das musst du doch in den Katalog eintragen, sonst werden 
	die noch verwechselt! \Regie{Entdeckt \Herbi, geht auf ihn zu und begrüßt ihn in der Bühnenmitte.} Ach hallo. Kann 
	ich Ihnen weiterhelfen?
	}
	\J{\Regie{Holt den Messier-Katalog und trägt Kometen ein.}
	}
	\H{Mein Name ist Herbi. Ich hätte hier gerne eingekauft.
	}
	\Lu{Luna Callisto mein Name. \Regie{Reicht ihm die Hand.} Suchen Sie etwas Bestimmtes?
	}
	\H{Ich möchte einen Planeten kaufen. Mein Heimatplanet ist leider auf die schiefe Bahn geraten \dots
	}
	\Lu{Woran hatten Sie denn da gedacht, Herbi?
	}
	\H{Naja. Also, ich \dots
	}
	\Lu{Gasriese, Eisriese oder Gesteinsplanet? Und {\it (zweideutig)} mögen Sie es lieber heiß oder kalt?
	}
	\H{\Regie{Verlegen.} Ähm. Also es ist schon etwas hei{\ss} zuhause. Hmm, aber deswegen bin ich nicht 
	hier. Also, es ist mir ja etwas peinlich, aber Sie sind ja Expertin, oder? Also, \ldots führen Sie auch Scheibenwelten? 
	Oder hohle Planeten?	
	}
	\Lu{\Regie{Lacht laut auf, dann ernst.} Nein. Haben Sie überhaupt das nötige Kleingeld?
	}\H{Nun, ich bin ja nur Professor \dots
	}
	\Lu{\Regie{Legt ihren Arm auf \textbf{Herbi}s Schulter und komplimentiert ihn nach rechts zu seinem Sitzplatz
	im Publikum, bleibt aber selbst auf der Bühne.} Setzen Sie sich doch und entspannen sich. Sie sind hier bei 
	\texttt{PLANETAMOS} in guten Händen.
	}
	\Lu{Am besten zeigen wir Ihnen erstmal unsere Kataloge, da werden Sie vielleicht fündig, Herbi. Mein Kollege 
	Jupi Mercury präsentiert Ihnen bestimmt gerne unser Portfolio.
	}
	\J{\Regie{Begeistert.} Oh ja, unsere Produktvielfalt. Davon kann ich ein Liedchen singen! \Regie{Geht zur Bühnenmitte.}
	}
	\R{\Herbi\ ab.}
	%
	
		
	\Musik{Orchestermusik: "In meiner Welt" aus dem Film "Aladdin" (Musik von A. Menken)}
	
	\medskip

	\J{{\color{teal}Sag, was kostet die Welt?\\
	Wir zwei könn’n sie erschaffen.\\
	Berge, Täler und Meere\\
	fantasievoll kombiniert.\\\\
	Träumt ihr von einer Welt?\\
	Seht nur her, lasst euch zeigen:\\
	Hochmoderne Planeten\\
	werden hier bei uns kreiert.\\\\
	Auf einer Welt, die wir bei \texttt{PLANETAMOS} bau'n,\\
	kann wirklich viel gescheh'n,\\
	und auch entsteh'n\\
	vielleicht gar neues Leben.}
	}
	\Lu{{\color{teal}Ob eine Welt, so schön und neu für euch gebaut,\\
	Leben auch möglich macht, \\
	Natur erwacht,\\
	hängt von ein paar Kleinigkeiten ab.}
	}
	\J{{\color{teal}Diese Kleinigkeiten sind wie folgt:}
	}
	\Lu{{\color{teal}Sucht euch erst einen Stern.\\
	Wollt ihr nicht beim Umkreisen\\
	brennen oder vereisen,\\
	wählt den richt’gen Abstand aus.}\\\\
	\Regie{Strahlend} {\color{teal}Ein Eisenkern}\dots
	} 
	\J{\Regie{Rezitativ} {\color{teal}\dots gibt ein Magnetfeld zum Schutz.}
	}
	\Lu{{\color{teal}Die Atmosphäre außen rum\dots}
	}
	\J{\Regie{Rezitativ} {\color{teal}\dots schafft ein Klima zum Wohlfühl’n.}
	}
	\Lu{{\color{teal}Damit sich Leben dann\\
	entwickeln kann,\\
	braucht es ganz viel Wasser auf der Welt.}
	}
	\J{\Regie{Strahlend} {\color{teal}Ein Feuerwerk\dots}
	}
	\Lu{\Regie{Rezitativ} {\color{teal}\dots bietet euch ein Vulkan.}
	}
	\J{\dots {\color{teal}mit Asche, Lava, Schwefel, Rauch.}
	}
	\Lu{\Regie{Rezitativ} {\color{teal}\dots regnet der auf den Boden.}
	}
	\LJ{ {\color{teal}Und wenn ihr gleich bestellt,\\
	gibt’s zu der Welt\\
	gratis einen vollen Mond dazu.}
	}
	\J{{\color{teal}Was ihr erträumt\dots}
	}
	\Lu{{\color{teal}Was ihr erträumt\dots}
	}
	\J{\Lyrics{\dots bau’n wir euch gern.}
	}
	\Lu{{\color{teal}\dots bau’n wir euch gern.}
	}
	\J{{\color{teal}Die neue Welt\dots}
	}
	\Lu{\Lyrics{\dots wie’s euch gefällt\dots}
	}
	\LJ{\Lyrics{\dots mit Mond und Stern.}
	}
	\J{\Regie{Schmeißt Konfetti und hält die Arme ausgestreckt nach oben.}
	}
	\Lu{\Regie{Verschränkt die Arme.}
	}
	\R{Freeze.}
\end{itemize}

\subsection{Himmelsmechanik}
\label{sec:himmelsmechanik}
\begin{itemize}
	\R{Auftritt \Vita\ von links. Türklingeln auf dem Klavier. Dadurch erwachen \Luna\ und \Jupi\ zu neuem Leben.}
	\V{Hallo? Guten Tag. Bin ich hier richtig bei \texttt{PLANETAMOS}? \Regie{Hängt ihren Hut am Hutständer auf.}
	}\J{\Regie{Schaut grübelnd ins Publikum.} Oh. Ja! \Regie{Wendet sich nach links zum Tafelglobus, schraubt daran 
	herum, putzt ihn usw..}
	}\Lu{Herzlich Willkomen bei \texttt{PLANETAMOS}, der Adresse für Planeten aller Art. \Regie{Geht nur wenige Schritte auf \Vita\ zu.} 
	Mein Name ist Luna Callisto. \Regie{Reicht ihr die Hand.}
	}\V{\Regie{Stolziert auf \Luna\ zu und schüttelt überschwänglich ihre Hand.} Freut mich! Ich bin das Leben. Nennen 
	Sie mich  Vita. \Regie{Stellt ihre Tasche auf einem Tisch, nimmt ihre Sonnenbrille ab und kramt in der Tasche 
	nach ihrem Brillenetui.}
	}\Lu{Das ich das noch erleben darf. Wie schön, dass  solch ein Leben in unseren kleinen Planetenladen 
	kommt.~\Regie{Kurzes Lachen.}~Mein Kollege Jupi Mercury und ich beraten Sie gerne.
	}\J{\Regie{Schaut auf, winkt kurz und nickt mit Schraubenzieher zwischen den Zähnen.}
	}\V{\Regie{Findet ihr Brillenetui.} Großartig!
	}\Lu{Was darf es sein, Vita?
	}\V{\Regie{Putzt ihre Sonnenbrille mit dem Brillenputztuch.} Mein Mann und ich möchten zusammen einen 
	Planeten kaufen, sind uns aber noch uneins, wie diese Welt aussehen soll. Am besten warten wir noch auf ihn. 
	Mortis parkt noch das Raumschiff. Aber Sie könnten mir schon zeigen, wie Planeten entstehen.
	}\Lu{Aber gerne doch. 
	}\V{\Regie{Fasst \Luna\ am Arm.} Haben Sie \Regie{Zögert} Baby-Planeten hier?
	}\Lu{Auf unsere hauseigene Planetenproduktion sind wir ganz besonders stolz. Wir geben unseren Planeten extra 
	viel Zeit, sich zu entwickeln und lassen sie viele Millionen Jahre im Sternenlicht reifen. Damit erreichen wir höchste Qualität.
	}\V{Wie aufregend! Woraus werden Planeten überhaupt hergestellt? Ich hab ja keine Ahnung.
	}\Lu{Wir nehmen natürlich nur die besten Zutaten aus der protoplanetaren Scheibe um sehr junge Sterne. Die 
	Sternproduktion haben wir ausgelagert, das können andere besser. Wir konzentrieren uns auf unsere Stärken 
	und kaufen die jungen Sterne auf dem Großmarkt ein.
	}\V{Ihre Arbeit beginnt also in den frühen Sternstunden?
	}\Lu{Ganz recht. Mit den Details der Planetenentstehung bin ich nicht so bewandert, da fragen wir besser meinen Kollegen 
	Jupi Mercury, der ist Planetologe.
	}\J{\Regie{Schreckt hoch.} Oh. Ja? \Regie{Säubert seine Hände von Kreide.}
	}\V{Das ist ja spannend. \Regie{Läuft zu Jupi.} Ich wollte schon immer einem Planetenwissenschaftler begegnen! 
	\Regie{Schnappt sich eine Hand von \Jupi\ und schüttelt sie.} Ich heiße Vita, das Leben. Freut mich Sie kennenzulernen,  
	Herr Doktor \dots
	}\J{Jupi reicht. Ich habe lediglich einen Abschluss als Master of the Universe. Wann ich meine Doktorarbeit fertigstellen 
	kann, steht noch in den Sternen. Die Experimente in der Planetologie dauern sehr lange, wissen Sie.
	}\Lu{\Regie{Geht nach hinten rechts und blättert in einem Katalog.}}
	\V{Dann erzählen Sie doch mal, wie Baby-Planeten sich entwickeln.}
	\J{Also, um Planeten herzustellen, beginnen wir mit einer protoplanetaren Scheibe um einen jungen Stern. Hier sehen 
	sie ein Bild von so einer Staubscheibe. Wie Sie sehen, ist die Scheibe in der Nähe des Sterns hell erleuchtet und wird 
	nach außen hin immer dunkler.}
	\V{Wie interessant. Und wie werden aus der Scheibe jetzt die Planeten?}
	\J{Zunächst einmal ist die protoplanetare Scheibe nicht überall gleich: Nah am Stern ist es warm, weiter weg kalt. Das sorgt 
	dafür, dass das Material in der Scheibe je nach Abstand entweder gasförmig oder fest ist. Nehmen wir z.B. diesen Ballon 
	mit Kohlenstoffdioxid, wie es auch in der Scheibe vorkommt.}
	
	\medskip
	
	{\bf Experiment:} \parbox[t]{11.7cm}{{\bf Luftballon in fl\"ussigem Stickstoff ($N_2^{\text{fl{\"{u}}ssig}}$), Kondensation}, \textit{cf.} App.~\ref{app:condensation}.}
	
	\medskip
	
	 \J{Bei der Temperatur hier im Planetenladen ist diese chemische
	 Verbindung gas\-förmig. In diesem Behälter aber ist flüssiger Stickstoff und der ist $-200\,^\circ\mathrm{C}$ kalt. Wenn ich 
	 den Ballon dort hinein drücke, kühlt das Kohlenstoffdioxid in ihm ab. \Regie{Jupi steckt den Ballon in $N_2$ 
	 und öffnet ihn dann.} Dadurch wird das Kohlenstoffdioxid zu Trockeneis. Das erstarrte Kohlenstoffdioxid ist ein feiner Schnee.}
	\V{Also ist in den äußeren Bereichen der Scheibe das Kohlenstoffdioxid fest und innen ein Gas?}
	\J{Ja genau. Das selbe gilt für Wasser. Gestein und Metall jedoch sind in der gesamten Scheibe fest in Form eines feinen 
	Staubes, so wie dieser hier.}
	
	\medskip
	
	{\bf Experiment: Zuckerwatte},  \textit{cf.} App.~\ref{app:cotton-candy}.
	
	\medskip
	
	 \J{Gemeinsam kreisen die Staubkörner in einer Scheibe um einen jungen Stern. 
	Wenn zwei Staubkörner aufeinandertreffen, kleben sie aneinander.}
	\V{Wie der Staub unterm Bett!}
	\J{Oh, ja. So entstehen nah am Stern Brocken aus Metall und Fels. Weiter außen, wo der z.B. das Kohlenstoffdioxid und 
	das Wasser fest gefroren sind, kommt ganz viel Eis dazu. Sehen Sie?}
	\V{\Regie{Hände an die Wangen.} Oooh, wie süß. Und was ist des Pudels Kern?
	}\J{Die Frage verstehe ich nicht. Pudel haben keine Kerne.
	}\V{Ich wollte fragen, was die Welt im Innersten zusammenhält?
	}\J{Achso. Also die kleinen Brocken kleben zusammen nachdem sie sich zufällig getroffen haben. Wenn die Brocken 
	größer werden, hält die Schwerkraft sie zusammen und sorgt dafür, dass sie schnell wachsen.
	}\V{\Regie{Entsetzt} Sie meinen, die Großen fressen die Kleinen? Sind Planeten etwa Kannibalen?
	}\J{So würde ich es nicht formulieren. 
	Aber wenn ein Planet dabei eine bestimmte Größe überschreitet, kann er ganz viel Gas an sich binden. Man nennt ihn 
	dann einen Gasriesen. \Regie{Reicht \Vita\ die Zuckerwatte.}
	}\V{\Regie{Probiert die Zuckerwatte.} Entzückend! Welche Planetengrößen bieten Sie denn an?
	}\Lu{\Regie{Blickt auf, stellt sich rechts neben \Vita\ und zeigt ihr die entsprechende Seite im Katalog.} Wir führen alle 
	Größen, von wenigen tausend bis hin zu über hunderttausend Kilometern im Durchmesser. Bedenken Sie, Vita, bei 
	doppeltem Durchmesser haben Sie viermal so viel Platz auf der Oberfläche.}
	\R{\Luna\ will \Vita\ einen möglichst großen Planeten andrehen, \Jupi\ versteht nicht...}
	\J{Aber auch doppelt so viel Schwerkraft!}
	\Lu{Dafür kann ein größerer Planet eine Atmosphäre halten.}
	\J{Aber wenn er zu groß ist, wird er zum Gasriesen!}
	\V{Mh. Aber so ein Gasriese hat keine feste Oberfläche, sagen Sie?}
	\J{Da haben Sie keinen Boden unter den Füßen. Mit einer festen Planetenkruste hat man deutlich mehr Möglichkeiten.}
	\Lu{Wie wäre es denn mit einem großen Gesteinsplaneten?}
	\V{Das hört sich gut an. Welchen Durchmesser können Sie denn empfehlen? \Regie{Steckt die Zuckerwatte zu den 
	anderen Baby-Planeten auf dem Tisch.}
	}\Lu{Das Modell Terra-Firma mit 12.700 Kilometer Durchmesser wird immer gern genommen.
	}\V{Na so was, 12.700 ist meine Lieblingszahl!
	}\Lu{Eine gute Wahl.}
	\V{\Regie{Sie legt einen Arm um \Luna\ und gestikuliert.} Wissen Sie, Luna, ich sonne mich für mein Leben gern. Mein 
	größter Traum ist es, in einer blühenden Landschaft zu liegen und das Prickeln wärmender Strahlung auf meiner Haut 
	zu genießen.
	}\Lu{Das Gefühl kennen wir bei \texttt{PLANETAMOS} von den Quartalsabrechnungen. Da sonnen wir uns gerne in unserem 
	Erfolg. \Regie{Kichert}
	}\J{Oh. Ja!
	}\V{Was ist denn für einen florierenden Planeten ratsam?
	}\Lu{Jupi, wie können wir dafür sorgen, dass sich das Leben auf einem Planeten wohlfühlt?
	}\J{\Regie{Stellt sich rechts neben \Luna.} Wir brauchen eine stetige Energiequelle in seiner Nähe, die Licht und Wärme 
	spendet. Dazu lassen wir den Planeten einfach um einen Stern kreisen. \Regie{Zu \Vita.} Es lebt sich recht angenehm 
	auf einem Planeten, der in geordneten Bahnen um einen Stern zieht.
	}\V{Aber können wir nicht gleich zwei Sterne nehmen? \Regie{Denkerpose.}
	}\Lu{Natürlich, Vita. Für einen geringen Aufpreis setzen wir Sie gerne in ein Doppelsternsystem.
	}\J{Ja. Wobei es da schwierig sein dürfte, eine stabile Umlaufbahn in Sternnähe zu finden.
	}\Lu{\Regie{Tadelt \Jupi\ mit einem Blick.} Dafür hätten Sie tagsüber zwei Schatten.
	}\J{\Regie{Hat den Blick von \Luna\ nicht bemerkt.} So eine Binärkonstellation würde zu erheblichen Schwankungen 
	im Strahlungsfluss auf dem Planeten führen.
	}\Lu{\Regie{Stößt \Jupi\ einen Ellbogen in die Rippen.} Mein Kollege möchte nur andeuten, dass das Leben mit zwei 
	Sternen viel abwechslungsreicher ist.
	}\V{\Regie{Schaut von einem zum andern und überlegt.} Mh. Vielleicht nehmen wir erstmal nur einen Stern. Wir müssen 
	uns ja noch steigern können.
	}\Lu{\Regie{Zerknirscht.} Wie Sie wünschen.
	}\V{Was für Sterne haben Sie denn vorrätig?}
	
	\medskip
	
	{\bf Experiment:} \parbox[t]{11cm}{{\bf leuchtender Ballon-Stern mit rotierendem Planeten}, \textit{cf.} App.~\ref{app:planet-pendulum}.}
	
	\medskip
	
	\J{\Regie{Steigt über den Tisch und bringt die Sternhülle in Stellung.}
	}\Lu{\Regie{Wieder in ihrem Element.} Das ganze Spektrum an Möglichkeiten! O, B, A, F, G und sogar K oder M-Sterne.
	}\V{Bitte was?}
	\J{Mit dieser Buchstabenfolge bezeichnen wir die verschiedenen Sternenmodelle. {\bf O}ffenbar {\bf B}enutzen 
	{\bf A}stronomen {\bf F}urchtbar {\bf G}erne {\bf K}omische {\bf M}erksätze.}
	\Lu{\Regie{Schreibt die Buchstabenfolge OBAFGKM untereinander auf einen Flipchart.}
	}\V{Und was ist der Unterschied?}
	\Lu{Nun, Vita, die Sternmodelle unterscheiden sich zuerst in ihrer Masse und deswegen in ihrer Temperatur und 
	Helligkeit. Die leichtesten, kühlsten und leuchtschwächsten Sterne nennen wir M-Sterne. 
	}\J{\Regie{Schaltet die Beleuchtung des Ballons auf rot.}  Bei diesen niedrigen Temperaturen von $3.000$ Grad Celsius 
	fangen Sterne an rot zu glühen. Wir nennen Sie rote Zwerge. \Regie{Hängt den Planeten ein an dem Faden.} Bei derart 
	kühlen und dunklen Sternen müssen wir den Planeten in eine kleine, enge Umlaufbahn bringen, damit es warm genug 
	ist und sich das Leben dort entfalten kann. \Regie{Bringt den Planeten auf eine kleine Umlaufbahn.}}

	\medskip

	\Musik{Musik: "Jupiter" aus der Orchestersuite "The Planets" (G. Holst)}
	
	\medskip
	
	\J{In der bewohnbaren Zone um einen Stern sind die Temperaturen genau richtig, um sich im Sternlicht zu sonnen. 
	}\V{Das kann ich mir lebhaft vorstellen.
	}\Lu{Durch den geringen Abstand sieht der Stern am Himmel ganz groß aus. Gefällt Ihnen der M-Stern, Vita?
	}\V{\Regie{Stemmt die Arme in die Hüfte.} Der sieht schon cool aus.
	}\J{Ja, allerdings neigen M-Sterne zu unregelmäßigen Ausbrüchen von Plasma. Außerdem ist der Planet so nah am 
	Stern dran, dass die Schwerkraft ihn fest im Griff hat. Der Planet kann sich dann nicht mehr frei drehen und wendet dem 
	Stern immer die gleiche Seite zu. Man nennt das gebundene Rotation. Auf der sternzugewandten Seite ist immer 
	Tag und es wird recht warm, während auf der dunklen Seite bitterkalte, ewige Nacht herrscht.
	}\V{Wie öde. Nein, ich brauche viel Abwechslung in meinem Leben. Ohne einen Tag-Nacht-Rhythmus verliere ich jedes Zeitgefühl.
	}\Lu{Ich verstehe. Wie wäre es stattdessen mit einem heißen O-Stern? 
	}\J{\Regie{Bläst den Ballon mit Druckluft auf und schaltet dabei von rot auf orange, weiß und schließlich blau.} Oh ja!  
	Dafür muss ich erstmal sehr viel Masse hinzufügen. Ein O-Stern wiegt tausendmal mehr als ein M-Stern. Entsprechend 
	sind O-Sterne sehr hell, heiß und glühen blau. 
	\Regie{Zusätzlicher Spot auf den Luftballon.}
	}\V{Diese bewohnbare Zone, von der Sie eben sprachen, ist dann weit von dem hellen O-Stern entfernt?
	}\J{Ganz genau. \Regie{Bringt den Planeten auf ein großes Orbit.} Übrigens fegt die starke Strahlung des Sterns den 
	meisten Staub weg. Da haben Sie freie Bahn für Ihren Planeten.}
	
	\medskip

	\Musik{Musik: "Jupiter" aus der Orchestersuite "The Planets" (G. Holst)}
	
	\medskip
	
	\Lu{Diesen O-Stern kann ich Ihnen wärmstens ans Herz legen, Vita.
	}\V{Ich könnte  mich schon dafür erwärmen. Wie lange ist der denn haltbar?
	}\J{Tja, diese O-Sterne leben nicht sehr lange. Das ist ein kurzes Vergnügen. Nach ein paar Millionen Jahren explodieren sie. 
	}\V{Eine furchtbare Vorstellung! Der Stern vergeht, wenn das Leben auf dem Planeten gerade in voller Blüte steht. 
	Das kommt nicht in Frage.
	}\Lu{Wie wäre es dann mit einem mittelschweren G-Stern, Vita? 
	}\J{Oh! Ja, das müsste gehen. Ich lasse wieder etwas Masse ab. \Regie{Lässt Luft aus dem Ballon und schaltet auf 
	blau und weiß.}
	}\V{Und G-Sterne leben lange genug und stoßen nicht ständig Masse aus, Jupi?
	}\J{Ja, genau. Da haben Sie erstmal zehn Milliarden Jahre Ruhe. Solange bleibt der G-Stern in einem stabilen Zustand. 
	\Regie{Bringt den Planeten auf eine Kreisbahn.} Wenn wir den Planeten in die bewohnbare Zone setzen, braucht er 
	genau pi mal zehn hoch sieben Sekunden um seinen Stern zu umkreisen. Wir nennen das ein JAHR.}
	
	\medskip

	\Musik{Musik: "Jupiter" aus der Orchestersuite "The Planets" (G. Holst)}
	
	\medskip
	
	\V{ Können Sie das noch einmal zusammenfassen?
	}\J{Ja, gerne.}

	\medskip
		
	\Musik{Orchestermusik: \parbox[t]{11.7cm}{"Probier's mal mit Gemütlichkeit" aus dem Film "Das Dschungelbuch" (Musik von T. Gilkyson)}}
	
	\medskip
	
	\J{{\color{teal}Probier‘s mal mit ‘nem roten Stern,\\
	doch kreis‘ um ihn nicht allzu fern,\\
	beim M-Stern gibt's gebund'ne Rotation.\\
	Das führt dazu, dass der Planet\\
	sich nicht mehr um sich selber dreht.\\
	Fürs Leben ist das nicht so die Option.\\\\
	Magst du es heißer und die Umlaufbahn weit,\\
	ein blauer O-Stern stünd‘ dafür bereit.\\
	Doch hast du nicht viel Spaß daran,\\
	ein O-Stern lebt nicht all zu lang.\\
	Dann macht es \enquote{Puff}, er explodiert\\
	und deine Welt wird ausradiert.\\
	Das willst du doch wohl nicht?\\
	Nein, nein, ein O-Stern ist hier nicht die richt‘ge Wahl,\\
	der wär‘ fatal.\\\\
	Probier mal für dein Ziel durchaus\\
	‘nen mittelschweren G-Stern aus.\\
	Ein G-Stern bietet alles, was du brauchst!\\
	Er lebt ein paar Milliarden Jahr‘\\
	und eignet sich ganz wunderbar\\
	für Leben auf der Welt, die ihn umkreist.\\
	Ein bisschen Wärme, ein bisschen Licht,\\
	da lebt man gerne, vergiss das nicht!\\
	Die Bienen summen in der Luft,\\
	es herrscht ein süßer Blumenduft.\\
	Das Leben wird in seiner Pracht\\
	erst möglich durch den Stern gemacht.\\
	Darum lass dir gesagt sein:\\
	Ja, so ein G-Stern ist zum Leben ideal,\\
	die beste Wahl!}
	}\V{Fantastisch. Der G-Stern ist gekauft. Genau das brauche ich.
	}\Lu{Sehr gerne. Wir besorgen Ihnen einen G-Stern, und zwar sternhagelgünstig.
	}\V{Da fällt mir noch was ein. Können sie für etwas Abwechslung und verschiedene Jahreszeiten sorgen?
	}\J{Jahreszeiten sind kein Problem. Dazu kippen wir einfach die Drehachse des Planeten. Schauen Sie. An 
	diesem Globus wird das deutlich. \Regie{Springt zum Globus.}
	}\V{\Regie{Schaut von rechts aus zu, holt eine Flasche aus ihrer Handtasche und trinkt.}
	}\Lu{\Regie{Führt den Scheinwerfer nach.}
	}\J{Die Drehachse des Planeten zeigt immer in die gleiche Richtung, selbst wenn dieser auf seiner Umlaufbahn 
	weiterzieht. Diese Hälfte \Regie{Deutet auf die obere Hälfte.} des Planeten nennen wir Nordhalbkugel, die andere 
	\Regie{Deutet auf die untere Hälfte.} heißt Südhalbkugel. Die Äquatorlinie \Regie{Dreht den Globus und zeichnet 
	dabei mit Kreide den Äquator ein.} trennt die beiden Halbkugel. Wenn der Stern senkrecht auf den Äquator strahlt 
	und das Licht beide Polargebiete streift, bekommen beide Halbkugeln gleicht viel Wärme ab und es ist Frühling.
	}\R{Alle drei Akteure auf der Bühne gefrieren.}

	\medskip

	\Musik{Orchestermusik: \parbox[t]{11.7cm}{"Frühling" aus den Violinkonzerten "Die vier Jahreszeiten" (A. Vivaldi)} }
	
	\medskip
	
        \R{Nach dem Ende der Musik machen alle drei Akteure normal weiter.
	}\J{Ein viertel Jahr später steht der Planet hier. \Regie{Schiebt den Globus nach hinten in die Mitte der B\"uhne.} Jetzt 
	herrscht um den Südpol herum Polarnacht. Am Nordpol hingegen geht der Stern Monate lang nicht unter. Insgesamt 
	bekommt die Nordhalbkugel hier viel mehr Strahlung ab als die Südhalbkugel. Auf der Nordhalbkugel ist dann Sommer.
	}\Lu{\Regie{Stellt sich dahin, wo der Globus im Frühling steht.}
        }\R{Alle drei Akteure auf der Bühne gefrieren.}
        
        \medskip
        
	\Musik{Orchestermusik: \parbox[t]{11.7cm}{"Sommer" aus den Violinkonzerten "Die vier Jahreszeiten" (A. Vivaldi)}}
	
	\medskip
	
	\R{Nach dem Ende der Musik machen alle drei Akteure normal weiter.}
	\J{Wenn der Planet weiter zieht, werden die Tage auf der Nordhalbkugel wieder kürzer, im Süden länger. 
	\Regie{Schiebt den Globus weiter nach rechts.} Irgendwann bekommen beide Halbkugeln wieder gleich viel Licht ab. 
	Überall sind Tage und Nächte gleich lang. Diese Tag-und-Nacht-Gleiche markiert den Herbstanfang.
	}\R{Alle drei Akteure auf der Bühne gefrieren.}
	
	\medskip
	
	\Musik{Orchestermusik: \parbox[t]{11.7cm}{"Herbst" aus den Violinkonzerten "Die vier Jahreszeiten" (A. Vivaldi)}}
	
	\medskip
	
	\R{Nach dem Ende der Musik machen alle drei Akteure normal weiter.}
	\Lu{Genau, und danach kommt hier auf der Nordhalbkugel der Winter. \Regie{Stellt sich an die entsprechende Stelle.}
	}\R{Alle drei Akteure auf der Bühne gefrieren.}
	
	\medskip
	
	\Musik{Orchestermusik: \parbox[t]{11.7cm}{"Winter" aus den Violinkonzerten "Die vier Jahreszeiten" (A. Vivaldi)}}
	
	\medskip
	
	\R{Nach dem Ende der Musik machen alle drei Akteure normal weiter.}
	\V{Und danach ist Karneval!}
	
	\medskip
	
	\Musik{Musik: "Viva Colonia" (H\"ohner)}
	
	\medskip
	
	\R{Alle 3 Akteure auf der B\"uhne tanzen zur Musik.}
	\V{Wow, phänomenal. Mein eigener Planet.  Und er wird voller Leben sein!}
	
	\medskip
	
	\Musik{Orchestermusik: \parbox[t]{11.7cm}{"Zum ersten Mal" aus dem Film ``Die Eiskönigin" (Musik von K. Anderson-Lopez und R. Lopez)}}
	
	\medskip
	
	\V{{\color{teal}Endlich hier, ein Traum wird wahr, alles hier ist planetar.\\
	So viele Planeten hab ich nie geseh’n.\\
	Ich habe so oft davon geträumt, die ein oder andere Chance versäumt.\\
	Endlich bring‘ ich Leben in die Welt!\\
	Magnetfeld, Atmosphäre, Wasser und Umlaufbahn,\\
	doch damit ist es lange nicht getan.\\\\
	Schaut nur her, ich will euch zeigen, diese Welt wird lebenswert.\\
	Grüne Bäume, bunte Blumen, das ist, was zu mir gehört.\\
	Hier die hohen Berge und dort das weite Meer\\
	es wird der Lebensraum für viele vereint in einer Welt.}\\
	\Regie{Gesprochen:} {\color{teal}Ich kann’s kaum erwarten, endlich loszulegen! Was soll ich nur als erstes tun?}\\
	\Regie{Gesungen:} {\color{teal}Sterne funkeln am Himmelszelt, ein Mond, der sich dazu gesellt.\\
	Die Eule hört man weither durch die Nacht.\\
	Die Sonne gibt durch ihre Kraft das Licht, das neues Leben schafft.\\
	Es blüht und grünt so grün auf weiter Flur.\\
	Vögel fliegen durch die Lüfte, die Biene ist in Aktion,\\
	die Raupe fliegt als Schmetterling davon.\\\\
	Schaut nur her, ich will euch zeigen, diese Welt wird lebenswert.\\
	Katzenbabys, Hundewelpen, das ist, was zu mir gehört.\\
	Die ganze Flora und Fauna ist fröhlich, bunt und schön.\\
	Es wird der Lebensraum für viele, ja, der Lebensraum für alle, vereint in einer Welt.}}
	\R{Freeze aller drei Akteure.}
\end{itemize}

\subsection{Mond und Meer}
\label{sec:mond}
\begin{itemize}
	\R{Auftritt \Mortis, verkleidet als der Tod, mit Kutte und Sichel, von der R\"uckseite des Auditoriums. Geht 
	durch das Publikum Richtung B\"uhne.}
	
	\medskip
	
	\Musik{Musik: \parbox[t]{13.2cm}{"Lied vom Tod" aus dem Film "Spiel mir das Lied vom Tod" (Musik von E.~Morricone)}}
	
	\medskip
	
	\M{\Regie{H\"angt seine Sense an den Hutständer, und nimmt dann die Kopfhörer ab. In dem Moment h\"ort die
	Musik auf. \Mortis\ geht langsam zum Klavier und lässt den Schicksalsschlag \{\kern-0.1em das Motiv, bestehend aus den 
	ersten vier Noten aus dem ersten Satz von Beethovens 5. Sinfonie, der Schicksalsinfonie\} spielen.}
	}\R{\Vita, \Luna\ und \Jupi\ lösen sich aus ihrer Totenstarre.}
	\V{\Regie{Strahlt über das ganze Gesicht.} Mortiii! Da bist du ja endlich. \Regie{Stürmt auf \Mortis\ zu, wirbelt um ihn 
	herum und umarmt ihn.} War es schwierig, einen Parkplatz zu finden? 
	}\M{Ahgh.
	}\LJ{\Regie{Tauschen angstvolle Blicke.}
	}\V{Bist du sicher, dass du das Raumschiff abgeschlossen hast?
	}\M{\Regie{Drückt auf den Raumschiffschlüssel.} \Musik{Autoabschließgeräusch}
	}\V{Luna, Jupi, darf ich vorstellen? Mein Mann: Mortis. \Regie{Hakt den Arm bei ihm unter.} Schatz, das sind 
	Luna und Jupi.
	}\Lu{ \Regie{Mit leichter Todesangst.} Willkommen bei \texttt{PLANETAMOS}. Wir haben todsicher den richtigen 
	Planeten für Sie. Darf ich Ihnen beiden etwas zu trinken anbieten?
	}\V{Wie aufmerksam von Ihnen. Für mich bitte einen Früchtetee mit allen Sorten die sie da haben. 
	\Regie{Zu \Mortis.} Du bist bestimmt todmüde, Schatz. Was nimmst du?
	}\M{Kaffee.
	}\Lu{Gerne doch. Zucker? Ein Schuss Ursuppe dazu?
	}\V{Ja, bitte!
	}\M{Schwarz.
	}\Lu{Natürlich. Jupi, kochst du unserer geschätzten Kundschaft die gewünschten Heiß\-getränke?
	}\J{\Regie{Schreckt auf.} Oh! Ja. Früchtetee. Schwarzer Kaffee. Bin schon unterwegs.
	\Regie{Ab B\"uhne rechts.}
	}\V{Stell dir vor, Liebling, ich habe Baby-Planeten gesehen! Total faszinierend. Ich habe schon mal ein 
	Planetenmodell ausgesucht und einen passenden Stern \ldots  um den er kreisen kann. Wir nehmen 
	einen Gesteinsplaneten und dazu einen wundervollen G-Stern. Durch die Kippung der Planetendrehachse 
	bekommen wir gratis alle Jahreszeiten dazu. Toll, nicht?
	}\M{Ahgh.
	}\Lu{Wo wir gerade bei der Drehachse und den Jahreszeiten sind, möchte ich Ihnen ein Angebot machen. 
	Wie wäre es mit ein oder zwei Monden für Ihren Planeten? \Regie{Holt einen Mond an einem Stab hervor.} 
	Unser Silver-Satellite-Special enthält einen extra großen Mond.
	}\M{Unnötig.
	}\V{Aber Schatz, ein Mond ist sooo romantisch! Da können wir stundenlang im Mond\-licht spazieren gehen.
	}\M{Ahgh.
	}\Lu{Monde sind nicht nur gut für romantische Flitterwochen, sondern haben auch einen praktischen Nutzen. 
	Sie stabilisieren die Drehachse ihres Planeten. Das sichert die Jahreszeiten gegen Störungen durch Schwankungen 
	der Achse.
	}\V{Das ist ja interessant. Erzählen Sie uns mehr über dieses mondäne Phänomen, Luna.
	}\Lu{Beim Silver-Satellite-Special ist der Abstand zwischen Planet und Mond 60 mal größer als der Planetenradius. 
	Diese Umlaufbahn führt zu gebundener Rotation. Vita, erinnern Sie sich, was mein Kollege Jupi über gebundene 
	Rotation bei M-Sternen erzählt hat?
	}\V{Oh, ja. Ich glaube, gebundene Rotation bedeutet hier, dass der Mond dem Planeten immer die gleiche Seite zuwendet.
	}\M{Todlangweilig.
	}\Lu{Gerne versuche ich, Ihnen diesen Mond anzudrehen, wenn Sie wünschen. \Regie{Dreht den Mond an und 
	wartet, bis  jemand diesen Wortwitz versteht.} Sie könnten dort ein eigenes Haus bauen, Mortis. Unabhängig von 
	der gebundenen Rotation ändert der Mond aber bei allen Editionen regelmäßig seine Gestalt.
	}\R{Die Kamera filmt die Mondsichel und projiziert das Bild auf die Leinwand.}
	\Lu{ \Regie{Hält den Mond jeweils an die richtige Stelle relativ zur Kamera, zeigt auf die Leinwand.} Hier können Sie 
	mitverfolgen, wie der Mond vom Planeten betrachtet seine Gestalt verändert. Steht der Mond zwischen Stern und 
	Planet, ist Neumond. Wenn er weiterzieht, wächst seine Sichel bis zum Vollmond an. Danach nimmt der Mond wieder ab.
	}\V{Das sieht bestimmt überirdisch aus. \Regie{Zu \Mortis.} Dann können wir in unserer Welt einen Mondkalender aufstellen, Mortis!
	}\M{Ahgh.
	}\V{Liefern Sie den Mond eigentlich separat?
	}\Lu{Nein, da machen wir uns das Leben leicht. Auf den noch jungen Planeten schießen wir einen etwas kleineren 
	Planeten. Bei dem gewaltigen Aufprall verschmelzen die beiden Planeten und schleudern Gesteinsmaterial in die 
	Umlaufbahn. Dieses verklumpt nach und nach zu einem Mond, der den Planeten umkreist.}
	\R{Animation von der Mondentstehung.\footnote{We used the film: \texttt{motion2\_25.avi} from the website: 
	\url{https://www.boulder.swri.edu/~robin/moonimpact/}.}}
	
	\V{Raffiniert.
	}\Lu{Ein weiterer Vorteil unseres innovativen Silver-Satellite-Special besteht in den Gezeitenkräften des Mondes 
	auf der Planetenoberfläche. Einerseits zerrt die Schwerkraft des Mondes an dem Planeten und andererseits 
	kreisen Mond und Planet um einen gemeinsamen Schwerpunkt, sodass die Fliehkraft an der mondabgewandten 
	Seite zieht. 
	Das ruft spektakuläre Effekte hervor, besonders an der Küste. Wir empfehlen dafür eine Flüssigkeit auf der Planetenoberfläche.
	}\V{Eine Flüssigkeit auf der festen Planetenkruste, sagen Sie? Wie wäre es mit \dots
	}\M{\dots Schwefelsäure?
	}\V{Nein, Liebling, keine ätzende Säure. Die zerstört noch unseren schönen Planeten. Ich wäre eher für \dots
	}\R{Auftritt \Jupi\  mit Tablett.}
	\M{\dots Quecksilber?
	}\J{Mercury? Hab ich da meinen Namen gehört?
	}\V{Nein, lieber nichts giftiges. Aber \dots 
	}\M{\dots glühende Lavaströme?
	}\V{Nein. Wasser. Das wäre schön.
	}\M{Ahgh.
	}\J{Oh ja! Wasser eröffnet viele Möglichkeit. \Regie{Ängstlich gegenüber \Mortis.}
	}\Lu{Bedienen Sie sich erstmal.
	}\M{\Regie{Kommt auf \Jupi\ zu, nimmt sich die schwarze Kaffeetasse und atmet tief ein.} Kaffee. \Regie{Genießt 
	seinen Kaffee und taut dabei auf.}
	}\Lu{Der Tee muss wohl noch etwas ziehen, um seine \Vita-lisierende Wirkung zu entfalten. Machen Sie es sich 
	schonmal gemütlich.
	}\R{\Luna, \Vita\ und \Mortis\ setzen sich.
	}\J{\Regie{Stellt das Tablett vor den anderen ab.} Leider ist Wasser auf einer Planeten\-ober\-fläche typischerweise 
	nicht flüssig. In der bewohnbaren Zone liegt Wasser meistens als Gas vor. Im äußeren Sonnensystem wird es 
	direkt zu Eis. Sie erinnern sich bestimmt daran, dass nur dort Eisplaneten entstehen können. Bei dem minimalen 
	Druck im Weltall kann es kein flüssiges Wasser geben. 
	}\Lu{Aber wir bieten Ihnen die ultimative Lösung.
	}\Regie{Verschwörerischer Blick zwischen \Luna\ und \Jupi.}
	\LJ{Eine Atmosphäre!
	}\M{\Regie{Bezogen auf den Kaffee.} Ahgh. Schwarz. Heiß. Lecker.
	}\Lu{Eine Luftschicht um den Planeten baut den nötigen Druck auf, um Wasser zu verflüssigen. Das schafft eine gute Atmosphäre. 
	}\J{
	Dürfte ich das an Ihrem Tee verdeutlichen, Vita?
	}\V{Kein Teema, Jupi.}
	
	\medskip
	
	{\bf Experiment: volle Teetasse unter der Vakuumglocke}, \textit{cf.}  App.~\ref{exp:tea-vacuum}.
	
	\medskip
	
	\J{\Regie{Stellt den Tee in die Vakuumglocke und pumpt die Luft ab.} Hier im Welten\-laden herrscht normaler 
	Atmosphärendruck und eine angenehme Raumtemperatur. Wasser ist da offensichtlich flüssig. Nun stelle ich 
	den Tee unter diese Glasglocke und sauge die Luft heraus. Die Temperatur bleibt dadurch unverändert, 
	aber der Druck sinkt. Es drückt nicht mehr so viel Luft auf den Tee und es entsteht ein luftleerer Raum.  Das 
	entspricht schon eher den Bedingungen im Weltall.}
	
	\medskip
	
	\Musik{Musik: "Aquarium" aus der Suite "Karneval der Tiere" (Musik von C. Saint-Saëns)}
	
	\medskip
	
	\J{Sehen Sie, das Wasser wird gasförmig! Es kocht, aber bei Zimmertemperatur! \Regie{Lässt Luft in die 
	Vakuumglocke, hebt die Glocke und reicht \Vita\ ihren Tee.}
	}\Lu{Bevor Ihr ganzer Tee verdampft, dürfen Sie ihn natürlich trinken, Vita.
	}\V{Beeindruckend! Ohne Atmosphäre würde also das ganze Wasser verdampfen?
	}\J{Oh, ja. Oder gefrieren, je nach Temperatur. Übrigens brauchen Sie eine wirklich dichte Atmosphäre. 
	Bei unserem Modell Mars-Krasni reicht der Atmosph\"arendruck zum Beispiel nicht aus für flüssiges Wasser auf der Oberfläche.}
	\Regie{Foto von Mars-Krasni einblenden.}
	\V{Woher nehmen Sie eigentlich das ganze Wasser für unseren Planeten?
	}\M{Gute Frage. Ohne Wasser kein Kaffee.
	}\Lu{Sehr richtig. Um die Ozeane zu erschaffen, lassen wir Eisklumpen aus dem äußeren Sonnensystem auf den Planeten prasseln. Das ist logistisch gesehen die einfachste Möglichkeit so viel Wasser auf Ihre Welt zu bringen.
	}\VM{Ahgh.
	}\Lu{Verzeihen Sie meine Neugier. Darf ich fragen, wo Sie zwei sich kennengelernt haben?
	}\V{\Regie{Schaut \Mortis\ verliebt an.} Weißt du noch, Mortiii? Damals im Restaurant am Ende des Universums?\footnote{A.d.V.: Diese Erörterung lässt sich beliebig ausschmücken. Wie wäre es mit einem Date in der Big Bang Burger Bar? Oder einer Konferenz für Disruptive Change Maker? Vielleicht sind die beiden doch Geschwister.} Da haben wir uns \dots}
	\M{\Regie{Lächelt.} \ldots unsterblich verliebt.}
	
	\medskip
	
	\Musik{Orchestermusik: \parbox[t]{11.7cm}{"If I never knew you" aus dem Film "Pocahontas" (Musik von J. Secada und Shanice)}}
	
	\medskip
	
	\M{\Lyrics{Seit du f{\"u}r mich da bist,\\
	seit du in mein Leben tratst,\\
	sp{\"u}re ich, wie wundersch{\"o}n das Universum ist.}}
	\V{\Lyrics{Wenn du mich in den Arm nimmst,\\
	wenn du mich h{\"a}ltst, dann werd' ich still.\\
	Am Ende finde ich in dir, die Ruhe, die ich brauch'.}}
	\VM{\Lyrics{Was ich lange nicht verstand,\\
	ahnte lediglich,\\
	habe ich jetzt klar erkannt:\\
	Ich bin nichts ohne dich.\\
	Denn du bist mein Begleiter.\\
	Ich kann immer auf dich bau'n.\\
	Alle Zeiten kann ich dir blind vertrau'n.}}
	\M{\Lyrics{Wir bauen eine Welt zum Leben auf,\\
	dort fühlen wir uns beide wohl.}
	}\V{\Lyrics{Stell dir nur vor, was wir schaffen in der Welt,\\
	kann erzählen von dem Spiel zwischen uns zwei'n,\\
	Altes weicht und Neues kann daraus gedeih'n.}
	}\VM{\Lyrics{Auf unserem Planeten ziehen wir zusammen ein.}
	}\M{\Lyrics{Anfangs w\"ust und leer.}
	}\V{\Lyrics{Doch wir schaffen mehr.}
	}\VM{\Lyrics{Tod und Leben, das perfekte Paar.}
	}\R{Freeze.}
	\Lu{\Regie{Gesprochen.} Sie beide sind f\"ureinander geschaffen, das ist sonnenklar.
	}\M{Geht Sie nichts an.
	}\R{Black. 15min Pause.}
\end{itemize}

\subsection{Atmosphäre}
\label{sec:atmosphaere}
\begin{itemize}
\item[] \Musik{Orchestermusik: \parbox[t]{11.7cm}{Overtüre: Instrumentales Medley der Lieder ``In meiner Welt", 
	``If I never knew you" und ``Zum ersten Mal"}}
	
	\medskip
	
	\R{Alle sitzen beisammen.}
	\M{\Regie{Zu \Luna.} Kleiner Tipp von mir. \Regie{Zu \Jupi.} Für ein längeres Leben: \Regie{ins Pub\-likum} 
	Mehr Gemüse essen. \Regie{Beißt in ein Radieschen und wirft weitere ins Publikum.}
	}\J{Oh. Ja.
	}\V{Seit wann verbreitest du Lebensweisheiten, Mortiii? \Regie{Kichert, zu \Jupi\ und \Luna.} Der Kaffee hat ihm 
	gut getan. Sie sollten seinem Ratschlag aber nur befolgen, wenn Sie gesund sterben wollen.
	}\Lu{Tja, man kann nur so lange von Luft und Liebe leben, bis einem die Luft ausgeht. Womit wir wieder auf die 
	Atmosphäre zurückkommen: Aus welchen Gasen soll sie bestehen? Wir hätten da \dots
	}\M{Stickstoff?
	}\V{Sauerstoff?
	}\M{Argon?
	}\V{$\mathrm{CO_2}$?
	}\M{Neon?
	}\V{Helium?
	}\M{Methan?
	}\Lu{Wir können auch verschiedene Gase mischen, wenn Sie möchten.
	}\V{Was gibt es denn bei einer Atmosphäre zu beachten, Jupi? \Regie{Stellt ihre Teetasse ab.}
	}\J{Es kommt vor allem auf den Anteil der sogenannten Treibhausgase an. Gerade das $\mathrm{CO_2}$, 
	also Kohlenstoffdioxid spielt eine wichtige Rolle. Bei gleicher Sonneneinstrahlung heizt sich eine Atmosphäre 
	aus $\mathrm{CO_2}$ schneller auf.
	}\Lu{$\mathrm{CO_2}$ sorgt auf so manchem Planeten für dicke Luft. Es wird heiß darüber diskutiert.}
	
	\medskip
	
	{\bf Experiment: Treibhauseffekt in Glasbehältern}, \textit{cf.} App.~\ref{app:treibhaus}.
	
	\medskip
	
	\J{Gerne erkläre ich Ihnen den Treibhauseffekt an diesen beiden Modellatmosphären hier. Die beiden 
	identischen Gesteinsplatten stellen zwei Planeten dar. Beide Planeten haben ihre eigene Atmosphäre, die 
	sich in den Glasbehältern befindet. Die rechte Atmosphäre besteht aus der guten Luft in unserem Planetenladen, 
	also einer Mischung hauptsächlich aus Stickstoff und Sauerstoff.
	}\Lu{Das ist unsere Standardatmosphäre für das Modell Terra-Firma.
	}\J{Die linke Atmosphäre besteht aus purem $\mathrm{CO_2}$. \Regie{Füllt CO$_2$ ein.}
	}\Lu{Die $\mathrm{CO_2}$-reiche Atmosphäre trägt den Namen Venus-Breeze.
	}\J{Ein Streichholz brennt auf Terra-Firma ganz normal, während es bei Venus-Breeze sofort erlischt. 
	\Regie{Hält ein brennendes Streichholz nacheinander in die Glasbehälter.} Nun bestrahle ich die beiden Planeten 
	mit Sternlicht und messe die Lufttemperaturen mit diesen Thermometern. Beide Atmosphären sind für Sternlicht 
	weitgehend durchlässig. Der Planet selbst nimmt das Sternlicht auf, erw\"armt sich, und sendet Wärmestrahlung ab. Diese 
	W\"armestrahlung wird bei Terra-Firma größtenteils durchgelassen. Bei Venus-Breeze hingegen hält das 
	$\mathrm{CO_2}$ die Wärme\-strah\-lung in der Atmosphäre gefangen. Wegen dieses Treibhauseffekts hängt die 
	mittlere globale Temperatur ganz empfindlich vom $\mathrm{CO_2}$-Gehalt in der Atmosphäre ab.
	}\VM{\Regie{Beugen sich vor, um die Thermometer besser sehen zu können.}
	}\J{\Regie{Verlegen} Oh, das dürfte eine ganze Weile dauern, bis der Effekt sichtbar wird.
	}\Lu{ Was halten Sie derweil von einem ausgedehnten Spaziergang am Strand, zwischen Ebbe und Flut?
	}\V{\Regie{Fasst \Mortis\ an den Händen und schaut ihm in die Augen.}  Da können wir uns richtig ausleben. Stell 
	dir vor. Nur wir zwei. Am Meer.
	}\M{Beim Sonnenuntergang?
	}\V{Im Abendrot.
	}\Lu{Ach, die Liebe ist tief wie das Meer. Um einen rosaroten, wildromantischen Sonnenuntergang kümmern wir uns gerne.
	}\J{Jawohl. Die Dämmerung färbt sich nämlich nicht automatisch rot. Dafür müssen wir eine Kleinigkeit hinzufügen.
	}\Lu{Für einen sternhagelgünstigen Aufpreis erhalten Sie leuchtende Abendröte und die Morgenröte gleich dazu. 
	Außerdem färbt sich durch unser kleines Extra der Himmel tagsüber so blau wie die Sehnsucht.
	}\J{Dafür bombardieren wir den Planeten einfach mit zahlreichen Felsbrocken.
	}\V{\Regie{Entsetzt} Bitte was? Meinen Sie das ernst, Luna?
	}\Lu{\Regie{Mit Blick zu \Mortis.} Todernst.
	}\J{Aber keine Sorge, wir nehmen vor allem kleine Brocken, so um die hundert Meter im Durchmesser. Die richten beim 
	Einschlag nur lokale Schäden an. Das ist die beste Möglichkeit, um Staub in die Atmosphäre zu bringen, der für rote 
	Sonnenuntergänge und blauen Taghimmel sorgt.
	}\V{Was hat das denn mit Sonnenuntergängen und dem Himmelblau zu tun?}
	
	\medskip
	
	{\bf Experiment:} \parbox[t]{11.7cm}{{\bf Rayleigh Streuung mit Aquarium und heller Lampe}, \textit{cf.} App.~\ref{app:rayleigh}.}
	
	\medskip
	
	\J{Das zeige ich Ihnen gerne an meinem Modell: Das mit Wasser gefüllte Aquarium stellt die Atmosphäre dar und 
	die Lampe die Sonne.  \Regie{Präsentiert das \E{Aquarium} und richtet den Scheinwerfer darauf.} Am Anfang ist im 
	Wasser, also der Atmosphäre, noch kein Staub drin, und das Licht fliegt ungehindert hindurch. Weder Himmelblau 
	noch Abendrot kann man sehen. Jetzt schauen Sie mal, was der Staub mit dem Sonnenlicht macht. \Regie{Schüttet 
	etwas Milch ins Aquarium.}
	}\VM{\Regie{Drehen dem Publikum den Rücken zu und legen den Arm um ihre Schultern.}}
	
	\medskip
	
	\Musik{Musik: "Force Theme" aus dem Film "Star Wars" (Musik von J. Williams)}
	
	\medskip
	
	\J{Wie man gut sieht streut der Staub, also die Milch, blaues Licht zur Seite heraus. Und rotes Licht strahlt geradeaus 
	ungehindert weiter.
	}\V{Aber wo kommen diese Farben her?
	}\J{Die steckten vorher schon im weißen Licht drin. Denn weißes Licht setzt sich aus allen Farben des 
	Regenbogens zusammen. Man braucht nur etwas, das die Farben voneinander trennt. Die kleinen Staubk\"orner  
	streuen das blaue Licht stärker in alle Richtungen als rotes Licht.
	}\V{Achso. Aber wieso sieht man denn jetzt das Rot nur abends und morgens? Die Atmosphäre ist doch den ganzen 
	Tag die gleiche.
	}\J{Das liegt am Weg, den das Licht von der Sonne aus durch die Atmosphäre nimmt. Wenn die Sonne tief am 
	Himmel steht, ist der Weg durch die Atmosphäre viel länger. Dann schafft es nur das rote Licht zu uns und das blaue 
	wird seitlich weggestreut.
	}\VM{\Regie{Wenden sich Arm in Arm \Jupi\ zu.}
	}\J{\Regie{Kippt mehr Milch ins Aquarium.} Oh, jetzt habe ich es mit der Romantik etwas übertrieben. Bei den 
	Einschlagkörpern kommt es sehr auf die Größe an. Ab einem Durchmesser von 400 Metern können sie Tsunamis 
	auslösen.
	}\V{Oh Nein.
	}\M{Ahgh. Gefällt mir.
	}\J{Ab ein paar Kilometern Durchmesser wirbelt der Einschlag so viel Staub auf, dass der Planet merklich abkühlt.
	}\VM{\Regie{Lösen sich voneinander.}
	}\V{Wie unangenehm.
	}\M{Sehr cool.
	}\J{Ein Felsbrocken von 20 Kilometern löst beim Aufprall wiederum ein großes Feuer aus und verdunkelt den ganzen Himmel.
	}\V{Entsetzlich!
	}\M{Ausgezeichnet!
	}\J{Bei einem Geschoss von 400 Kilometern verdampfen sämtliche Ozeane. Da empfehle ich ihnen, unter der 
	Oberfläche zu bleiben.
	}\M{Ahgh. \Regie{Denkerpose}
	}\V{Iiiiih! \Regie{Handgeste: Affe sieht nichts. H\"alt sich die Augen zu.} 
	}\Lu{Aufhören! \Regie{Vorwurfsvoll;  Handgeste: Affe hört nichts.}
	}\J{Uups. \Regie{Handgeste: Affe sagt nichts.}
	}\R{Alle lassen die Arme sinken.}
	\Lu{\Regie{Stellt sich zwischen \Mortis\ und \Vita.} Ich mache Ihnen ein Angebot. Erst schießen wir ein großes 
	planetenähnliches Objekt auf Ihre Welt drauf und formen aus den Überresten einen Mond. Dann lassen wir 
	ein paar mittelgroße Eisklumpen aus dem äußeren Sonnensystem auf ihn fallen und erschaffen so die Ozeane. 
	Später lassen wir dann nur noch kleine Felsbrocken auf den Planeten einprasseln für die rote Dämmerung. 
	Und die kleinen Staubkörner lassen wir beim Eindringen in die Atmosphäre als Sternschnuppen verglühen.
	}\V{\Regie{Sehnsüchtig} Wir bekommen Sternschnuppen?
	}\VM{Einverstanden.
	}\Lu{In Ordnung, dann bereiten wir Ihnen ein großes Bombardement. \Regie{Erinnert sich.} Was macht eigentlich der 
	\E{Treibhauseffekt}, Jupi?
	}\J{Oh! Jaaaa. Sehen Sie selbst. Die Fakten sprechen für sich.
	}\VM{\Regie{Beugen sich vor, um die Thermometer besser ablesen zu können.}
	}\V{\Regie{Liest die Thermometer ab.} Die Atmosphäre aus $\mathrm{CO_2}$ ist jetzt \_ Grad Celsius wärmer als die Ladenluft.
	}\J{Sie sagen es. Übrigens tragen auch Methan und gewöhnlicher Wasserdampf zum Treibhauseffekt bei.
	}\V{Was bedeutet das denn für uns? \Regie{Kramt nach ihrem Notizblock.}
	}\J{Wenn Ihre Lebewesen beispielsweise viel Kohle und Öl verbrennen, dann steigt der  $\mathrm{CO_2}$-Gehalt 
	der Atmosphäre an und mit ihm die mittlere globale Temperatur. Das führt zum weltweiten Artensterben und anderen Effekten.
	}\M{Grausam gut, dieser Treibhauseffekt. Er ist ein Geschenk.
	}\V{\Regie{Haut \Mortis\ mit ihrem Notizblock.} Nein, Mortis, aus! Ich muss mir das aufschreiben: NICHT ZU 
	VIEL $\mathrm{CO_2}$ FREISETZEN.
	}\J{Tja, das ist ein sehr empfindliches Gleichgewicht. Wenn Sie stattdessen den gan\-zen Planeten mit Pflanzen 
	und anderen Lebewesen eindecken, die der Luft $\mathrm{CO_2}$ entziehen, dann sinkt die Temperatur. Falls 
	Sie es übertreiben, wächst eine dicke Eisschicht auf der ganzen Planetenoberfläche. Das Eis spiegelt das 
	wärmende Ster\-nenlicht stärker, dadurch würde es noch kälter bis zur \dots
	}\M{\dots Todeskälte. Der Winter naht. \Regie{Diabolisches Grinsen.}
	}\V{\Regie{Hysterisch.} Meine kleinen Gänseblümchen! Wie können wir sie retten? Wo bekommen wir jetzt 
	genug $\mathrm{CO_2}$ her für einen natürlichen Treibhauseffekt?
	}\Lu{Keine Panik, Vita. Denn wir haben die perfekte Lösung parat.
	}\R{Verschwörerischer Blick zwischen \Luna\ und \Jupi.}
	\LJ{Vulkane!
	}\V{Nein!
	}\M{Doch!
	}\V{Oooh!
	}\Lu{Vulkane befördern genug $\mathrm{CO_2}$ in die Atmosphäre. Dadurch kann Vulkanismus sogar eine globale 
	Eiskruste zum Schmelzen bringen.
	}\M{Ich mag Vulkane.
	}\V{Ist das denn lebensnotwendig? Vulkanausbrüche machen immer so viel Dreck und Ärger.
	}\J{Oh, ja, das stimmt. Kurzfristig bringen Vulkane eine gewisse Todesgefahr mit sich. Langfristig sind vulkanische 
	Aktivitäten für den Planeten jedoch lebenswichtig. Vulkane fördern nämlich maßgeblich die Bildung der Atmosphäre, 
	indem sie das Ausgasen des Planeten beschleunigen. Außerdem setzen die das im verwitterten Gestein gefangene 
	CO2 wieder frei. Ich führe Ihnen das einfach mal vor. \Regie{Geht zum \E{Vulkan}-Experiment und zieht sich Handschuhe 
	an.}}
	
	\medskip
	
	{\bf Experiment:} \parbox[t]{11.8cm}{{\bf Vulkanexperiment, N$_2^{\text{fl\"ussig}}$ in Flasche mit kleiner \"Offnung}, \textit{cf.} App.~\ref{app:vulkan}.}
	
	\medskip
	
	\Lu{Tief im Inneren eines Gesteinsplaneten sind viele chemische Stoffe gebunden, zum Beispiel Kohlenstoff. Den 
	können Sie gut gebrauchen, oder, Vita?
	}\V{\Regie{Nickt, holt eine Nagelpfeile aus ihrer Handtasche und widmet sich ihren Fingernägeln.}
	}\J{Um den Effekt darzustellen, bringe ich das Gas in die Gesteins\-schichten ein. \Regie{Schüttet flüssigen Stickstoff in 
	einen Messbecher, an dessen Hinterseite Gesteins\-schichten dargestellt sind.}
	}\Lu{Das Gas bleibt im Gestein, wie Sie sehen. Doch durch Risse und Spalten im Gestein können flüssige 
	Gesteinsströme das Gas zur Oberfläche mitnehmen, wo es durch Vulkane in die Atmosphäre gelangt.
	}\J{\Regie{Setzt ein Röhrchen mit Korken in den Messbecher ein und hofft auf eine  Stickstofffontäne}}
	
	\medskip
	
	\Musik{Musik: Sinfonie Nr. 9 "Aus der Neuen Welt", 4. Satz (A. Dvorak)}
	
	\medskip
	
	\M{Vulkane sind wundervoll. Ich will Vulkane.
	}\Lu{Sehr gerne, so viele Sie möchten, Mortis.
	}\M{Viele.
	}\V{\Regie{Kaut kurz an ihren Fingernägeln.} Aber nicht zu viele, Liebling.
	}\M{\Regie{Lacht auf.}
	}\Lu{\Regie{Verschwörerisch zu \Jupi.} Ich glaube, wir haben bei ihm den richtigen Ton getroffen.
	}\M{\Regie{zeigt mit ausgestrecktem Arm auf den Pianisten} Du da! Spiel mir das Lied vom Tod.}
	
	\medskip
	
	\Musik{Musik:} \parbox[t]{13.3cm}{\textit{(Der Pianist spielt den)} \Musik{"Imperial March" aus dem Film "Star Wars" (Musik von J. Williams)}}
	
	\medskip
	
	\M{Nicht doch. Das andere.}
	
	\medskip
	
	\Musik{Orchestermusik: "What a Wonderful World" (Musik von B. Thiele 
	und G. D. Weiss.)}
	
	\medskip
	
	\M{\Lyrics{Ich seh' Wüsten, heiß, nur Trockenheit,\\
	Der Pol vereist, kalt und verschneit.\\
	Und ich denk' so bei mir:\\
	Wunderbar, diese Welt.\\\\
	Ich seh' Sturm und Flut, Taifun, Orkan,\\
	Lava und Glut von dem Vulkan.\\
	Und ich denke für mich:\\
	Wunderbar, diese Welt.\\\\
	In Meeres dunklen Tiefen, in Himmels blauen Höh'n,\\
	Auf jedem Fleck des Planeten kann Schreckliches gescheh'n.\\
	Ein Virus, so klein, fast unsichtbar.\\
	Überall wartet Todesgefahr.\\\\
	Ich seh' Dürre im Land, Feuer und Rauch.\\
	Angst und Beklemmung, Einsamkeit auch.\\
	Und ich denke für mich: Wunderbar, diese Welt.\\
	Ja, ich denke für mich: Wunderbar, diese Welt.}}
\end{itemize}

\subsection{Planetenoberfläche}
\label{sec:planetenoberflaeche}
\begin{itemize}
    \R{\Mortis\ beendet sein Lied. Das Publikum ist begeistert. Während der tosende Applaus abebbt, beginnen 
    Tischtennisbälle auf ihn zu fallen.}
    \M{Aua. \Regie{Zieht Kapuze an.}}
    \R{\Mortis\ geht einen Schritt zur Seite. Die Bälle folgen ihm.}
    \J{Oh. Das ist die kosmische Strahlung. Das Dach ist wohl undicht...}
    \V{Die komische Strahlung? Ich finde die gar nicht so lustig.}
    \J{{\it Kosmische} Strahlung. Das sind sehr schnelle geladene Teilchen, die den Planeten treffen. Die 
    st\"arkste Quelle dafür ist der Wind des Sterns, um den der Planet kreist. Die {\it kosmische} Strahlung 
    schadet Lebewesen. Zu viel davon kann tödlich sein.}
    \M{Tödliche Strahlung. Das gefällt mir! Ich nehme gleich zwei!}
    \V{Aber Mortis, du musst dich doch schützen.}
    \Lu{Kein Problem. Wir haben da das perfekte Produkt für Sie: ein globales Magentfeld, hier dargestellt 
    durch diesen Regenschirm. \Regie{\Luna\ gibt \Mortis\ einen Regenschirm.}}
    \V{Super! So etwas brauchen wir.}
    \Lu{Schön, dass wir Sie überzeugen konnten. Lassen Sie uns die Details vertraglich festhalten. Ich bereite 
    den Vertrag schon mal vor. \Regie{Geht ab.}}
    \V{\Regie{Zu \Jupi} Wie funktioniert denn so ein Magnetfeld?}
    \J{Es lenkt die Strahlung ab.}
    
    \medskip
    
    {\bf Experiment: Helmholtzspulen,} \textit{cf.} App.~\ref{exp:Helmholtz}.

    \medskip    
    
    \J{Das kann man hier sehr schön sehen. Dieses Gerät erzeugt einen Strahl aus geladenen Teilchen. Das ist 
    diese blaue Linie. Und jetzt achten Sie mal drauf, was passiert, wenn ich das Magnetfeld einschalte.}
    \V{Ah, also lenken wir mit dem Magnetfeld die kosmische Strahlung von unserem Planeten einfach nur weg?}
    \J{Ja genau!}
    \M{\Regie{Enttäuscht} Ein Magnetfeld schützt vor tödlicher Strahlung? Todtraurig.}
	\J{Nun ja, selbst mit einem Magnetfeld sind Sie nicht rund herum geschützt. In den Polarregionen schaffen 
	es die schnellen Teilchen oft, tief in die Atmosphäre einzudringen. Das führt zu schönen Polarlichtern. 
	Zwischendurch können Sie das Magnetfeld auch für ein paar Jahre ausschalten oder umpolen, um mit der 
	einfallenden kosmischen Strahlung ein paar Lebewesen umzubringen.}
	\M{\Regie{Gedankenversunken} Kosmische Strahlung. Wir könnten sie es tun lassen.}
	\V{Also bauen Sie uns einen riesigen Stabmagneten in den Planeten ein?}
	\J{Nicht ganz, hehe. Mit einem Stabmagneten würde das Umpolen nicht funktionieren. Außerdem würde 
	der Magnet gar nicht magnetisch bleiben. Im Planeteninneren ist es so heiß, dass keine dauerhalfte 
	Magnetisierung möglich ist.}
	
	\medskip
	
	{\bf Experiment: Curie-Temperatur,} \textit{cf.} App.~\ref{exp:Curie}.

	\medskip
		
	\J{Schauen Sie sich einmal dieses Stück Nickel  an. Es wird von einem Magneten im Wasser angezogen. 
	Jetzt erhitzen wir es. Wie Sie sehen, ist das Nickel  nicht mehr magnetisch und löst sich vom Magneten. Ebenso 
	funktioniert auch ein Magnet nicht mehr, wenn er zu heiß wird.}
	\V{Und wenn das Stück Nickel ein-zwei Mal hin und her geschwungen ist, hat es sich weit genug abgekühlt 
	um wieder magnetisch zu sein?}
	\J{Genau so ist es.}
        \V{Und wie erschaffen Sie nun ein weltumspannendes Magnetfeld, wenn ein einfacher Magnet nicht funktioniert? 
        Wie schalten Sie es ein?}
        \M{Und wieder ab?}
	\J{Eigentlich soll ich Ihnen das nicht sagen. Luna ist sehr streng, was Betriebsgeheimnisse angeht. 
	Versprechen Sie mir bitte, dass Sie niemandem verraten, was ich Ihnen jetzt erkläre. Kein Sterbenswörtchen.}
	\V{Versprochen!}
	\M{Ich schweige wie ein Grab.}
	\J{Nun gut. Im Inneren des Planeten herrscht wie gesagt eine sehr hohe Temperatur. Dadurch entsteht Plasma.}
	
	\medskip
	
	{\bf Experiment: H\"ornertrafo}, \textit{cf.} App.~\ref{exp:Hoerner}.
	
	\medskip
	
	\J{Wir können so ein Plasma auch hier mit Blitzen erzeugen. Wie Sie sehen, steigt das heiße 
	Plasma nach oben. Im Planeten geschieht das gleiche, sodass das Plasma ständig in Bewegung bleibt. Mit dem 
	Plasma fließen große elektrische Ströme. Diese Ströme erzeugen das Magnetfeld.}
	\Lu{\Regie{Kommt mit Vertrag auf die Bühne; böse.} Jupi, hast du etwa verraten wie wir das Magnetfeld machen?}
	\J{Nein.
	}\Lu{Doch.
	}\J{Ooh.}
	\Lu{Wir haben doch darüber gesprochen! Das Magnetfeld wird aus betriebs\-wirt\-schaft\-lichen Gründen nicht genauer 
	beschrieben. \Regie{Strenger Blick zu \Jupi.}
	}\J{Aber warum?
	}\Lu{\Regie{Zu \Vita\ und \Mortis.} Nur hier bei \texttt{PLANETAMOS} bekommen Sie ein globales Magnetfeld. Wir sind der 
	einzige Planetenladen, der Ihnen magnetischen Schutz bietet. Das soll auch so bleiben.
	}\J{Wie eigenartig, dass die Konkurrenz globale Magnetfelder noch nicht versteht. Vielleicht sollte ich denen mal 
	ein paar Tipps \dots
	}\Lu{\Regie{Autoritär zu \Jupi.} Untersteh dich!
	}\J{Oh. Na gut.
	}\Lu{\Regie{Zu \Vita\ und \Mortis.} Verzeihung. Lassen Sie uns gemeinsam den Kaufvertrag durchgehen. \Regie{Öffnet die Mappe.}}
	
	\medskip
	
	\Musik{Orchestermusik: "Dream a little dream of me" (Musik von F. Andre \& W. Schwandt)}
	
	\medskip
	
	\Lu{\Lyrics{Sie wünschen was zum Leben,\\
	doch nicht jeder Planet kann das geben.\\
	Sie schaffen was Besonderes an,\\
	wo das Leben blühen kann!\\\\
	Ein G-Stern ist notwendig,\\
	darum kreist ihr Planet dann beständig.\\
	Mit leicht gekippter Achse gibt‘s frei\\
	Jahreszeiten gleich dabei.\\\\
	Die Stabilisierung der Achse,\\
	vor Schwankung verschont,\\
	auch Ebbe und Flut in den Meeren:\\
	das schafft ein Mond.\\\\
	Die Atmosphäre baut dann\\
	den Druck auf, damit Wasser auch fließen kann.\\
	Fürs Klima schauen wir noch dabei\\
	ganz genau aufs CO2.}
	}\Lu{\Regie{Gesprochen} Für rosarote, wildromantische Sonnenuntergänge sorgen wir durch ein großes 
	Bombardement Ihres Planeten. So wird die Atmosphäre bestäubt. Außerdem bekommen Sie so unzählige 
	Sternschnuppen gratis dazu!
	}\M{Und meine Vulkane?
	}\Lu{\Lyrics{Damit es auf Ihrem Planeten\\
	nicht allzu stark friert,\\
	braucht's manche aktiven Vulkane.\\
	Das ist notiert.\\\\
	Es schützt Sie ein Magnetfeld\\
	vor Strahlung aus dem All, die auf Sie fällt.\\
	Das ist es, was in dem Vertrag steht.\\
	Alles das ist Ihr Planet.}
	}\Lu{Was für ein Wetter hätten Sie denn gerne auf Ihrem gemeinsamen Planeten?
	}\V{Regenbögen!
	}\M{Wirbelstürme!
	}\J{Mit der gegebenen Atmosphäre, dem ganzen Wasser und der Strahlungsenergie des G-Sterns bekommen Sie beides ganz automatisch.
	}\Lu{Also für einen sternhagelgünstigen Aufpreis.
	}\J{Für den Regenbogen reichen ein paar Wassertropfen in der Atmosphäre.
	}\V{Ahgh. \Regie{Nimmt einen Fächer aus ihrer Tasche, fächert sich Luft zu, folgt dem weiteren Verkaufsgespräch 
	und wird dabei zunehmend sauer.}
	}\Lu{Bei den Wirbelstürmen haben wir etwas ganz besonderes auf Lager. Das wird Ihnen gefallen, Mortis.
	}\J{Du meinst bestimmt die \dots
	}\LJ{Feuertornados.
	}\Lu{Das sind keine gewöhnlichen Windhosen.
	}\J{Keine alltäglichen Staubteufel.
	}\Lu{Wir sprechen von flammenden Wirbelstürmen.
	}\M{Ich bin Feuer und Flamme. Zeigen Sie mir Ihre Feuertornados.}
	
	\medskip
	
	{\bf Experiment: Feuertornado}, \textit{cf.} App.~\ref{exp:firetornado}.
	
	\medskip
	
	\J{Sehr gern. Hier drüben. \Regie{Geht zu den \E{Feuertornados}.} Allgemein leben Wir\-belstürme davon, dass 
	warme Luft nach oben steigt. Beim Feuertornado wird die Luft unten durch ein Feuer erwärmt. Die Drehung des Käfigs
	und damit der Luft stabilisiert die Flammen.
	}\Lu{Lassen Sie sich von Jupis brennender Begeisterung anstecken, Mortis.
	}\J{\Regie{Entzündet die \E{Feuertornados}.}}
	
	\medskip
	
	\Musik{Musik: "Sauron's Theme" aus dem Film "Der Herr der Ringe" (H. Shore)}
	
	\medskip
	
	\M{Die Erklärung \textit{leuchtet} ein. Doch der Tornado ist sehr klein.
	}\Lu{Das sind nur Ansichtsexemplare. Wir führen alle Größen von winzig bis gewaltig.
	}\V{\Regie{Völlig aufgebracht und stocksauer, rastet aus,  klappt ihren Fächer zu und pfeffert ihn 
	gegen \Mortis.} Auf keinen Fall! Du zerstörst unseren gemeinsamen Planeten! Ein Feuertornado 
	kommt mir nicht in die Welt. Immer denkst du nur an dich, Mortis. \dots
	}\R{Alle reden und schreien durcheinander.}
	\M{Momentan mal. Es ist doch nur ein kleiner Wirbelsturm. \dots
	}\Lu{Ich bitte Sie. Das war doch nicht so gemeint. \dots
	}\J{Wir haben auch harmlose Feuertornados. Machen Sie sich keine Sorgen. Wir kriegen das hin.
	}\V{Sie wollen uns doch nur ihren Schrott andrehen! Ich unterschreibe ihren Vertrag nicht. Mortis, wir gehen.
	}\M{Beruhig dich wieder. Denk an deine Hundewelpen. 
	}\Lu{Was erlauben Sie sich? Das ist der beste Planetenladen den Sie finden können. Ich sollte Ihnen Hausverbot erteilen.
	}\J{Gute Idee. Ich widme mich dann wieder meiner Kometensammlung.
	}\M{\Regie{Sehr laut:} {\bf Totenstille!} Wir nehmen keine Feuertornados.
	}
	\R{Alle verstummen und lassen die Arme sinken. Ab jetzt sprechen sie wieder ein\-zeln.}
	\V{Ach Liebling. \Regie{Fällt \Mortis\ schluchzend um den Hals.}
	}\Lu{\Regie{Verschränkt die Arme.}
	}\M{Ich unterschreibe. Geld spielt keine Rolle.
	}\Lu{\Regie{Dollarzeichen in den Augen und ein breites Lächeln auf den Lippen.} Sehr erfreut. \Regie{Reicht \Mortis\ 
	die Mappe.}
	}\V{Danke. \Regie{Löst sich von \Mortis.}
	}\M{\Regie{Unterschreibt, murmelt dabei.} Mors, Mortis f.
	}\J{Wissen Sie was? Ich mache ihnen eine hinreißende Fjord-Landschaft auf der Nordhalbkugel. Ihr Planet 
	wird der Schönste von allen sein.
	}\Lu{Das wird er. \Regie{Reicht \Vita\ ein Taschentuch.}
	}\V{\Regie{Putzt sich ausgiebig die Nase.}
	}\J{Wollen Sie bei der Landschaftsbildung zuschauen?
	}\VM{Gerne.
	}\J{Ja, die Landschaft bilden wir hier drüben.}
	
	\medskip
	
	{\bf Experiment: Sandbox}, \textit{cf.} App.~\ref{exp:Sandbox}.
	
	\medskip
	
	\R{Alle versammeln sich um die \E{Sandbox}.}
	\Lu{Am Anfang ist der Planet wüst und leer. Mit verschiedenen Techniken lassen sich Berge, Täler und Meere 
	fantasievoll kombinieren. Einerseits tragen Wind und Wetter zur Landschaftsbildung bei. Andererseits haben 
	auch Prozesse im Inneren des Planeten Einfluss auf die Oberflächenstruktur.
	}\J{Beispielweise füllen sich unter der Planetenoberfläche manchmal große Blasen mit flüssigem Gestein. 
	Wenn so eine Blase anwächst, entsteht an der Oberfläche ein Vulkankegel. \Regie{Bläst den Luftballon auf.}
	Bei einem Vulkanausbruch entleert sich diese Blase und hinterlässt einen Krater. Der kann sich dann mit 
	Regenwasser füllen.}	
	\M{Wie wär's mit einer Sintflut?
	}\J{Oh, Moment. Das haben wir gleich. \Regie{Flutet die \E{Sandbox}.}
	}\V{Sehr inspirierend. Ein Unterwasservulkan scheint mir das ideale soziokulturelle Umfeld für meine ersten Lebewesen zu sein.
	}\Lu{Eins noch. \Regie{Legt beiden einen Arm um die Schultern.} Vita, Mortis, wir brauchen noch einen Namen für Ihre Welt. 
	}\V{Natürlich, wir müssen unserem Baby-Planeten einen Namen geben, Liebling.
	}\M{Ahgh.
	}\Lu{Wie wäre es mit einem vielsagenden Akronym? Einer Abkürzung?
	}\V{Mh. \Regie{Grübelt} Irgendwas mit \Regie{Grübelt} {\bf E}rlebnis {\bf R}esort \Regie{Grübelt} für \dots
	}\M{\Regie{Grübelt} \dots {\bf D}rama \Regie{Grübelt} und \dots
	}\V{{\bf E}volution!
	}\Lu{{\bf E}rlebnis {\bf R}esort für {\bf D}rama und {\bf E}volution. {\bf E}.{\bf R}.{\bf D}.{\bf E}. Erde.
	}\V{Erde klingt gut.
	}\J{Oh, Ja. Der Name \enquote{{\bf E}rlebnis {\bf R}esort für {\bf D}rama und {\bf E}volution} verdeutlich das 
	empfindliche Gleichgewicht auf Ihrem Planeten.
	}\Lu{Denn seien wir mal ehrlich: Das Leben ist immer lebensgefährlich.
	}\VM{\Regie{Halten sich aneinander fest und lachen.}
	}\M{Ich lach mich tot.
	}\V{Es lebe die Evolution. Viva la evolución!}
	
	\medskip
	
	\Musik{Orchestermusik:\,\parbox[t]{10.5cm}{``Der ewige Kreis" aus dem Film "Der König der Löwen" (Musik  von E. John)}}
	
	\medskip
	
	\V{\Lyrics{Nun erhalten wir unser Ergebnis,\\
	es wird sich um die Sonne dreh’n.\\
	ein so glorreiches Werk, das man hier sehen kann,\\
	und Dank euch können wir es versteh’n.}
	}\Lu{\Lyrics{Das Leben hier wird ein Wunder,\\
	alles neu, alles endlos und weit.\\
	Oben Wolken so weiß,\\
	und der Erdkern so heiß,\\
	mittendrin steht das Leben bereit.}
	}\J{\Lyrics{Und im ewigen Kreis\\
	Zieht die Welt ihre Bahnen.\\
	Dem Gesetz der Physik\\
	Ist sie geweiht.}
	}\V{\Lyrics{Sie ist nun ein Teil\\
	Dieses Universums\\
	und ihr Orbit\\
	ein ewiger Kreis.}
	}\R{Black. Spot auf \Mortis.}
	\M{\Regie{Kopfabschneidergeste} ENDE!}	
\end{itemize}

\centerline{\bf ENDE}


\section{Experiments}
\label{app:experimente}
Here we present a brief  description of the experiments incorporated into the Phyusical.

\subsection{Experiments Regarding Celestial Mechanics}
See Section~\ref{sec:himmelsmechanik-e} and Section~\ref{sec:himmelsmechanik}.

\subsubsection{Condensation}
\label{app:condensation}
This experiment demonstrates the condensation of a gaseous material into a solid. These are the two states of matter, solid 
and gas, present in the prototempplanetary cloud prior to the formation of planets. A liquid phase is not realized due to the low 
pressures in space. The chemical components there can roughly be classified into three groups: gases, ices and refectory 
elements (\textit{i.e.} ``metal and rock"). While the gases (H, He and noble gases) stay gaseous throughout the whole disk and
the refractories remain in the solid phase, the state of the ices (water, carbon dioxide, methane, and ammonia) depends on the 
local temperature of the protoplanetary disc and thus on the distance to the central star. The distance beyond which they freeze 
out is known as the snow- or ice-line. Hence in the inner region, close to the central star, only rocky planets form while in the 
outer regions icy planets/moons and gas giants are present \cite{liassauer-de-Pater}.

\paragraph{Material}
Balloons, a small dewar with an opening large enough to at least fit in partially the inflated balloon, so it can cool down, 
liquid nitrogen, carbon dioxide, scissors, safety glasses, gloves, tongs.

\paragraph{Presentation}
The balloon is filled with carbon dioxide gas and is sealed. It represents the potentially icy components of the 
protoplanetary disk.  At 
room temperature carbon dioxide is gaseous corresponding to the condition of the disk within the snow-line. The balloon 
is then with help of the tongs inserted into the dewar, which is filled with liquid nitrogen. This cools the gas down wherefore 
it condensates to dry ice representing the state of the ices beyond the snow line. The condensation may take a while. The 
process is completed, when the balloon has reached minimal volume. Using the scissors the balloon is opened and the 
solid nature of the filling is revealed.

\paragraph{Safety}
Liquid nitrogen is very cold and can lead to serious burns. Always wear protection and avoid contact with skin and clothes. Pressurised cans of carbon dioxide may require prior instructions.

 \subsubsection{Cotton Candy}
\label{app:cotton-candy}
This experiment illustrates the formation of the first macroscopic chunks of matter from dust. In the protoplanetary 
disk, dust particles, whose chemical composition depends on the distance to the central star, aggregate to loose 
lumps of material by intermolecular forces. They are not very dense and can fall apart easily by e.g. shear or 
collisions. We identify the intermolecular forces with the stickiness of the sugar, and the loose nature of the dust lumps 
with the fluffiness of the cotton candy. Later, when the lumps have grown massive enough, gravity takes over, 
compresses the material and accelerates the growth. Note that the exact process between condensation and gravity 
induced growth is not well understood yet \cite{liassauer-de-Pater}.

\paragraph{Material}
Cotton candy machine, sugar or crushed coloured candies, wooden sticks.

\paragraph{Presentation}
The cotton candy machine is heated up (our device needs 5\,min of preparation) and then filled with either sugar or 
crushed coloured sweets, which result in a dyed cotton candy. With a wooden stick the experimentor creates a portion 
of cotton candy by moving it through the disk of the running machine, while explaining the analogy. The process can take 
a while.  See Fig.~\ref{fig:cotton-candy}.

\paragraph{Safety}
Parts of the machine can become hot. Consult the manual of your device. Eating too many sweets is unhealthy.

\subsubsection{Luminous Balloon as  a Star with Orbiting Planet}
\label{app:planet-pendulum}
This experiment is used to demonstrate the habitable zone of different stars. Depending on the luminosity of the star 
the orbit (or better orbitable  range) representing the habitable zone of the planet is closer to or further away from 
the star.

\paragraph{Materials:} The star is here represented by a large balloon attached  to a stand about 1.7m  high. The 
balloon can be inflated by a compressed air line and deflated by a small valve. A lamp is placed inside the balloon, 
which can change the color of the star during the experiment. The orbiting planet is represented by a small sphere, 
which is wrapped in aluminum foil for better visibility, and attached to a long string fixed to the ceiling above the star. 
(In our case the lecture hall ceiling is about 6m high.) This pendulum can then be pushed into a circular motion, to show 
the orbit of the planet around the star.

\paragraph{Presentation:} The balloon is inflated to a small size. Simultaneously the color switching lamp is turned 
on with a red color, representing a low mass M-type star. The planet is held closely to the star and is given a small push, 
initiating a  circular orbit in the close habitable zone of the star. Afterwards the balloon is inflated to a large size, the 
color of the lamp can be switched according to the size of the balloon from red to yellow (to white) ending with blue, 
finally representing a heavy O-type star. Here the habitable zone lies further away from the star, so the planet is sent onto 
a larger radius orbit by holding it further away from the star and giving it a stronger push to the side. Finally the balloon 
is deflated to a size between the M- and O-type star size, the color is switched to yellow representing a G-type star like our sun. 
The radius of the orbit lies between the two previous orbits.

A demonstration of the experiment is shown in Fig.~\ref{fig:orbiting-planet} in the script, Sec.~\ref{sec:himmelsmechanik-e}.

\paragraph{Remarks:}
Inflating the balloon is faster than deflating it. It should be tested to push the pendulum by the correct amount and direction beforehand.

\subsubsection{Demonstrating the Seasons with a Blackboard-Globe and a Spotlight}

A blackboard globe is moved at a fixed distance around a bright spotlight representing the sun. The globe is tilted.
This shows the four seasons are a result of the tilted Earth's axis of rotation with respect to the Earth's orbital plane
around the sun.

\paragraph{Materials:}
The globe, about 40cm in diameter, is mounted with a titled axis on a thin wooden pedestal, it can rotate about this axis. The
globe has a blackboard surface, such that it is possible to draw on it with chalk. The pedestal is placed on a mechanic's 
creeper, so it can easily be moved around. This corresponds to the orbit of the Earth around the sun. The mechanic's 
creeper is connected with a fixed length rope to  the tripod of the spotlight. Thus we approximate the weakly eccentric
elliptical orbit by a circular orbit. The height of the lamp is adjusted to the globe's.

\paragraph{Presentation:}
Place the chalk on a fixed point of the equator and rotate the globe, thus marking the equator for the entire globe.
Then demonstrate the rotation of the Earth around the sun by moving the globe around the lamp in a constant distance. Rotate 
the direction the lamp is pointing accordingly, such that it always shines at the globe. It is advisable to focus on one hemisphere 
to show the different amounts of incoming solar radiation and the resulting seasons for this hemisphere. Furthermore, demonstrate 
the phenomena of polar night and day by showing that at certain positions no light shines on one of the poles at all.

Depending on how much time is allowed for this demo, it is also possible to show the changes in the lengths of night and day by 
marking one position on the globe and showing for how long in one rotation of the globe at a certain position with respect to the 
lamp this point is illuminated.

\begin{figure}[t]
	\centering
	\includegraphics[width=0.75\textwidth]{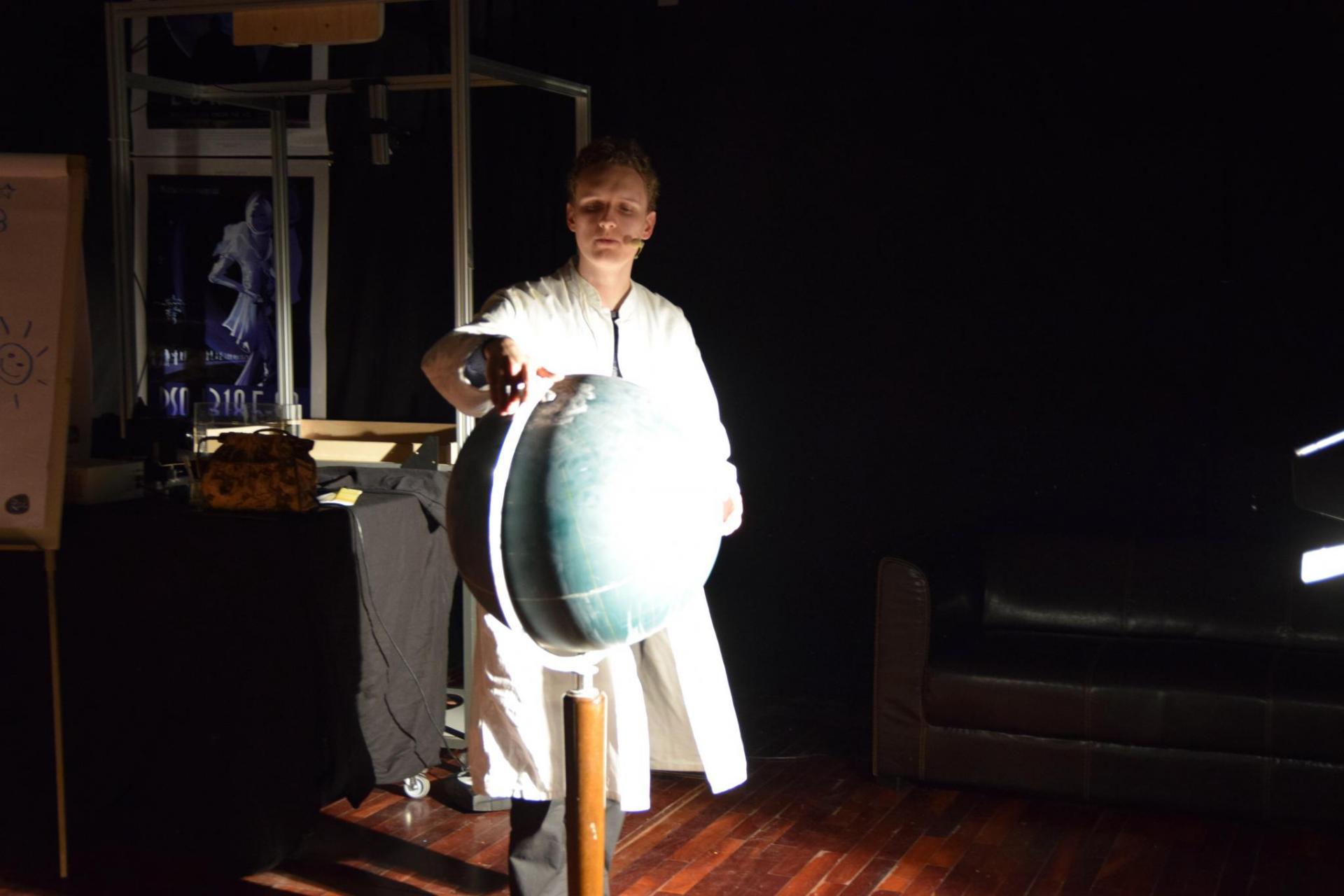}
	\caption{\small David Ohse as Jupi performing the blackboard globe demonstration.}
	\label{fig_blackboardglobe}
\end{figure}

Fig. \ref{fig_blackboardglobe} shows a photo of this demonstration from the show.

\paragraph{Safety:} Take care not to point the lamp directly at the audience.


\subsection{Experiments Regarding the Moon and Oceans}
See Section~\ref{sec:mond-engl} and Section~\ref{sec:mond}.
\subsubsection{Phases of the Moon}
\label{exp:Mondphasen}
The experiment’s objective is to demonstrate the Moon phases.

\paragraph{Materials:} The Moon is represented by a sphere made of paper mâché about 25cm in diameter. This sphere 
is fastened to a fishing rod, allowing the Moon to be held at a distance. A camera takes earth’s perspective, filming the Moon. 
This image is projected onto a large screen for the audience. A spotlight or a bright lamp symbolizes the sun. See 
Fig.~\ref{fig:moon-phases}.

\paragraph{Presentation:} Two persons are necessary to perform this demonstration. One of them handles the camera, the 
other stands right behind the camera person and holds the rod, so that the Moon is in front of the camera. The spotlight is 
positioned at a fixed point illuminating the camera (Earth) and the Moon, and is not moved during the experiment. For the 
demonstration, both persons have to rotate simultaneously with the camera (Earth) always watching the paper sphere (Moon).

The experiment starts by showing the new Moon position: the Moon is held between the spotlight (Sun) and the camera (Earth).
Thus the Moon appears dark; the lamp only illuminates its backside. 

Then, the camera (Earth) and the paper sphere (Moon) are rotated by 90° in the same direction. The spotlight (Sun) now shines 
on the side of the paper sphere (Moon), causing half of the visible sphere to appear bright (from the camera's perspective).
The shift demonstrates the waxing crescent Moon to a quarter Moon.

Earth and Moon are further rotated by another 90°, the spotlight (Sun) is aligned with the camera (Earth) pointing at the paper sphere.
The side of the Moon facing the Earth (camera) is now fully illuminated resulting in a full moon. During this step, it is crucial to 
hold the moon slightly higher than the camera, thus avoiding shadows of the camera and camera person on the moon's surface. 

For the waning Moon, the Earth and the Moon are turned further, to let the Sun now illuminate the other half of the Moon.

\subsubsection{Boiling Tea in a Vacuum}
\label{exp:tea-vacuum}

The phase diagram of water shows that water in its liquid phase can only exist in a certain range of temperature and 
pressure. Thus, one necessary condition for a planet to have liquid water, and thus support life, is having an atmosphere 
dense enough to produce suitable pressure values at the surface. To demonstrate this, the lowering of the boiling point 
of water in a (low-pressure) vacuum chamber was used.

\begin{figure}[h!]
	\centering
	\includegraphics[width=0.75\textwidth]{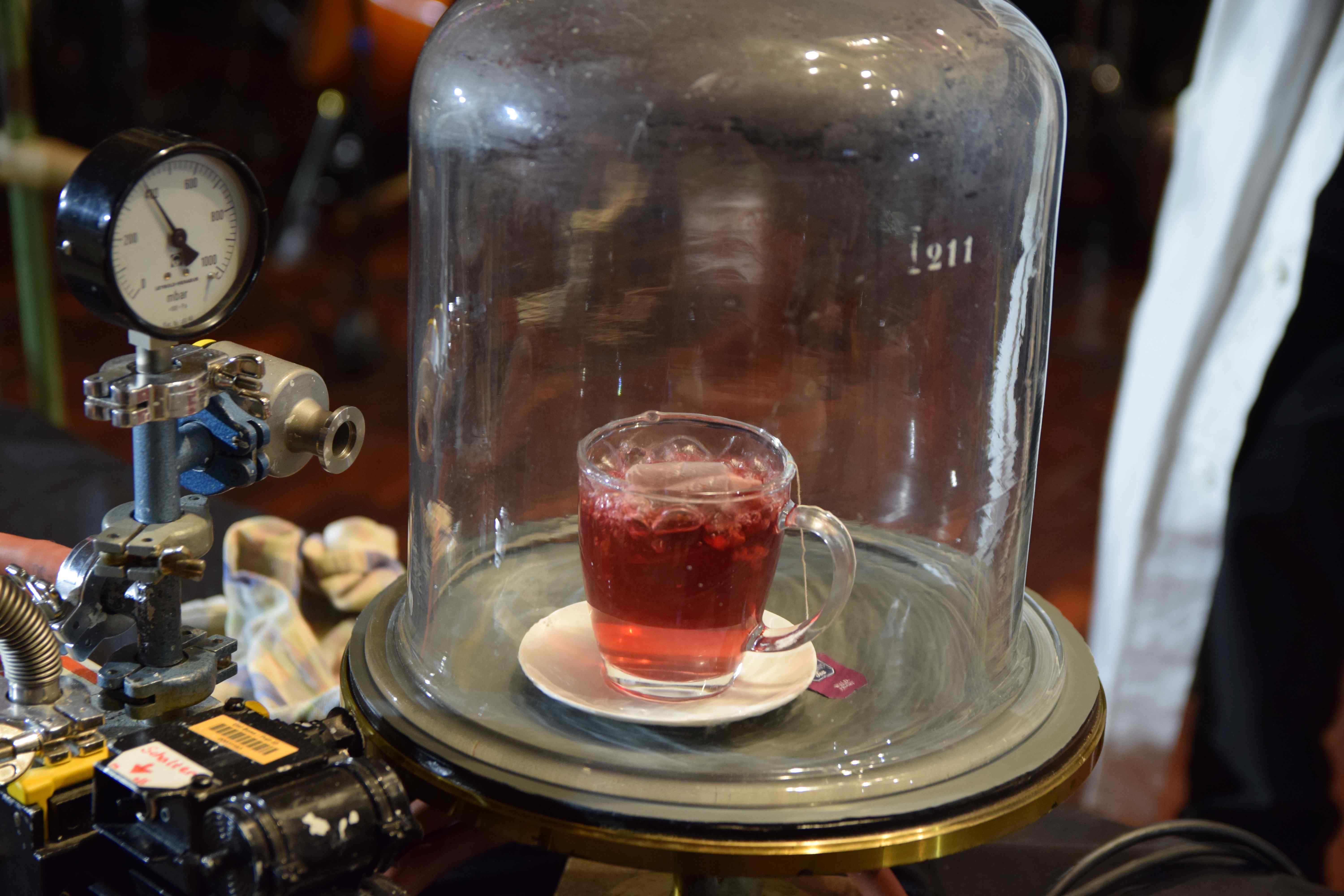}
	\caption{\small Vacuum chamber with pressure gauge at the back. The tea water is boiling as the pressure has been
	sufficiently lowered. The connection to the vacuum pump is below the saucer.}
	\label{fig:vakuumglocke}
\end{figure}

\paragraph{History:} In essence the physics of the experiment is described by the Clausius-Clapeyron-relation. The effect 
can be experienced when heating water at higher altitudes and also can be used to measure pressure. 

\paragraph{Materials:} The experiment requires a glass of hot water as well as a vacuum pump and chamber.

\paragraph{Presentation:}  Backstage water is boiled in a simple electric kettle. After letting the water cool off to about \SIrange{80}{90}{\celsius} it is 
brought on stage. The audience is shown that the water is not boiling before placing it in the vacuum chamber. Depending 
on the temperature of the water and the vacuum pump the water will begin to boil on evacuation after a few seconds.

\paragraph{Safety:}
When performing the experiment be careful when carrying the hot water onto the stage. Then take care that no water can 
be sucked into the vacuum pump as it may cause damage to it. For instance, in the experiment as shown in 
Fig.~\ref{fig:vakuumglocke} the tube of the vacuum pump enters the chamber from below. As the boiling water might spill 
from the glass we put a saucer below the glass to prevent this from happening.


\subsection{Experiments Regarding the Atmosphere}
See Section~\ref{sec:atmosphaere-engl} and Section~\ref{sec:atmosphaere}.
\subsubsection{Greenhouse Effect}
\label{app:treibhaus}
The effect of greenhouse gases, here carbon dioxide (CO$_2$), on the temperature of the atmosphere is demonstrated 
to show the importance for the habitability of a planet. For this, the temperature of two model atmospheres is 
compared.

\paragraph{Materials}
The experiment consists of two glass vessels (aquaria), which stand for one atmosphere each. In each of the 
vessels a large piece of black cardboard is inserted, which represents the surface of a planet. This absorbs the 
infalling light and emits it approximately as a black body, at a lower frequency, \textit{i.e.} in the infrared, as the 
planetary surfaces do. One of the vessels is filled with carbon dioxide, the other contains the air of the lecture 
hall. Both are covered with a wooden lid with a hole in the middle to insert a thermometer, which are read out 
on a joint digital display next to the aquaria. Additionally, two strong lamps are needed to impose radiation on 
the model atmospheres.

\begin{figure}[t]
	\centering
	\includegraphics[width=0.75\textwidth]{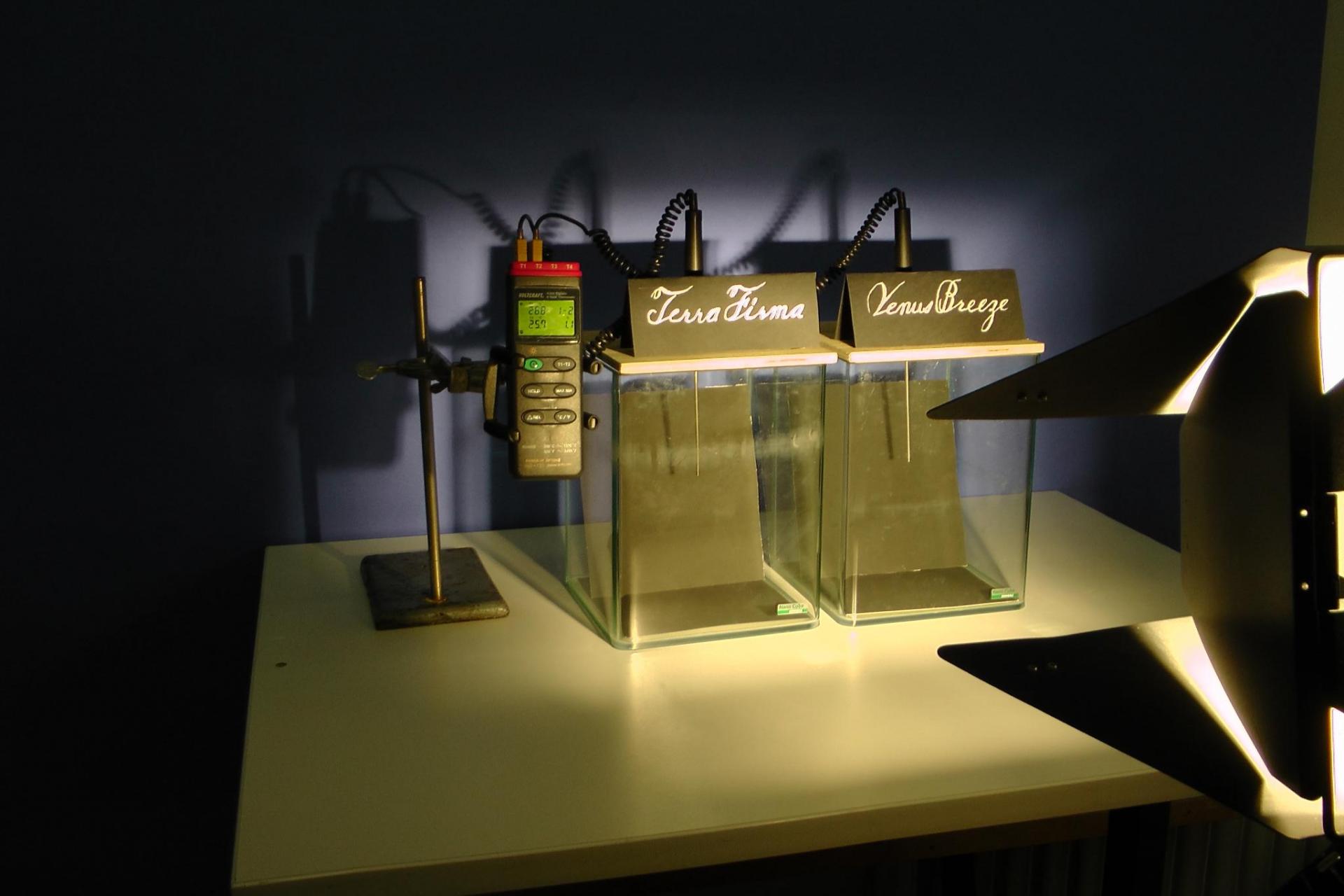}
	\caption{\small The greenhouse effect experiment. The atmosphere in \enquote{Terra Firma} consists of air, 
	in \enquote{Venus Breeze} it is dominantly carbon dioxide. The thin wires in the aquaria are the sensors for 
	the thermometers. On the left the digital temperature is read out, which is then via a camera projected on to
	the large screen for the audience.}
	\label{fig:aquaria}
\end{figure}

\paragraph{Presentation}
The carbon dioxide is introduced into one of the vessels; this can be done before or directly at the start of the
presentation. The thermometers are read off, with the audience watching. They should both be showing the same
temperature, i.e. the room temperature of the lecture hall. Then the lamps are turned on and directed at the aquaria. 
It takes some time for an observable effect, which is why our show proceeds after turning on the lamps and returns 
to this experiment later. After approximately \SI{10}{\min} it is possible to measure a noticeable difference in 
the temperature of the two model planets. Fig.~\ref{fig:aquaria} shows a photo of this experiment.

\paragraph{Remarks}
\begin{itemize}
\item Keep in mind, that when using carbon dioxide from a pressured gas tank it may be colder than the air in the other 
vessel. In this case it is advisable to fill one of the vessels with carbon dioxide before the presentation starts. 

\item One can show the difference in the atmospheres by inserting a burning match in each vessel.

\item The effects can be amplified by adding humidity to the CO$_2$ atmosphere, for example by placing a cup of water 
in this aquarium. H$_2$O is also a greenhouse gas. The humidity can be reduced in the other aquarium by adding a cup of salt.
\end{itemize}

\subsubsection{Rayleigh Scattering with an Aquarium and a Bright Light}
\label{app:rayleigh}
This experiment is used to explain why the sky is blue during the day and why it turns red at sunset. The 
wavelength-dependent Rayleigh scattering is demonstrated on a model atmosphere. 

\paragraph{Materials} For the experiment an aquarium about 30cm long and a bright light are needed. It is 
important, that the aquarium is not quadratic but has a rectangular base, with a long and a short side. During 
the experiment the aquarium is filled with water. Small droplets of milk can be added to the water with a pipette. 

\paragraph{Presentation} 
To start, the aquarium is only filled with water and the short side of the aquarium is illuminated by the bright light. 
This represents a clean or pure atmosphere, with no small scattering centers, with incident sunlight. Then small droplets 
of milk are added to the water. This illustrates an atmosphere, which is no longer clean but contains small particles 
of order or smaller than the wavelength of the incident light. On these the sunlight scatters. After the milk droplets 
and water are well mixed, the color of the light emitted at the different sides of the aquarium is observed. The long 
side of the model atmosphere shows a bluish cast, while the short side of the model atmosphere, away from 
the incident light, appears reddish. The long side of the aquarium here represents the situation where the observer 
is not directly looking into the sun and sees a blue sky, while the short side of the aquarium represents the situation 
at sunset, where the observer is looking towards the sun and the sunlight has a longer path through the atmosphere.

The experiment is depicted in Fig.~\ref{fig:milk-sunset} in the script, Sec. \ref{sec:atmosphaere-engl}.

\paragraph{Remarks} 
\begin{itemize}
\item It is important not to use too much milk, since otherwise the effect is not visible. For larger droplets it becomes Mie-scattering. 
\item Furthermore 
the aquarium should be cleaned before the experiment. 
\item If the experiment is filmed for better viewing by the audience, 
both sides of the aquarium should be filmed at the same time to show the difference. If only the long side is filmed, 
the white balance of the camera might correct the blue-white surface of the model atmosphere to a white surface. 

\item This experiment can also be conducted at home with a flashlight (torch) and an oven dish in the correct dimensions 
(one short and one long side).
\end{itemize}

\subsubsection{Volcano Experiment -- Liquid Nitrogen Fountain}
\label{app:vulkan}

A fountain of liquid nitrogen forms an impressive cloud. This experiment is more of a spectacular visualization than an actual 
physical analogy for a volcano. In the show we also use it to explain how minerals are transported from the earth's interior to the surface, though admittedly this interpretation of the experiment is slightly far-fetched.

\paragraph{Materials}
For the experiment a cold resistant bottle with stopper is used. A (metal) tube of several millimetres in diameter has been tightly fitted through a hole in the stopper. The tube should nearly reach the bottom of the bottle. In addition some heat reservoir like sand is required within the bottle prior to filling the liquid nitrogen in. Without the heat reservoir the bottle quickly assumes the temperature of liquid nitrogen which in consequence stops boiling.

\paragraph{Presentation}
Fill the bottle with liquid nitrogen. You might use a funnel and the help of another person for convenience. Immediately close the bottle with the stopper. Now the boiling liquid nitrogen builds up a high pressure (when evaporating it increases its volume by a factor 694) within the bottle and ejects the liquid from the bottom through the tube. In open air the liquid nitrogen evaporates instantly cooling the surrounding air and in particular the contained water vapour. The cloud of condensed water is what makes the fountain visible.
A photo of the experiment is shown given in Fig.~\ref{fig:nitrogen_volcano} in the script (Sec.~\ref{sec:planetenoberflaeche-engl}).

\paragraph{Safety}
Liquid nitrogen is very cold ($\SI{77}{\kelvin}$). Do not burn yourself. Take the usual safety precautions you should have received before getting access to liquid nitrogen. Take care not to point the fountain at a person (including yourself) and ensure that everyone maintains a save distance of several meters to the experiment.

\medskip

\subsection{Experiments Regarding the Planetary Surface}
\subsubsection{Helmholtz Coils}
\label{exp:Helmholtz}

\begin{figure}[t]
	\centering
	\includegraphics[width=0.447\textwidth]{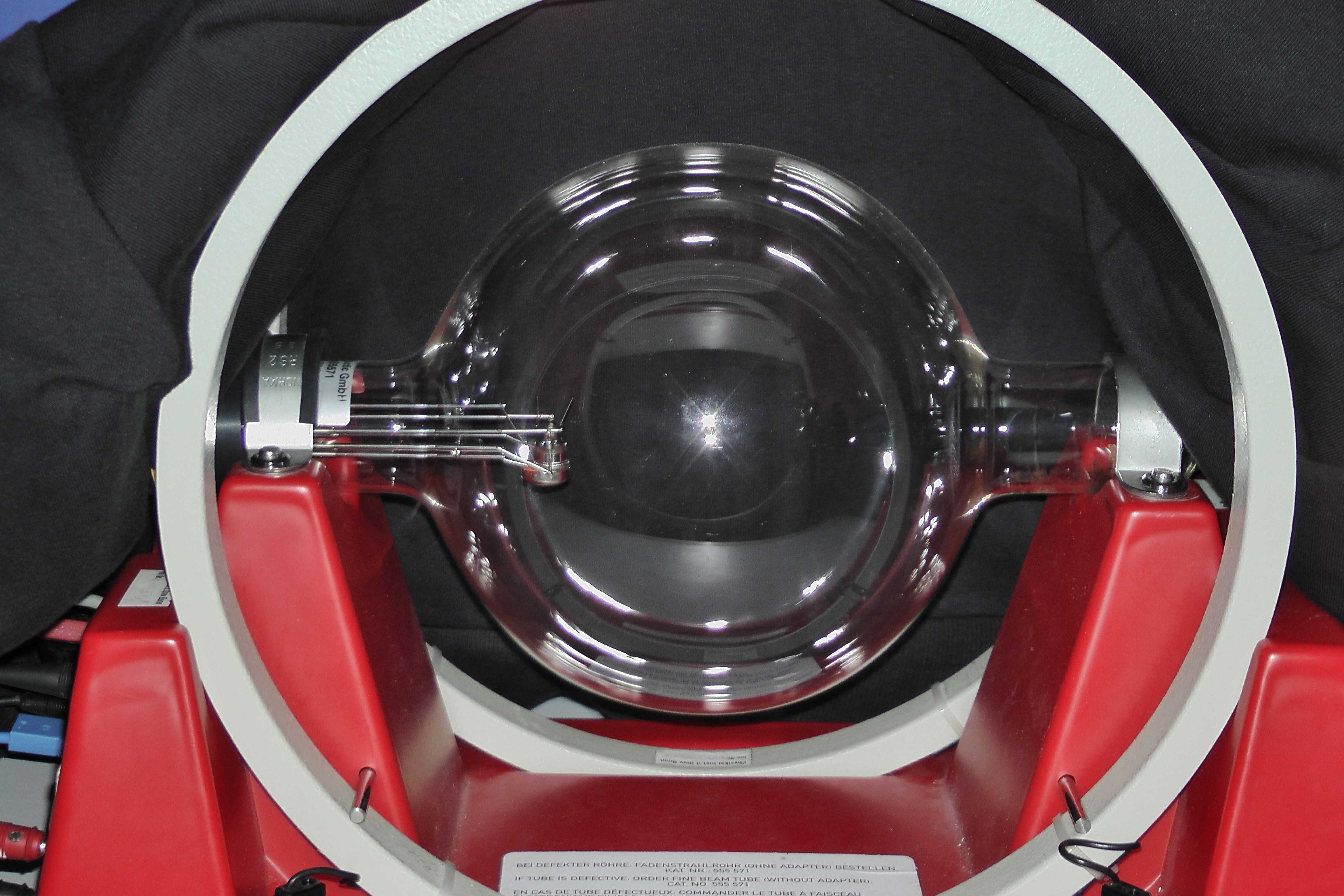} \includegraphics[width=0.53\textwidth]{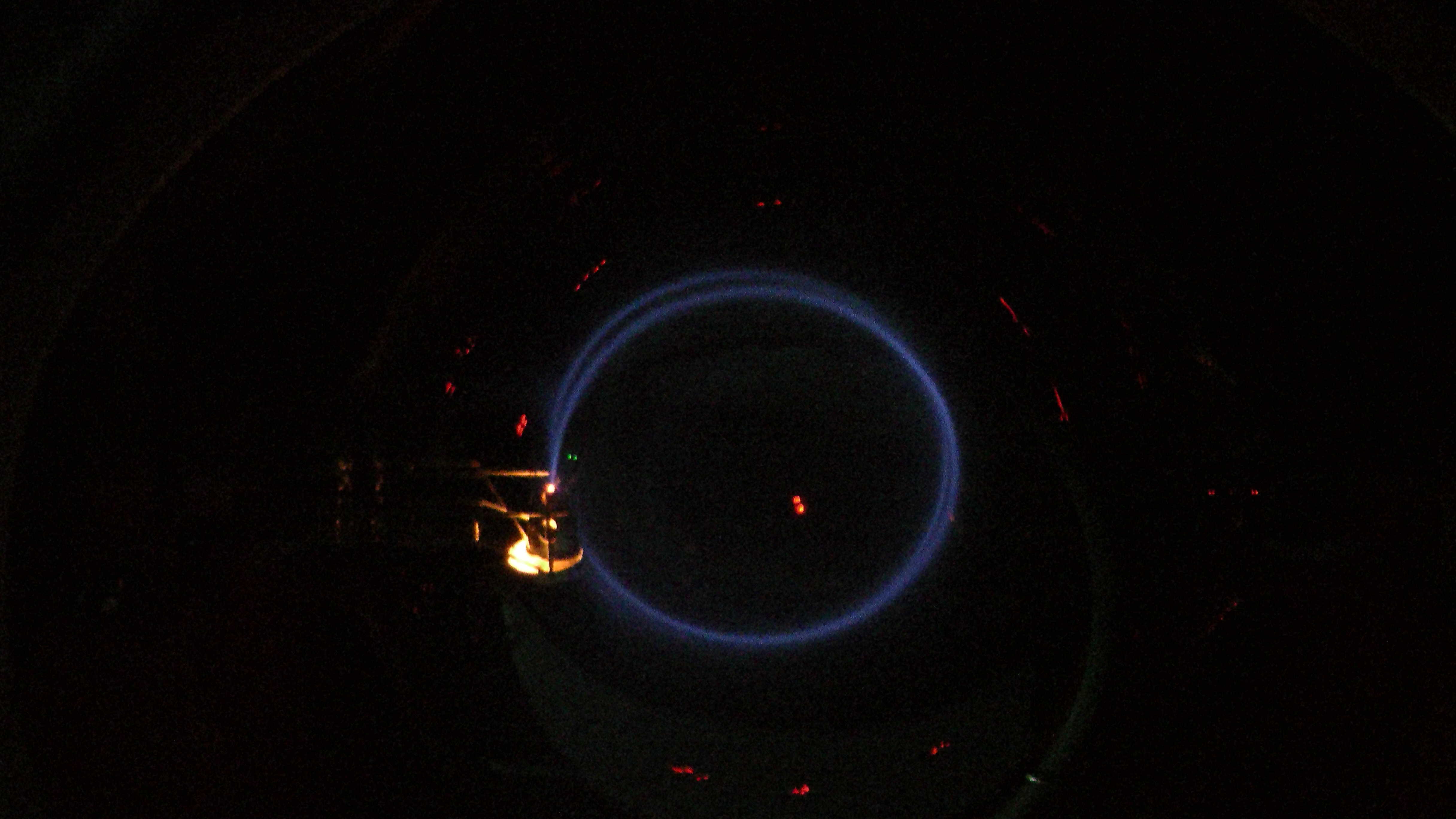}
	\caption{\small On the left, vacuum tube with electron gun inside and one set of coils upfront. On the right an electron ring inside
	the vacuum tube. }\label{fig:Helmholtz}
	\end{figure}

The Helmholtz coils used here were described in detail in Ref.~\cite{Dreiner:2016enl}.
The presentation in the show here is slightly different, as the emphasis is not on
introducing the electron, but instead to show the deflection of charged particles in a magnetic field.
See Fig.~\ref{fig:Helmholtz}.

\subsubsection{Curie Temperature}
\label{exp:Curie}

This experiment is used to explain why the magnetic field of the Earth cannot possibly be caused by a permanent magnet. The reason is that 
the interior of the earth is very hot (up to approximately $\SI{6000}{\kelvin}$) and every material loses its possible magnetic properties above 
some specific Curie temperature (Cobalt has the highest known Curie temperature with $\SI{1400}{\kelvin}$).

\paragraph{Theoretical and Historical Background}
Permanent magnetic properties are a consequence of collective behaviour, described for instance by the Ising model in a simple way. The 
material is considered to consist out of many (on the order of the number of atoms) tiny magnets or spins which favour a parallel alignment. 
These spins can exist in two different phases. At low temperatures the energetically favoured state of macroscopic order is assumed, 
thus all spins (or at least a majority) are aligned in the same direction making an overall magnetization possible. In practice magnetic materials 
form so called Weiss domains, each perfectly ordered within itself and susceptible to magnetic attraction, but disordered with respect to each 
other. The material therefore has to undergo additional treatment in order to become permanently magnetic. Above the critical temperature 
$T_C$ a state of high entropy with randomly oriented spins is preferred yielding zero net magnetization and a loss of attraction by another 
magnet. $T_C$ is named Curie temperature after Pierre Curie who discovered the phase transition in 1895.

\paragraph{Materials}
For the experiment a glass of water is required, to the inner side of which a permanent magnet has been glued. The water protects the magnet 
from heating up. The centerpiece of the experiment consists of a small piece of nickel (about $\SI{1}{\centi\meter}$ long and several millimeters 
wide and thick) wired to a tripod forming a pendulum. In addition a gas blow-torch is required for heating. Nickel has a Curie temperature of 
$T_C=\SI{627}{\kelvin}$.

\paragraph{Presentation}
In the beginning the piece of nickel, attracted by the magnet, is significantly deflected from its neutral pendulum position and rests just outside the
water container. When heated, the nickel loses its magnetization, is no longer attracted by the magnet and swings freely.  It then cools and with a bit 
of luck it is again sufficiently magnetized when it approaches the vessel and remains in the deflected position. Do not be disappointed if it does not 
come high enough on its way back, the experiment still served its purpose. Simply give the piece of nickel or the wire a push  and it will reconnect 
with the magnet. The main take away message is that the piece of nickel loses its permanent magnetic properties when heated and regains them 
when cooled, though of course it makes for a nicer show effect if no additional push is needed. A photo of the experiment is shown in 
Fig.~\ref{fig:curie_temperature} in the script, Sec.~\ref{sec:planetenoberflaeche-engl}.

\paragraph{Safety}
Open fire is very hot and so are the end piece of the blow torch, the piece of nickel and the lower parts of the wire. Do not burn yourself. Bear in mind 
that the heated parts remain hot for several minutes even though this might no longer be visible. Some of the authors speak from personal experience.

\subsubsection{Jacobs Ladder Experiment}
\label{exp:Hoerner}

\begin{figure}[t]
	 \centering
	 \includegraphics[width=0.75\textwidth]{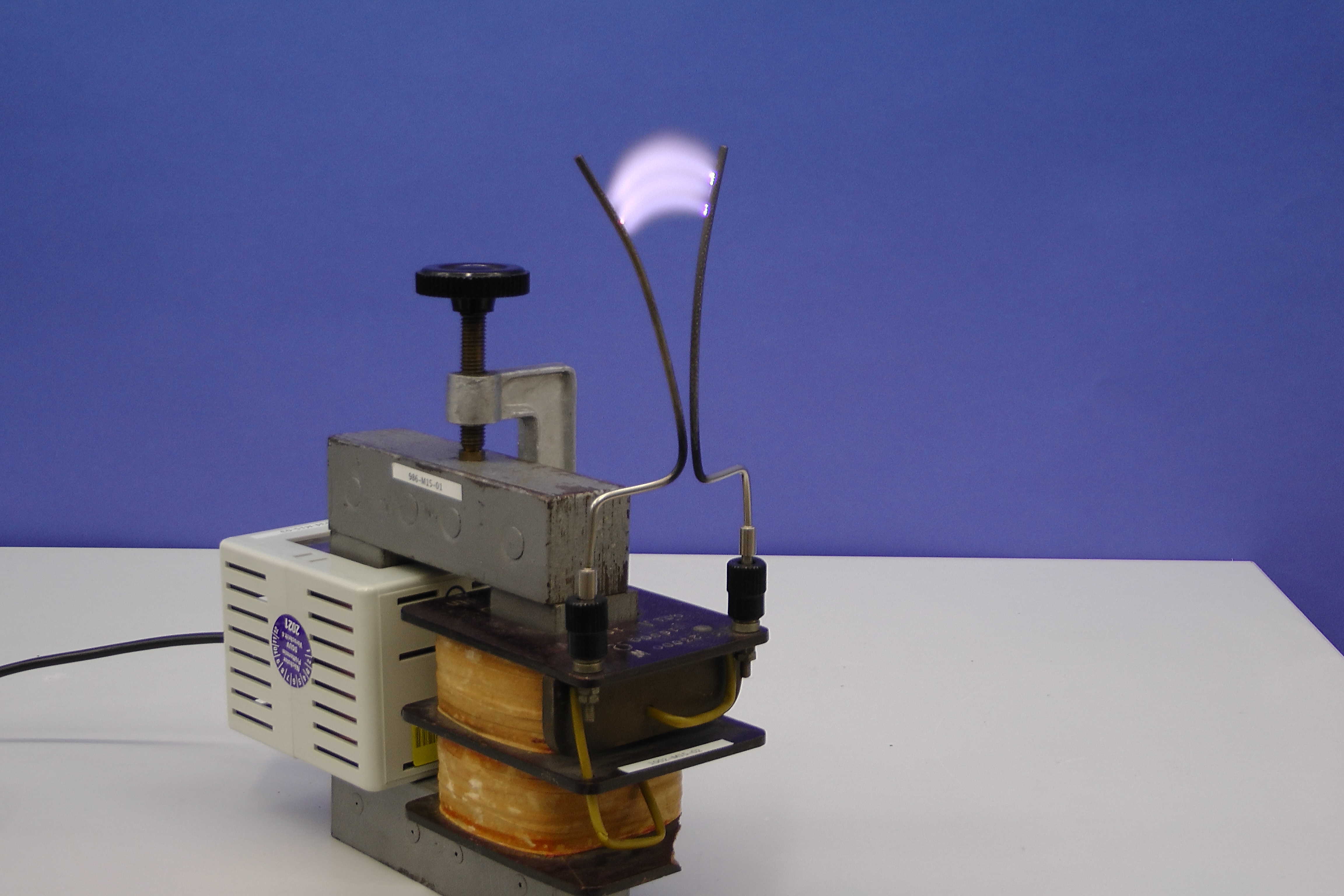}
	 \caption{Jacob's ladder experiment, as used in our show.}
	 \label{fig:jacobs-ladder}
	 \end{figure}

The Jacob's ladder experiment, see Fig.~\ref{fig:jacobs-ladder}, used here was described in detail in 
Ref.~\cite{Dreiner:2016enl}. Here we do not use it to demonstrate how to detect ionizing radiation, but 
instead to show a hot plasma, as might exist inside the Earth's core.

\subsubsection{Fire Tornado}
\label{exp:firetornado}

The fire tornado experiment used in this show is identical to that described in Ref.~\cite{Dreiner:2016enl}.

\subsubsection{Sandbox}
\label{exp:Sandbox}

The augmented reality (AR) sandbox used at the end of the show is more of an eye-catcher than an 
experiment. It is used to briefly explain how volcanos form craters.

\begin{figure}[h!]
	\centering
	\includegraphics[width=0.75\textwidth]{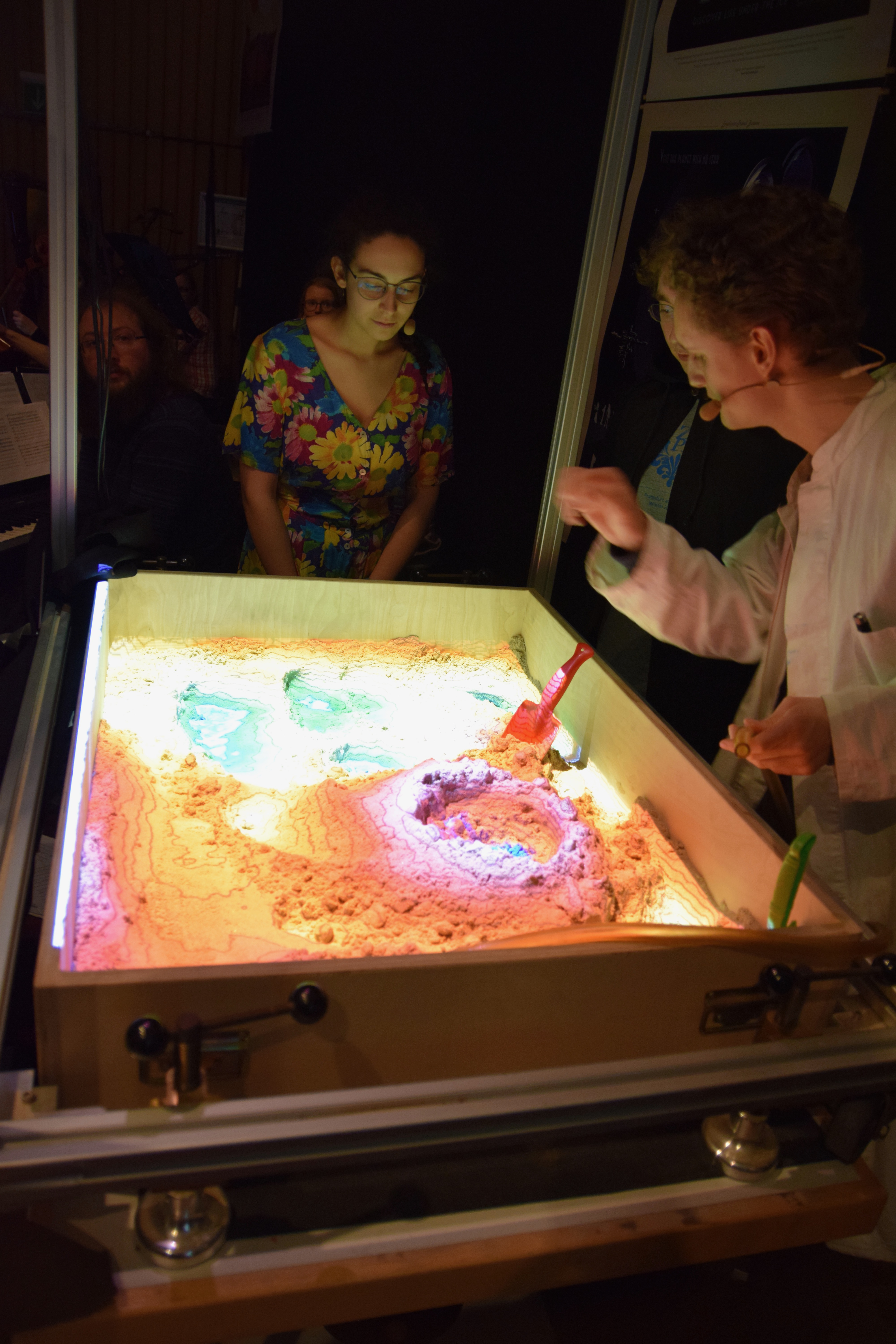}
	\caption{\small The projector above the box projects height contour lines onto the sand. The crater left by the punctured balloon can be seen
	on the front right.}
\end{figure}

\paragraph{History:}
The AR sandbox was developed by the UC Davis, W.M. Keck Center for Active Visualization in the Earth 
Sciences (\url{http://www.keckcaves.org/}), supported by the National Science Foundation as a way to 
teach Earth science concepts. The underlying software is free to use and they kindly provide detailed 
building instructions on their website (\url{https://arsandbox.ucdavis.edu/}).

\paragraph{Material:}
The AR sandbox continually reads the height profile of the sand using a \enquote{Microsoft Kinect} sensor 
which is mounted above the box. Contour lines of constant height are projected onto the sand using a projector, 
which is located above the sandbox. Additionally it is able to simulate water, which can be added to the 
observed simulation via gestures in the projection space.

To setup the AR sandbox one needs a sandbox, a rack which is able to hold a video projector and the 
\enquote{Microsoft Kinect} sensor, a computer to run the simulation as well as a mouse and keyboard to 
set up the program. To calibrate the simulation one also requires a flat cover for the box and a calibration 
tool, which can be built with a CD, a rod and a bit of paper. Please refer to 
\url{https://arsandbox.ucdavis.edu/} for building instructions and required components.

For the demonstration of the volcano we use a balloon buried under the sand, which can be inflated using 
a tube attached to the balloon and sticking out of the sand.

\paragraph{Presentation:}
In our show the actors pretend to form the landscape of the planet being bought. It is also briefly explained 
how a volcano forms a crater: A balloon buried under the sand is inflated using a tube; this forms a mountain. 
As the balloon is deflated the it leaves a crater in the sand, as in its expanded form it pushed some of the overlying 
sand outwards. This is similar to how a crater is formed by the outflowing magma.

\paragraph{Safety:}
The rack holding the projector and sensor should be quite robust. Apart from the obvious danger of it falling 
over, the calibration depends on the relative position of box and sensor. Especially since playing with the 
sandbox was quite popular with the audience after the show, a stable rack prevents the  need for frequent
re-calibration.


\subsection{Extra Experiments -- not used}
\subsubsection{Planeterella}
Jean Lilensten has developed a plasma physics demonstration experiment to show among other things polar lights.
He has discussed this device extensively in the literature, see Refs.~ \cite{Planeterella1,Lilensten:2012ku,Planeterella3}. It would
be worthwhile to show this effect in our show. However, we had some problems getting the device
to work well, especially so that it is bright enough to show on-stage to a large audience.

\subsubsection{Tsunami}
The tsunami demo-experiment we use consists of a plexiglas aquarium, which is about 3m long, 40cm high and 30cm wide.
On one side  we have installed a ramp, to simulate a beach run-up. This can be seen on the right in the photo. We could now 
create simple surface waves and see how they run up the beach. On the left we could also vertically insert a plexiglas plate. 
Behind this we could fill the water higher than the rest of the aquarium. Then we could pull out the separator rapidly, leaving a 
raised water column, similar to an earthquake or if a large comet hits the ocean. When the resulting disturbance reached the 
beach we got a very much larger wave.

\begin{figure}[htb]
	\centering
	\includegraphics[width=0.75\textwidth]{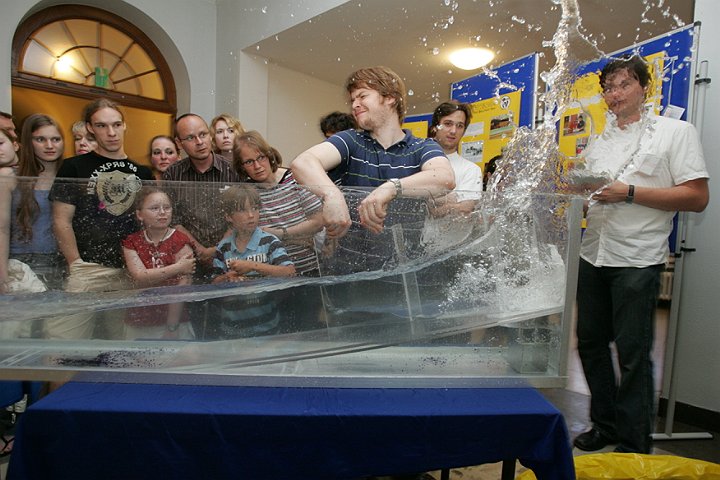}
	\caption{\small The tsunami experiment in action! Photo: V. Lannert.}
\end{figure}

\end{appendix}

\end{document}